\documentclass[twocolumn,aps,prd,longbibliography]{revtex4-2}
\usepackage{dcolumn}
\usepackage{enumerate}
\usepackage{natbib}
\usepackage{amsmath}
\usepackage{amssymb}
\usepackage{graphicx}
\usepackage{graphics}

\begin{document}

\title{Interplay between geostrophic vortices and inertial waves in precession-driven turbulence}
\author{F. Pizzi}
\affiliation{Institute of Fluid Dynamics, Helmholtz-Zentrum Dresden-Rossendorf, Bautzner Landstrasse 400, D-01328 Dresden, Germany}%
\affiliation{Department of Aerodynamics and Fluid Mechanics, Brandenburg University of Technology, Cottbus-Senftenberg, 03046 Cottbus, Germany}
 \email{f.pizzi@hzdr.de}
\author{G. Mamatsashvili}
\affiliation{Institute of Fluid Dynamics, Helmholtz-Zentrum Dresden-Rossendorf, Bautzner Landstrasse 400, D-01328 Dresden, Germany}%
\affiliation{E. Kharadze Georgian National Astrophysical Observatory, Abastumani 0301, Georgia}
\author{A.~J. Barker}
\affiliation{Department of Applied Mathematics, School of Mathematics, University of Leeds, Leeds, LS2 9JT, UK}
\author{A. Giesecke}
\author{F. Stefani}
\affiliation{Institute of Fluid Dynamics, Helmholtz-Zentrum Dresden-Rossendorf, Bautzner Landstrasse 400, D-01328 Dresden, Germany}%



\begin{abstract}
The properties of rotating turbulence driven by precession are studied using direct numerical simulations and analysis of the underlying dynamical processes in Fourier space. The study is carried out in the local rotating coordinate frame, where precession gives rise to a background shear flow, which becomes linearly unstable and breaks down into turbulence. We observe that this precession-driven turbulence is in general characterized by coexisting two dimensional (2D) columnar vortices and three dimensional (3D) inertial waves, whose relative energies depend on the precession parameter $Po$. The vortices resemble the typical condensates of geostrophic turbulence, are aligned along the rotation axis (with zero wavenumber in this direction, $k_z=0$) and are fed by the 3D waves through nonlinear transfer of energy, while the waves (with $k_z\neq0$) in turn are directly fed by the precessional instability of the background flow. The vortices themselves undergo inverse cascade of energy and exhibit anisotropy in Fourier space. For small $Po<0.1$ and sufficiently high Reynolds numbers, the typical regime for most geo-and astrophysical applications, the flow exhibits strongly oscillatory (bursty) evolution due to the alternation of vortices and small-scale waves. On the other hand, at larger $Po>0.1$ turbulence is quasi-steady with only mild fluctuations, the coexisting columnar vortices and waves in this state give rise to a split (simultaneous inverse and forward) cascade. Increasing the precession magnitude causes a reinforcement of waves relative to vortices with the energy spectra approaching the Kolmogorov scaling and, therefore, the precession mechanism counteracts the effects of the rotation.
\end{abstract}

\maketitle

%

\section{\label{sec:level1}Introduction}

Rotating turbulence is an ubiquitous phenomenon in a broad context ranging from astrophysical and geophysical flows \cite{Barnes2001, Cho2008, Knobloch1981, Zahn1992} to industrial applications \cite{Dumitrescu2004, He2021}. Understanding the impact of rotation on the turbulence dynamics is far from trivial due to the complexity of the nonlinear processes involved. In general, when a fluid is subjected to rotational motion, the nonlinear interactions are affected by the Coriolis force whose strength is quantified by the Rossby number, $Ro$ (the ratio of the advection time-scale to the rotation time-scale) and the Reynolds number, $Re$ (the ratio of advective to viscous time-scale). If the Coriolis force is strong enough the formation of coherent columnar vortices occurs inside the fluid flow. This phenomenon has been observed in experimental campaigns for several systems such as oscillating grids \cite{Hopfinger1982}, for decaying turbulence \cite{Staplehurst2008, Lamriben2011}, forced turbulence \cite{Campagne2014, Gallet2014, Gallet2015}, and turbulent convection \cite{Kunnen2010,Fernando1991}. Also numerical simulations have been instrumental to analyze such tendency in a myriad of cases \cite{Bardina1985, Mansour1991, Mansour1992, Hossain1994, Yeung1994, Smith1996, Chen2003}, making use of large eddy simulations \cite{Squires1993, Bartello1994, Yang2004} and even turbulence models \cite{Elena1996, Spalart1997}.

The emergence of columnar vortices aligned along the flow rotation axis is accompanied by inertial waves which are inherent to rotating fluids. Their frequency magnitude ranges between zero and twice the rotation rate $\Omega$ of the objects \cite{greenspan_book}. The dynamics and mutual couplings between these two basic types of modes largely depend on $Re$ and $Ro$. The results of the asymptotic analysis at $Ro \ll 1$ indicate that three-dimensional (3D) inertial waves and two-dimensional (2D) vortices are essentially decoupled and evolve independently: vortices undergo inverse cascade, while the wave energy cascades forward through resonant wave interactions in the regime of weakly nonlinear inertial wave turbulence \cite{Greenspan1969, Smith1999, Galtier2003, Alexakis2018}. However, this picture does not carry over to moderate Rossby numbers $Ro\sim 0.1$, where the situation is much more complex, since 3D inertial waves (the so called fast modes) and 2D vortices (also called slow modes) can coexist and be dynamically coupled. In this case, asymptotic analysis cannot be used and more complex mathematical models have been proposed to explain the geostrophic vortices-wave interaction, such as the quartic instability \cite{Brunet2020} or near-resonant instability \cite{Lereun2020}. However  progress can be made mainly by numerical simulations. Several works are devoted to the study of these two manifolds and their interactions for forced rotating turbulence \cite{Smith1999, Muller2007, Buzzicotti2018, Khlifi2018} and also for convective and rotating turbulence \cite{Rubio2014, Knobloch1998}.

Indeed other forcing mechanisms have been shown to be characterized by this interplay of vortices and waves, such as elliptical instabilities \cite{Barker2013, Favier2015, Barker2016b, Lereun_2017, Lereun_2019} and tidal forcing \cite{Barker2010, Barker2016, Lereun2018, Morize_2010, Tilgner2007}. In this respect, the precession-driven dynamics represents a possible candidate for the development of both 3D waves with embedded 2D vortices \cite{Khlifi2018, Salhi2020} but so far these studies does not investigate a wide range of governing parameters. Other works were devoted to the stability analysis of the precession flows \cite{Salhi2009}. 

The modified local Cartesian model of a precession-driven flow was proposed by Mason and Kerswell \cite{Mason2002} and later used by Barker \cite{Barker2016} to study its nonlinear evolution. In the first paper, rigid and stress-free axial boundaries in the vertical direction were used , while in the second paper an unbounded precessional flow was considered in the planetary context, employing the decomposition of perturbations into shearing waves. In this paper, we follow primarily the approach of Barker \cite{Barker2016}, who analyzed the occurrence of vortices, function of the precession parameter (Poincar\'{e} number), including energy spectrum and dissipation properties. The main advantage of the local model is that it allows high-resolution study of linear and nonlinear dynamical processes in precession-driven flows, which is much more challenging in global models. Also, this model allows to focus only on the dynamics of the bulk flow itself avoiding the complications due to boundary layers. This is important for gaining a deeper understanding of perturbation evolution in unbounded precessional flows and then, comparing with the global simulations, for pinning down specific effects caused by boundaries.  

In this paper, we continue this path and investigate in detail the underlying dynamical processes in the turbulence of precessional flow in the local model. We decompose perturbations into 2D and 3D manifolds and analyse their dynamics and interplay in Fourier space. Our main goal is to address and clarify several key questions: (i) how the presence and properties of columnar vortices depend on the precession strength and Reynolds number (here defined as the inverse  Ekman number), (ii) what are the mechanisms for the formation of the columnar vortices in precessing driven flows, in particular, how their dynamics are affected by precessional instability of inertial waves, that is, if there are effective nonlinear transfers (coupling) between vortices and the waves; (iii) what are the dominant nonlinear processes (channels) in this vortex-wave system, i.e., the interaction of 2D-3D modes or 2D-2D modes (inverse cascade), (iv) in terms of total shell-average spectral analysis, what type of cascades (inverse, forward) occur and what kind of spectra characterize precessional flows.

The paper is organized as follows: in Section \ref{sec:math} the local model and governing equations in physical and Fourier space are presented, and numerical methods introduced. Section \ref{sec:general} presents general evolution of the volume-averaged kinetic energy and dynamical terms as well as flow structure. In Section \ref{subsec:2d_3d} we investigate the nonlinear dynamics of 2D vortices and 3D inertial waves and nonlinear interaction between them in Fourier space. In this section we also characterize turbulent dissipation as a function of precession parameter $Po$. Discussions and the future perspectives are presented in Section \ref{sec:conclusions}.

\section{Model and equations}\label{sec:math}

We consider a precessional flow in a local rotating Cartesian coordinate frame (also referred to as the `mantle frame' of a precessing planet) in which the mean total angular velocity of fluid rotation $\boldsymbol{\Omega}=\Omega\boldsymbol{e}_z$ is directed along the $z$-axis. In this frame, the equations of motion for an incompressible viscous fluid take the form (see a detailed derivation in Refs. \cite{Mason2002, Barker2016}): 
\begin{multline}\label{eq:ns}
\frac{ \partial \boldsymbol{U}} {\partial t} + \boldsymbol{U} \cdot \boldsymbol{\nabla} \boldsymbol{U} + 2\Omega(\boldsymbol{e}_z+\boldsymbol{\epsilon}(t))\times\boldsymbol{U}= \\ 
-\frac{1}{\rho}\boldsymbol{\nabla} P +  \nu\boldsymbol{\nabla}^{2}\boldsymbol{U}+2z\Omega^2\boldsymbol{\epsilon}(t),
\end{multline}
\begin{equation}
\boldsymbol{\nabla} \cdot \boldsymbol{U} = 0,
\end{equation}
where $\boldsymbol{U}$ is the velocity in this frame, $\rho$ is the spatially uniform density and $P$ is the modified pressure equal to the sum of thermal pressure and the centrifugal potential. The last two terms on the left-hand side in the brackets are the Coriolis and the Poincar\'{e} forces, respectively, and $\boldsymbol{\epsilon}(t)=Po({\rm cos}(\Omega t), -{\rm sin}(\Omega t), 0)^T$ is the precession vector with $Po$ being the Poincar\'{e} number characterizing the strength of the precession force. The last term on the right-hand side is the second part of the precession force with vertical shear, which is the main cause of hydrodynamic instability in the system, refereed to as the precessional instability \cite{Kerswell1993}. $\nu$ is the constant kinematic viscosity. 

The basic precessional shear flow in this local frame represents an unbounded horizontal flow with a linear shear along the vertical $z$-axis and oscillating in time $t$, i.e., $\boldsymbol{U}_b=(U_{bx}(z,t), U_{by}(z,t), 0)$ with the components given by \cite{Mason2002, Barker2016}
$$ \boldsymbol{U}_b= -2 \Omega\: Po \: \left(
  \begin{array}{c c c}
     0 & 0 & \sin(\Omega t)\\
     0 & 0 & \cos(\Omega t)\\
     0 & 0 & 0\\
  \end{array} \right)
  \left( \begin{array}{c}
  x \\
  y \\
  z\\
  \end{array} \right) \equiv M \: \boldsymbol{r},$$
where $\boldsymbol{r}=(x,y,z)$ is the local position vector. Our local model deals with perturbations to this basic flow,  $\boldsymbol{u}=\boldsymbol{U}-\boldsymbol{U}_{b}$, for which from Eq. (\ref{eq:ns}) we obtain the governing equation:
\begin{multline}\label{eq:ns_pert}
\frac{\partial \boldsymbol{u}}{\partial t} + \boldsymbol{u} \cdot \boldsymbol{\nabla u} = - \frac{1}{\rho}\boldsymbol{\nabla}P+\nu \boldsymbol{\nabla}^{2} \boldsymbol{u} - 2\Omega \boldsymbol{e}_z \times \boldsymbol{u} - \\2\Omega \boldsymbol{\varepsilon}(t) \times \boldsymbol{u} - M\boldsymbol{u} -M\boldsymbol{r} \cdot \nabla \boldsymbol{u}, 
\end{multline}
where the last three terms on the rhs are related to precession and proportional to $Po$. The flow field is confined in a cubic box with the same length $L$ in each direction,  $L_x=L_y=L_z=L$. In other words, both horizontal and vertical aspect ratios of the box are chosen to be equal to one in this paper. Varying these aspect ratios affects the linearly unstable modes that can be excited in the flow and the properties of the vortices \cite{Barker2013}. 

Below we use the non-dimensionalization of the variables by taking $\Omega^{-1}$ as the unit of time, box size $L$ as the unit of length, $\Omega L$ as the unit of velocity, $\rho L^2\Omega^2$ as the unit of pressure and perturbation kinetic energy density $E=\rho \boldsymbol{u}^2/2$. The key parameters governing a precession-driven flow are the Reynolds number (inverse Ekman) defined as 
\[
Re = \frac{\Omega L^{2}}{\nu}
\]
and the Poincar\'{e} number $Po$ introduced above.

\begin{figure}
\centering
\includegraphics[scale=0.5]{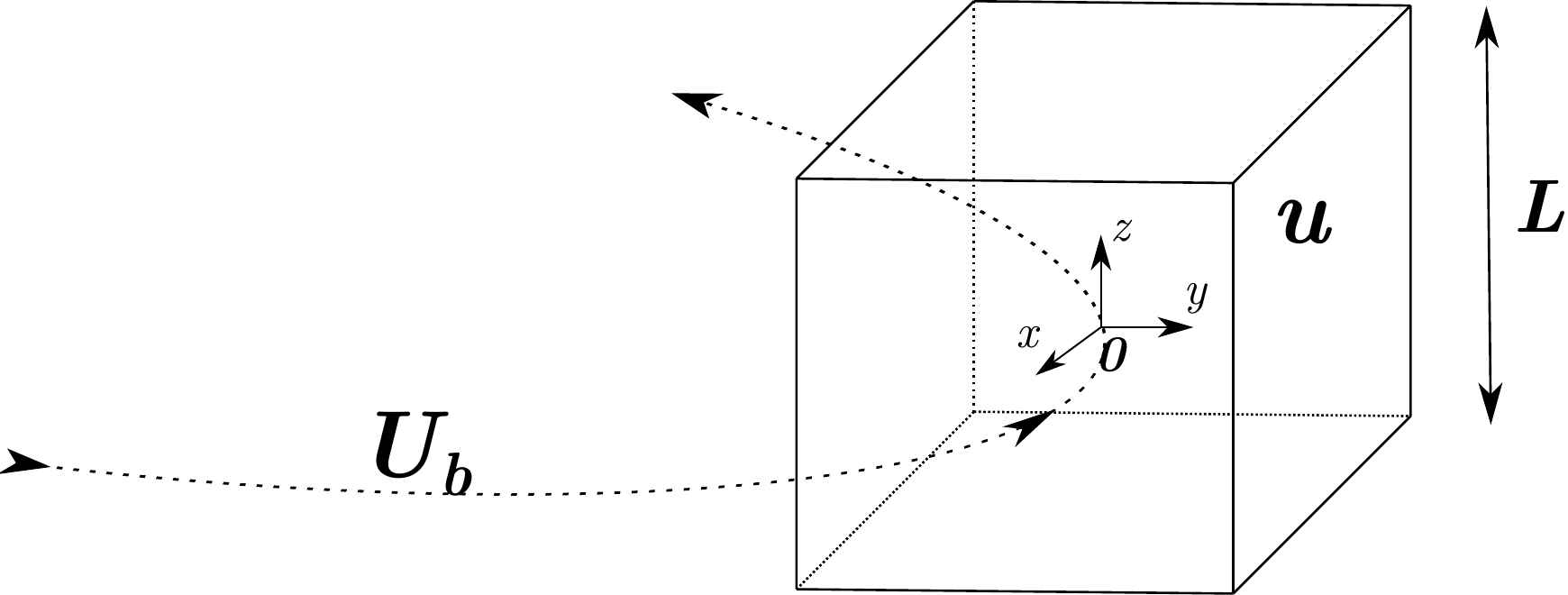}
\caption{Sketch of the periodic cubic domain with length $L$ in each direction where the base flow inside it is $\boldsymbol{U}_b$ with superimposed perturbation velocity $\boldsymbol{u}$.}\label{fig:0}
\end{figure}

\begin{table}
 \begin{tabular}{c c c} 
    &  $Re=10^{3.5}$ &  \\
 \hline \hline
 $Po$ & $N$ & $\langle E \rangle$ \\ 
 \hline
  $0.01$ & 64 & - \\ 
 
 $0.025$ & 64 & - \\
 
 $0.05$ & 64 & - \\
 
 $0.075$ & 64 & - \\ 
 
 $0.1$ & 64 & - \\
 
 $0.125$ & 64 & - \\
 
 $0.15$ & 64 & - \\ 

 $0.175$ & 64 &  - \\

 $0.2$ & 64 & - \\

 $0.225$ & 64 & -  \\ 

 $0.25$ & 64 & -  \\
 
 $0.3$ & 64 & $6.09 \times 10^{-5}$  \\
 \hline  \hline
\end{tabular}
\begin{tabular}{c c c} 
 & $Re=10^{4}$ & \\
 \hline \hline
 $Po$ & $N$ & $\langle E \rangle$\\ 
 \hline
  $0.01$ & $64$ & - \\ 
 
  $0.025$ & 64 & -\\
 
  $0.05$ & 64 & - \\
 
  $0.075$ & 128 & - \\ 
 
  $0.1$ & 128 & - \\
 
  $0.125$ & 128 & $1.89 \times 10^{-5}$  \\
 
  $0.15$ & 128 & $6.04 \times 10^{-5}$  \\ 

  $0.175$ & 128 & $1.42 \times 10^{-4}$   \\

  $0.2$ & 128 & $3.54 \times 10^{-4}$   \\

  $0.225$ & 128 & $6.11 \times 10^{-4}$   \\ 

  $0.25$ & 128 & $9.94 \times 10^{-4}$   \\
 
  $0.3$ & 128 & $2.10 \times 10^{-3}$   \\
 \hline  \hline
\end{tabular}

\vspace{0.4cm}

\begin{tabular}{c c c} 
 &  $Re=10^{4.5}$  & \\
 \hline \hline
 $Po$ & $N$ & $\langle E \rangle$\\  
 \hline
 $0.01$ & 128 & -\\ 
 
 $0.025$ & 128 & -\\
 
 $0.05$ & 128 & -\\
 
 $0.075$ & 128 & $3.82 \times 10^{-5}$ \\ 
 
 $0.1$ & 128 & $1.13 \times 10^{-4}$  \\
 
 $0.125$ & 128 & $4.93 \times 10^{-4}$ \\
 
 $0.15$ & 256 & $1.30 \times 10^{-3}$ \\ 

 $0.175$ & 256 & $2.30 \times 10^{-3}$  \\

 $0.2$ & 256 & $5.40 \times 10^{-3}$  \\

 $0.225$ & 256 & $6.20 \times 10^{-3}$ \\ 

 $0.25$ & 256 & $6.40 \times 10^{-3}$ \\
 
 $0.3$ & 256 & $9.00 \times 10^{-3}$  \\  
 
 $0.5$ & 256 & $1.25 \times 10^{-2}$  \\
 \hline  \hline
\end{tabular}
\vspace{0.1cm}
\begin{tabular}{c c c} 
 & $Re=10^{5}$ &   \\
 \hline \hline
 $Po$ & $N$ & $\langle E \rangle$ \\ 
 \hline
 $0.01$ & 256 & - \\ 
 
 $0.025$ & 256 & - \\
 
 $0.05$ & 256 & $4.55 \times 10^{-5}$ \\
 
 $0.075$ & 256 & $2.37 \times 10^{-4}$ \\ 
 
 $0.1$ & 256 & $1.10 \times 10^{-3}$  \\
 
 $0.125$ & 256 & $5.90 \times 10^{-3}$ \\
 
 $0.15$ & 256 & $6.80 \times 10^{-3}$ \\ 

 $0.175$ & 256 & $7.50 \times 10^{-3}$ \\

 $0.2$ & 256 & $8.90 \times 10^{-3}$ \\

 $0.225$ & 256 & $1.02 \times 10^{-2}$ \\ 

 $0.25$ & 256 & $1.14 \times 10^{-2}$ \\
 
 $0.3$ & 256 & $1.15 \times 10^{-2}$ \\
 \hline  \hline
\end{tabular}
  \caption{List of all simulations performed in the present work. Each subtable corresponds to a specific Reynolds number and various Poincar\'{e} numbers $Po$ (first column). The second column shows numerical resolution $N$ (before dealiasing), which is the same in each direction, $N_x=N_y=N_z=N$ (total number of point is $N^{3}$). The third column shows the time- and volume-averaged kinetic energy $\langle E \rangle$. Runs marked with hyphen are not sustained and quickly decay. Notice that for $Re=10^{4.5}$ we have run also a simulation at very large $Po=0.5$.}
\end{table}

\subsection{Governing equations in Fourier space}

Our main goal is to perform the spectral analysis of precession-driven turbulence in Fourier (wavenumber $\boldsymbol{k}$-) space in order to understand dynamical processes (energy injection and nonlinear transfers) underlying its sustenance and evolution. To this end, following \cite{Barker2013, Barker2016}, we decompose the perturbations into spatial Fourier modes (shearing waves) with time-dependent wavevectors $\boldsymbol{k}(t)$,
\begin{equation}\label{eq:fourier}
f\left(\boldsymbol{r}, t\right) = \sum_{\boldsymbol{k}} \bar{f} \left( \boldsymbol{k}(t),t\right) e^{i\boldsymbol{k}(t)\cdot \boldsymbol{r}},
\end{equation}
where $f \equiv (\boldsymbol{u},P)$ and their Fourier transforms are $\bar{f} \equiv (\bar{\boldsymbol{u}},\bar{P})$. In the transformation (\ref{eq:fourier}), the wavevector of modes oscillates in time, 
\begin{equation}\label{eq:wavevector}
\boldsymbol{k}(t)=(k_{x0}, k_{y0}, k_{z0}+2Po(-k_{x0}{\rm cos}(t)+k_{y0}{\rm sin}(t)))^T,
\end{equation}
about its constant average value $\langle \boldsymbol{k}(t) \rangle=(k_{x0},k_{y0},k_{z0})$ due to the periodic time-variation of the basic precessional flow $\boldsymbol{U}_b$. Substituting Eq. (\ref{eq:fourier}) into Eq. (\ref{eq:ns_pert}) and taking into account the above non-dimensionalization, we obtain the following equation governing the evolution of velocity amplitude
\begin{multline}\label{eq:fourier_ns}
\frac{d\bar{\boldsymbol{u}}}{dt}  = -i\boldsymbol{k}(t)\bar{P} - \frac{k^2}{Re}\bar{\boldsymbol{u}} - 2 \boldsymbol{e}_z\times\bar{\boldsymbol{u}} \\ -2 \boldsymbol{\varepsilon} \left(t \right) \times \bar{\boldsymbol{u}} - M(t) \bar{\boldsymbol{u}} + \boldsymbol{Q} \: ,
\end{multline}
\begin{equation}
\boldsymbol{k}(t)\cdot \bar{\boldsymbol{u}} = 0.
\end{equation}
Note that the wavevector $\boldsymbol{k}(t)$ as given by expression (\ref{eq:wavevector}) satisfies the ordinary differential equation
\begin{eqnarray}\label{eq:fourier_k}
\frac{d \boldsymbol{k}}{dt}  = - M^{T} \boldsymbol{k}
\end{eqnarray}
and as a result the last term on the rhs of Eq. (\ref{eq:ns_pert}) related to the basic flow has disappeared when substituting (\ref{eq:fourier}) into it. The term $\boldsymbol{Q}(\boldsymbol{k},t)$ on the rhs of Eq. (\ref{eq:fourier_ns}) represents the Fourier transform of the nonlinear advection term $\boldsymbol{u} \cdot \boldsymbol{\nabla u}=\nabla\cdot(\boldsymbol{u}\boldsymbol{u})$ in the original Eq. (\ref{eq:ns_pert}) and is given by convolution \citep{George2016, Buzzicotti2018} 
\begin{eqnarray}
Q_m(\boldsymbol{k},t)=-i\sum_n\sum_{\boldsymbol{k}'} k_n\bar{u}_m(\boldsymbol{k}',t)\bar{u}_n(\boldsymbol{k}-\boldsymbol{k}',t),
\end{eqnarray}
where the indices $(m,n)=(x,y,z)$. This term describes the net effect of nonlinear triadic interactions (transfers) among a mode $\boldsymbol{k}$ with two others $\boldsymbol{k}-\boldsymbol{k}'$ and $\boldsymbol{k}'$ and thus plays a key role in turbulence dynamics.

Multiplying both sides of Eq. (\ref{eq:fourier_ns}) by the complex conjugate of spectral velocity $\bar{\boldsymbol{u}}^{\ast}$, the contribution from Coriolis and part of the Poincar\'{e} force in the total kinetic energy of a mode cancel out, since they do not do any work on the flow, $\bar{\boldsymbol{u}}^{\ast}\cdot (2\boldsymbol{e}_z \times \bar{\boldsymbol{u}}-2 \boldsymbol{\varepsilon}\left(t \right) \times \bar{\boldsymbol{u}})=0$, and as a result we obtain the equation for the (non-dimensional) spectral kinetic energy density $E=|\bar{\boldsymbol{u}}|^2/2$ in Fourier space as
\begin{equation}\label{eq:energy_fourier}
\frac{dE}{dt}=\underbrace{\frac{1}{2}\left[\bar{\boldsymbol{u}}^{\ast} \left(\boldsymbol{M} \bar{\boldsymbol{u}}\right) + \bar{\boldsymbol{u}} \left(\boldsymbol{M} \bar{\boldsymbol{u}}\right)^{\ast}\right]}_{injection}  +  \underbrace{\frac{1}{2}\left[\bar{\boldsymbol{u}}^{\ast} \boldsymbol{Q} + \bar{\boldsymbol{u}} \boldsymbol{Q}^{\ast}\right]}_{nonlinear~transfer} - \underbrace{\frac{2k^{2}}{Re}E}_{dissipation}.
\end{equation}
The pressure term also cancels out since $\bar{\boldsymbol{u}}^{\ast}\cdot\boldsymbol{k}(t)\bar{P}=0$. Thus, the rhs of Eq. (\ref{eq:energy_fourier}) contains three main terms:
\begin{itemize}
\item
{\bf Injection}  
$$
A\equiv\frac{1}{2}\left[\bar{\boldsymbol{u}}^{\ast} \left(\boldsymbol{M} \bar{\boldsymbol{u}}\right) + \bar{\boldsymbol{u}} \left(\boldsymbol{M} \bar{\boldsymbol{u}}\right)^{\ast}\right],
$$
which is of linear origin, being determined by the matrix $\boldsymbol{M}$, i.e., by the precessing background flow and describes energy exchange between the perturbations and that flow. If $A>0$, kinetic energy is injected from the flow into inertial wave modes and hence they grow, which is basically due to precessional instability \cite{Kerswell1993, Mason2002, Naing2011, Barker2016}, whereas at $A<0$ modes give energy to the flow and decay.
\item
{\bf Nonlinear transfer}
\[
NL\equiv \frac{1}{2}\left[\bar{\boldsymbol{u}}^{\ast} \boldsymbol{Q} + \bar{\boldsymbol{u}} \boldsymbol{Q}^{\ast}\right]
\]
describes transfer (cascade) of spectral kinetic energy among modes with different wavenumbers in Fourier space due to nonlinearity. The net effect of this term in the spectral energy budget summed over all wavenumbers is zero i.e.,
\[
\sum_{\boldsymbol{k}} NL(\boldsymbol{k},t) = 0,
\]
which follows from vanishing of the nonlinear advection term
in the total kinetic energy equation integrated in physical space. Thus, the main effect of the nonlinear term is only to redistribute energy among modes that is injected from the basic flow due to $A$, while keeping the total spectral kinetic energy summed over all wavenumbers unchanged. Although the nonlinear transfers $NL$ produce no net energy for perturbations, they play a central role in the turbulence dynamics together with the injection term $A$. The latter is thus the only source of new energy for perturbations drawn from the infinite reservoir of the background precessional flow. Due to this, below we focus on these two main dynamical terms -- linear injection and nonlinear transfer functions, compute their spectra and analyse how they operate in Fourier space in the presence of precession instability using the tools of Refs. \cite{George2014, George2016}. 
\item
{\bf Viscous dissipation}
\[
D\equiv - \frac{2k^{2}}{Re}E
\]
is negative definite and describes the dissipation of kinetic energy due to viscosity. 
\end{itemize}

\subsection{2D-3D decomposition}\label{subsec:2d_3d}

In the present section we follow a widely used approach in the theory of rotating anisotropic turbulence \cite{Smith1999, Rubio2014, Buzzicotti2018, Khlifi2018, Alexakis2018, Barker2013} and  decompose the flow field into 2D and 3D modes in Fourier space to better characterize this anisotropy between horizontal and vertical motions. This choice is motivated by the observation of two main types of perturbations: vortices, which are essentially 2D structures, and 3D inertial waves in rotating turbulent flows with external forcing such as libration, elliptical instability \cite{Lereun_2017, Barker2013}, precession \cite{Khlifi2018, Barker2016} and other artificial types of forcing concentrated at a particular wavenumber \cite{Muller2007, Buzzicotti2018, Sesha2020}. The 2D vortical modes, also called \textit{slow} (geostrophic) modes, have dominant horizontal velocity over the vertical one and are almost uniform, or aligned along the $z-$axis, i.e., their wavenumber parallel to this axis is zero $k_z=0$. This slow manifold is also referred to as 2D and three-component (2D3C) field in the literature, since it varies only in the horizontal $(x,y)$-plane perpendicular to the rotation axis, but still involves all three components of velocity with the horizontal one being dominant. On the other hand, 3D inertial wave modes,  called \textit{fast} (with nonzero frequency $\omega=\pm 2\Omega k_z/k$) modes, have comparable horizontal and vertical velocities and vary along $z$-axis, i.e., parallel wavenumber is nonzero $k_z \neq 0$. In the present case of the basic precessional flow, the $k_z(t)$ wavenumber of modes oscillates in time according to Eq. (\ref{eq:wavevector}), so we classify 2D and 3D modes as having $\langle k_z(t) \rangle=k_{z0}=0$ and $k_{z0} \neq 0$, respectively. Specifically, these two mode manifolds are
\begin{equation}
\Psi_{2D}=\left\lbrace \boldsymbol{k} \: | \: k_{x}, k_{y}; k_{z}=0 \right\rbrace,~~~\Psi_{3D}=\left\lbrace \boldsymbol{k} \: | \: k_{x}, k_{y}; k_{z} \neq 0\right\rbrace.
\end{equation}
and the spectral velocities for 2D vortices and 3D inertial waves can be defined as
\begin{equation}
\bar{\boldsymbol{u}}(\boldsymbol{k})=\left\{ 
  \begin{array}{ c l }
   \bar{\boldsymbol{u}}_{2D}(\boldsymbol{k})  & \quad \textrm{if } \quad \boldsymbol{k} \in \Psi_{2D}\\
    \bar{\boldsymbol{u}}_{3D}(\boldsymbol{k}) & \quad \textrm{if } \quad \boldsymbol{k} \in \Psi_{3D}.
  \end{array}
\right.
\end{equation}  
Note that the definition of 2D manifolds here differs from the Taylor-Proudman problem since it does not necessarily have vanishing vertical flow. Indeed, velocity for both the 3D and 2D modes can be decomposed in turn into respective horizontal $\bar{\boldsymbol{u}}_h=(\bar{u}_x, \bar{u}_y, 0)$ and vertical $\bar{u}_z$ components.\\
Using the general Eq. (\ref{eq:energy_fourier}), separate equations can be written for 2D and 3D mode spectral energies defined, respectively, as $E_{2D}=|\bar{\boldsymbol{u}}_{2D}|^2/2$ and $E_{3D}=|\bar{\boldsymbol{u}}_{3D}|^2/2$ \cite{Buzzicotti2018, Barker2013, Biferale2017},
\begin{equation}\label{eq:e2d}
\frac{dE_{2D}}{dt} = A_{2D} + \underbrace{NL_{2D2D} + NL_{3D2D}}_{NL_{2D}} +  D_{2D},
\end{equation}
\begin{equation}\label{eq:e3d}
\frac{dE_{3D}}{dt} = A_{3D} + \underbrace{NL_{3D3D} + NL_{2D3D}}_{NL_{3D}} + D_{3D}.
\end{equation}
Since injection $A$ and dissipation $D$ terms are of linear origin, they act for 2D and 3D modes separately, i.e., 
\[
A_{2D}=A(\boldsymbol{k}),~~~D_{2D}=D(\boldsymbol{k}),~~~{\rm for}~~~\boldsymbol{k} \in \Psi_{2D}~~~~~~~~~~
\]
\[
A_{3D}=A(\boldsymbol{k}),~~~D_{3D}=D(\boldsymbol{k}),~~~{\rm for}~~~\boldsymbol{k} \in \Psi_{3D}.~~~~~~~~~~ 
\]
However, the nonlinear transfers can act only among modes which lie respectively within the slow or the fast manifold, that is, nonlinear interactions separately among 2D-2D modes (vortex-vortex),
\[
NL_{2D2D}=-\bar{\boldsymbol{u}}_{2D} \cdot \overline{\left(\boldsymbol{u}_{2D} \cdot \boldsymbol{\nabla} \boldsymbol{u}_{2D}\right)},~~
\]
and among 3D-3D modes (wave-wave),
\[
NL_{3D3D}=-\bar{\boldsymbol{u}}_{3D} \cdot \overline{\left(\boldsymbol{u}_{3D} \cdot \boldsymbol{\nabla}\boldsymbol{u}_{3D}\right)},\]
(long bars denote Fourier transforms) as well as between these two manifolds, that is, nonlinear cross interactions/couplings between 2D and 3D modes (vortex-wave) \cite{Biferale2016, Buzzicotti2018},
\[
NL_{3D2D}=-\bar{\boldsymbol{u}}_{2D} \cdot \overline{\left(\boldsymbol{u}_{3D} \cdot \boldsymbol{\nabla} \boldsymbol{u}_{3D} \right)}. \quad
\]
\[ 
NL_{2D3D}=-\bar{\boldsymbol{u}}_{3D} \cdot \overline{\left(\boldsymbol{u}_{2D} \cdot \boldsymbol{\nabla} \boldsymbol{u}_{3D} \right)}-\bar{\boldsymbol{u}}_{3D} \cdot \overline{\left(\boldsymbol{u}_{3D} \cdot \boldsymbol{\nabla} \boldsymbol{u}_{2D} \right)}.
\]
In this case, in a triad, 2D modes can receive/lose energy via nonlinear interaction of two 3D modes with opposite signs of $k_z$, while 3D modes can receive/lose energy via interaction of only 2D and another 3D mode (interaction between 2D and 2D modes obviously cannot feed 3D modes). 

The 2D-2D and 3D-3D nonlinear transfers are each conservative, $\sum_{\boldsymbol{k}} NL_{2D2D}=0$ and $\sum_{\boldsymbol{k}} NL_{3D3D}=0$, whereas the cross transfer terms $NL_{2D3D}$ and $NL_{3D2D}$ are not, but their sum is conservative, as the net effect of these terms summed over all wavenumbers, as it should be, are equal in absolute value but have opposite signs i.e., 
\begin{equation}\label{eq:sum_nl_2d_3d}
\sum_{\boldsymbol{k} \in \Psi_{3D}}  NL_{3D2D} \left( \boldsymbol{k} \right) = - \sum_{\boldsymbol{k} \in \Psi_{2D}}  NL_{2D3D} \left( \boldsymbol{k} \right).
\end{equation}
Below we analyse the action of these injection and transfer terms for different $Po$ and $Re$ (Table I). 

In the following we will mostly use shell-averages of these spectra, which are defined in the standard way as\cite{Alexakis2018} 
\[
{\boldsymbol{f}(k)}= \sum_{k\leq |\boldsymbol{k}| \leq k+\Delta k}f(\boldsymbol{k}),
\]
for each spectral quantity $f\equiv (E, A, NL, D)$. For 3D modes $\boldsymbol{k} \in \Psi_{3D}$ the summation is done over spherical shells with radii $k=(k_x^2+k_y^2+k_z^2)^{1/2}$ while for 2D modes, having $\boldsymbol{k} \in \Psi_{2D}$, the summation is over rings in the $(k_x,k_y)-$plane with radii $k=(k_x^2+k_y^2)^{1/2}$. When we plot spectra for 2D and 3D quantities, we implicitly assume each depends on its respective wavenumber magnitude $k$ and ignore any anisotropy in each spectrum.

\subsection{Numerical method}

We solve Eq. (\ref{eq:ns_pert}) using the pseudo-spectral code SNOOPY \cite{lesur2005, lesur2007} which is a general-purpose code, solving HD and MHD equations, including shear, rotation, weak compressibility, and several other physical effects. The Fourier transforms are computed using the FFTW3 library. Nonlinear terms are computed using a pseudospectral algorithm with antialiasing 3/2-rule. The original version of the code has been modified \cite{Barker2016} to include precessional forcing and hence variables are decomposed in terms of shearing waves with periodically time-varying wavevector (Eq. \ref{eq:wavevector}) due to the basic shear flow $\boldsymbol{U}_b$ induced by precession. In this way, the shearing-periodic boundary conditions in the local domain (which are in fact fully periodic in the frame co-moving with the basic flow) are naturally satisfied in the code.  
\begin{figure}
\centering
\includegraphics[trim=1.cm 0cm 0cm 0cm,scale=0.50]{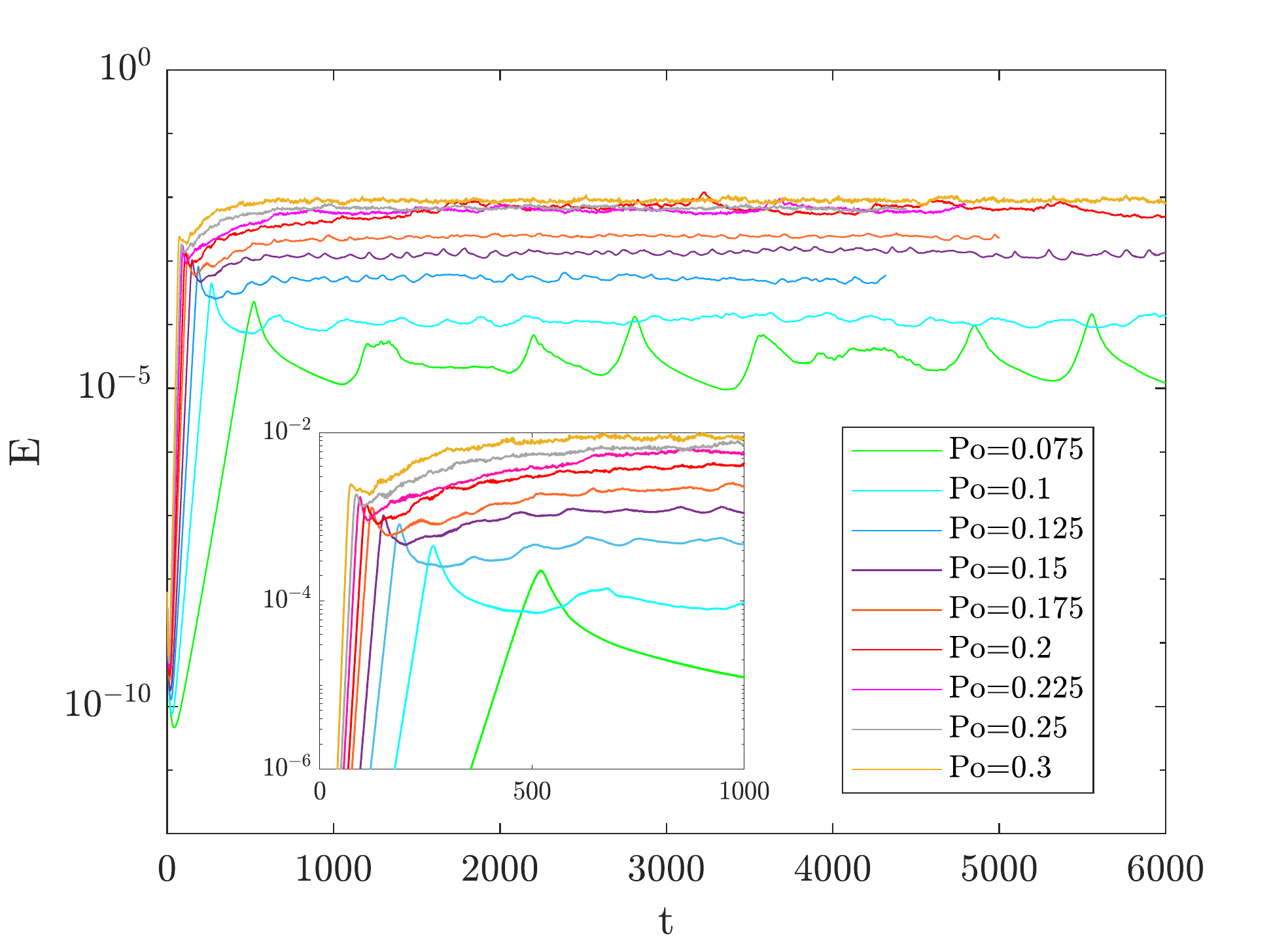}
\caption{Time-evolution of the volume-averaged total (2D+3D) kinetic energy for $Re=10^{4.5}$ and different $Po$. The impact of the precession parameter on the energy evolution is clearly seen, which is characterized by quasi-periodic bursts at small $Po=0.075$ and gradually becomes statistically steady turbulence with minor fluctuations and increasing amplitude as $Po$ increases. Inset panel zooms in the initial exponential (appearing as linear in logarithmic $y$-axis) growth and early saturation phases.}\label{fig:0b}
\end{figure}
\begin{figure}
\centering
\includegraphics[trim=1.2cm 0cm 0cm 0cm,scale=0.53]{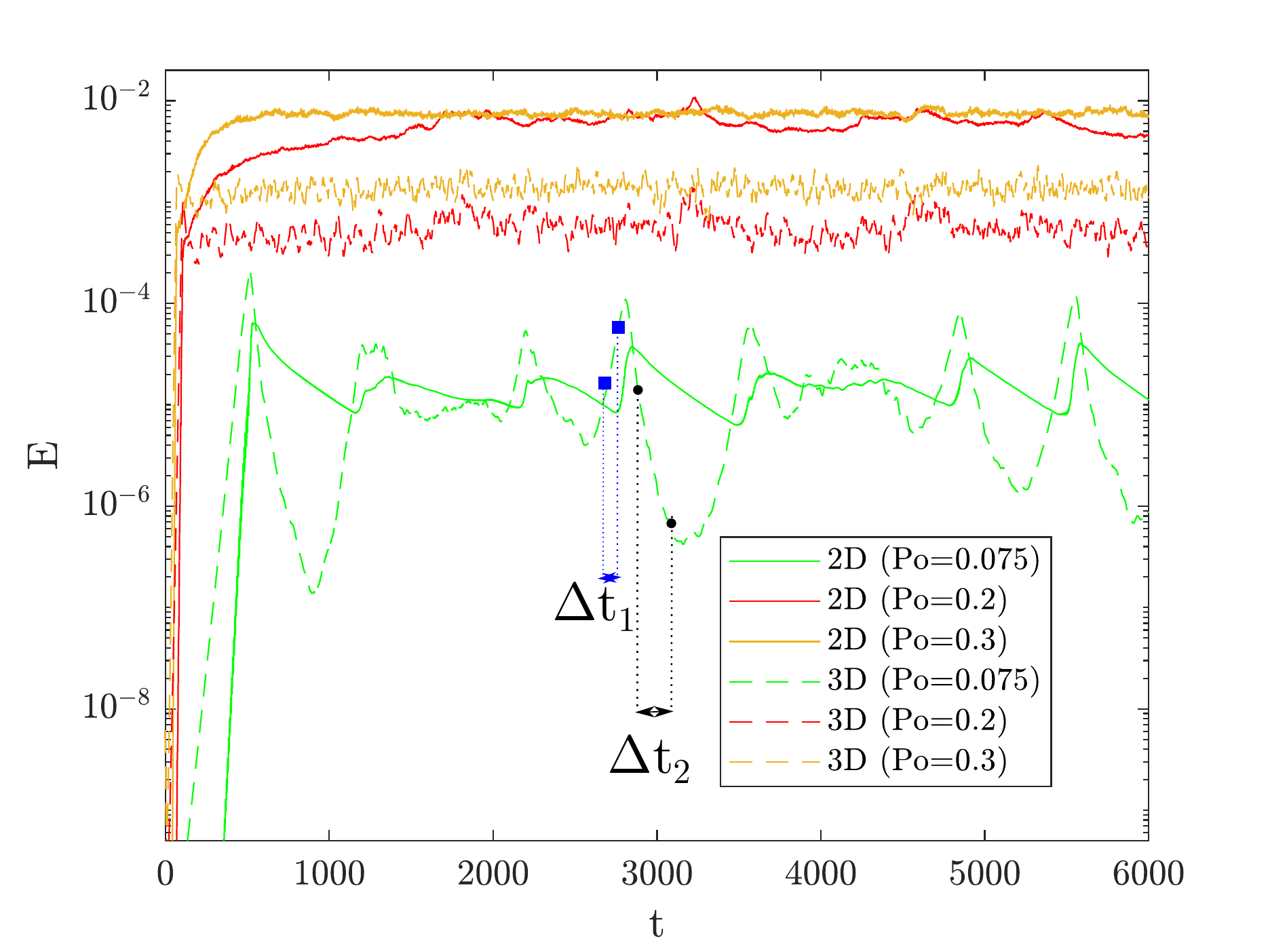}
\caption{Time-evolution of the volume-averaged kinetic energies for 2D vortices, $\langle E_{2D}\rangle$ (solid), and 3D inertial waves, $\langle E_{3D}\rangle$ (dashed), for $Re=10^{4.5}$ and three different precession parameters representative of three characteristic regimes shown in Fig. \ref{fig:0b}: bursts at weak ($Po=0.075$) and quasi-steady turbulence at medium ($Po=0.2$) and strong ($Po=0.3$) precessions. Two intervals $\Delta t_1$ (from $t=2690$ to $t=2770$) in State 1 and $\Delta t_2$ (from $t=2880$ to $t=3080$) in State 2 denote those time intervals over which spectral analysis is performed for these two different states.} \label{fig:1a}
\end{figure}
\begin{figure}
\centering
\includegraphics[trim=1.cm 0cm 0cm 0cm,scale=0.27]{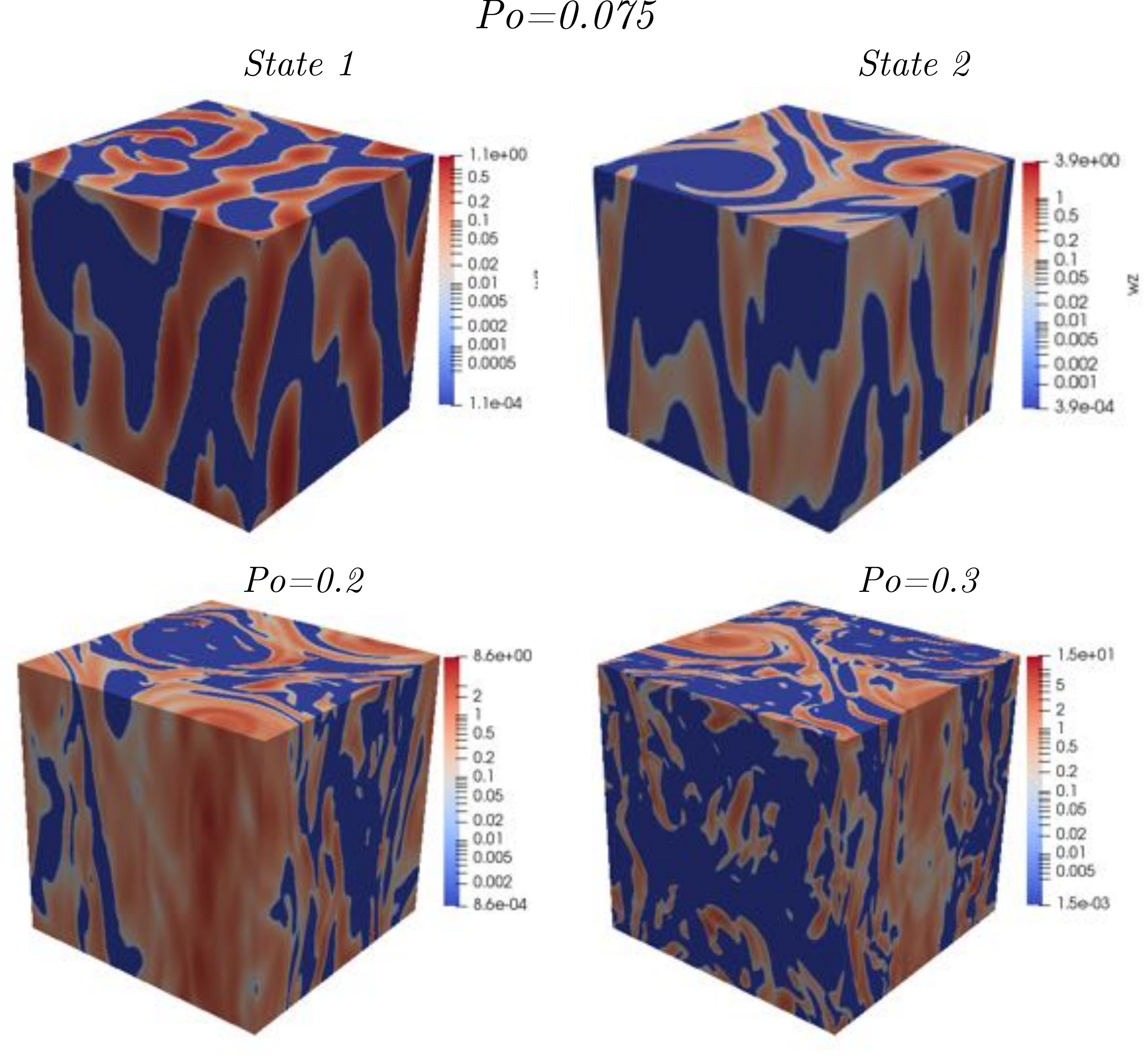}
\caption{Snapshot of the vertical component of vorticity, $\omega_{z}=(\boldsymbol{\nabla}\times\boldsymbol{u})_z$ (log-scale), in physical space for three characteristic precession parameters: $Po=0.075, 0.2, 0.3$ and $Re=10^{4.5}$ in the saturated state (at $t=3000$). The upper two boxes show the structures in State 1 (left), dominated by vertically-varying 3D inertial waves, and State 2 (right), dominated by 2D vortices nearly uniform along the $z$-axis. Large-scale 2D columnar vortices are also evident together with turbulent field of waves at $Po=0.2$ and 0.3.}\label{fig:snapshot_wz}
\end{figure}

\section{General features of the precession-driven turbulence}\label{sec:general}

The simulations performed in this paper for different pairs of $(Po,Re)$ are listed in Table I. All runs start with initial small random noise perturbations of velocity imposed on the basic flow. In Figure \ref{fig:0b}, we plot the time evolution of the volume-averaged kinetic energy, which is equal to the sum of energies of 2D and 3D modes over all wavenumbers, $\langle E\rangle=\sum_{\boldsymbol{k}}E=\sum_{\boldsymbol{k}}(E_{2D}+E_{3D})$, for several precession parameters $Po$ and at an intermediate Reynolds number $Re=10^{4.5}$. In the initial linear regime, the energy grows exponentially corresponding to dominant 3D inertial waves being excited first by the precessional instability \cite{Kerswell1993, Mason2002, Naing2011, Barker2016} (see inset in Fig. \ref{fig:0b}). In the given range of $Po$, the growth rate of the precession instability increases with $Po$. After about several hundreds of orbital times the exponential growth attains sufficient amplitude for nonlinearity to come into play and cause the instability to saturate with higher amplitudes and shorter saturation times for larger $Po$. Depending on $Po$, the saturated states are qualitatively different, exhibiting statistically steady turbulence at higher $Po \gtrsim 0.1$ with only minor fluctuations, whereas strong quasi-periodic oscillations (bursts) are observed at small $Po \lesssim 0.1$ with more than an order of magnitude variations in the kinetic energy. This temporal behavior of the volume-averaged kinetic energy in the nonlinear state of the precessional instability with $Po$ is consistent with previous related local studies \cite{Barker2016, Khlifi2018}. Below we interpret this temporal evolution of the total kinetic energy in terms of the dynamics of 2D vortices and 3D waves and their interplay. 

\begin{figure}[t!]
\centering
\includegraphics[trim=0.75cm 0cm 0cm 0cm,scale=0.45]{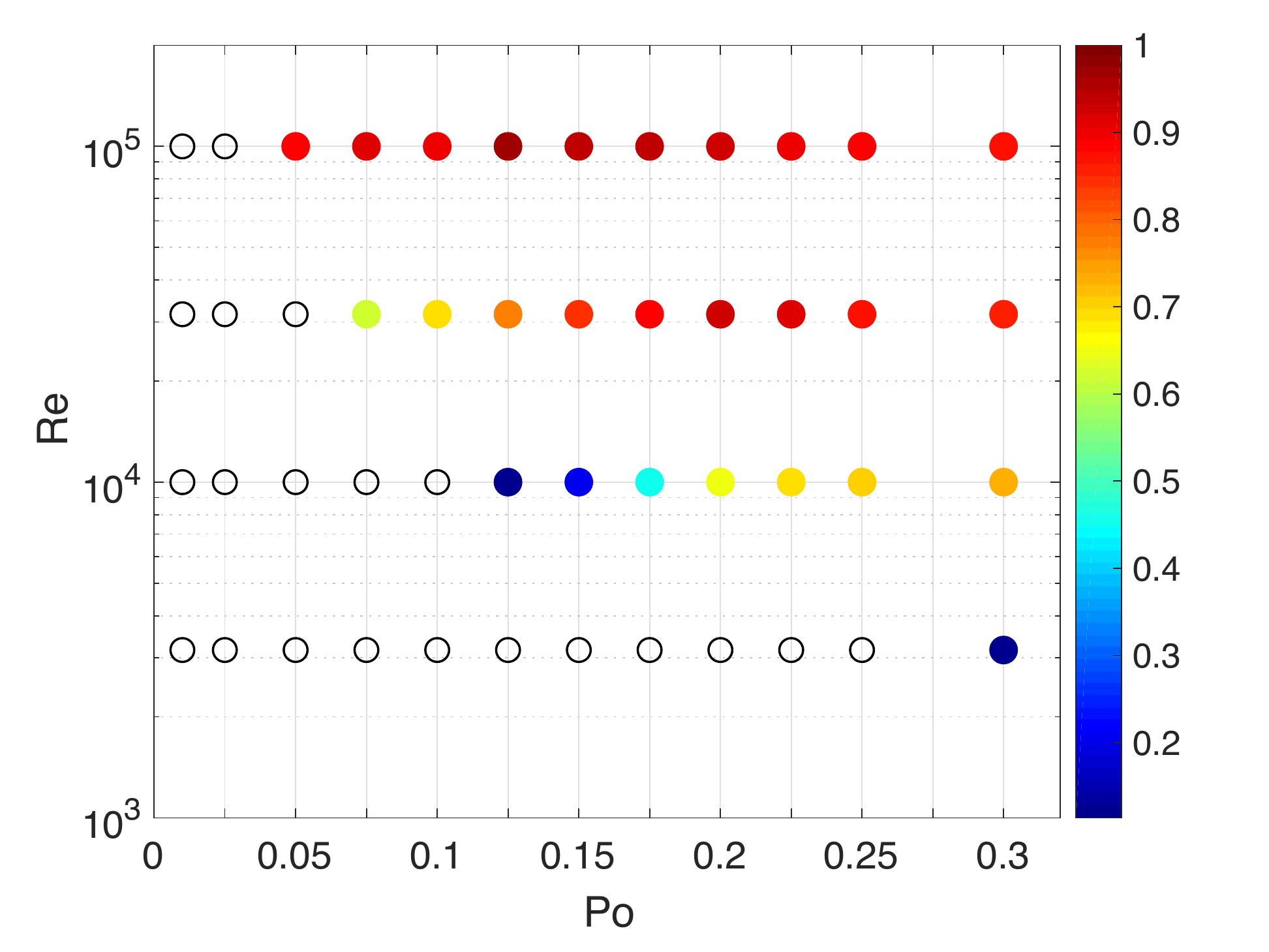}
\caption{Regime diagram in the $(Po,Re)-$plane. The colors represent the ratio of time-averaged 2D to total (2D+3D) energies, $\langle E_{2D}\rangle/\langle E\rangle$, in the saturated state, while the empty points correspond to the cases stable to precessional instability when perturbations decay.}\label{fig:0a}
\end{figure}

\begin{figure}
\centering
\includegraphics[scale=0.54]{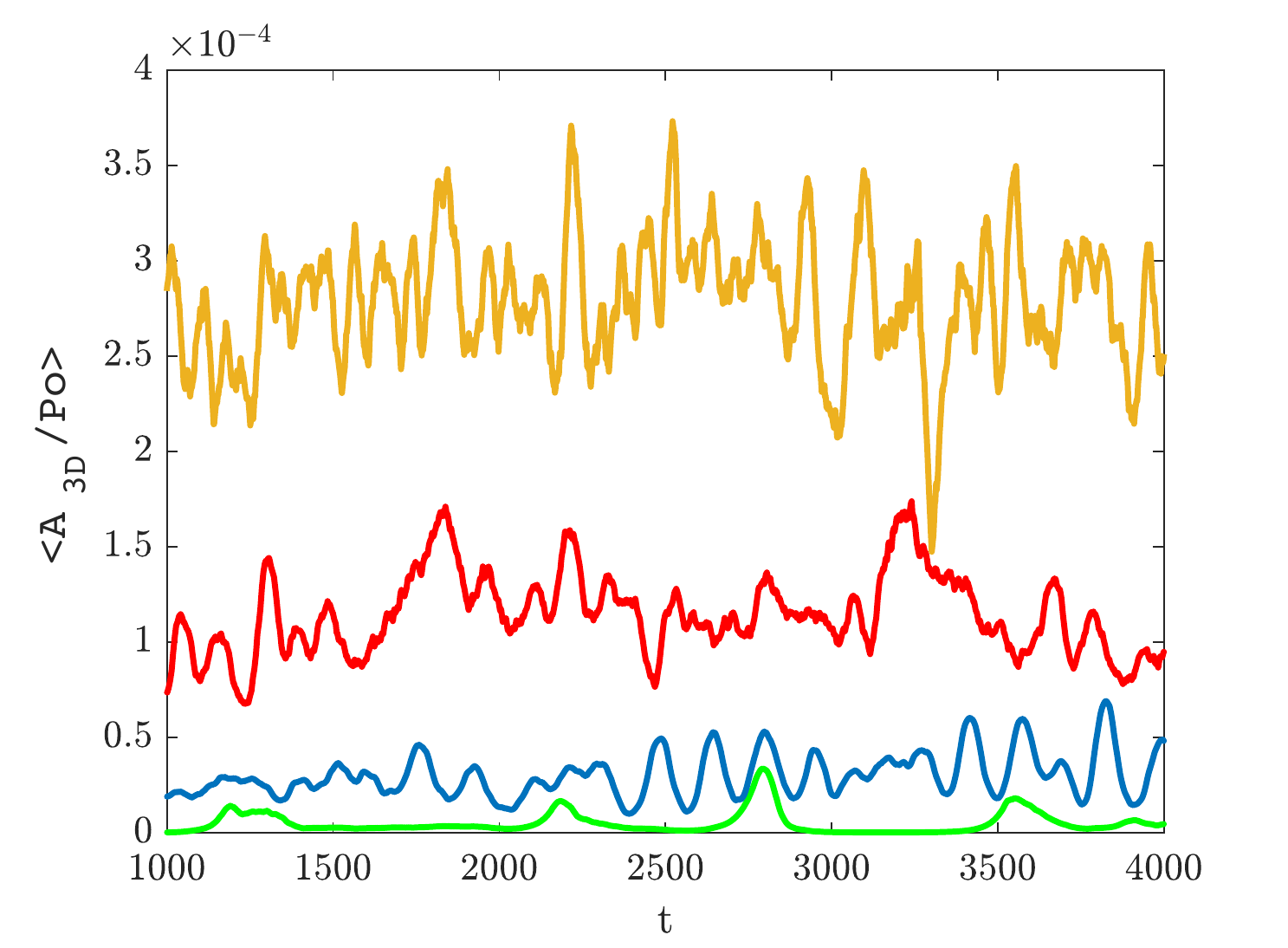}
\includegraphics[scale=0.54]{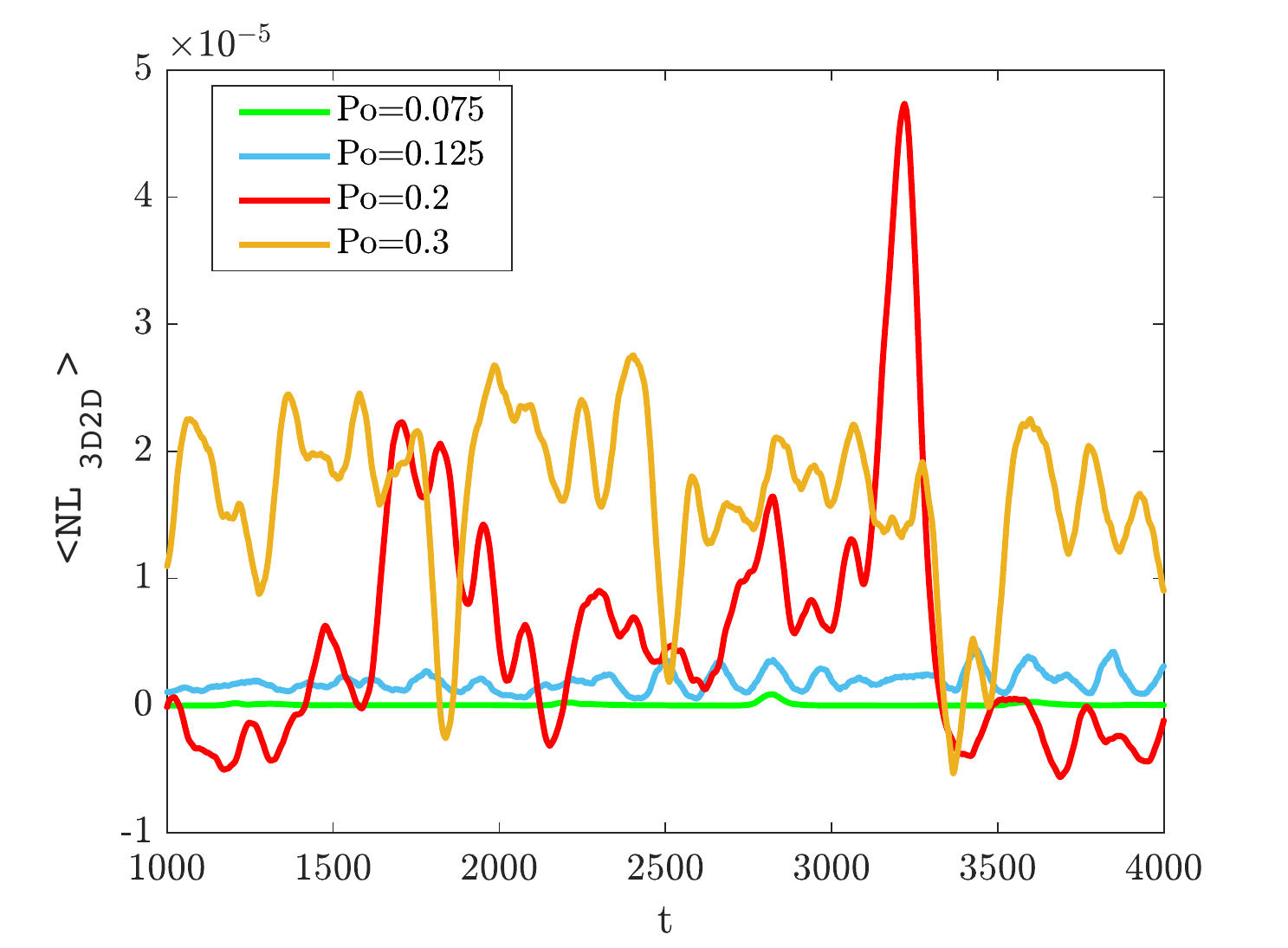}
\includegraphics[scale=0.54]{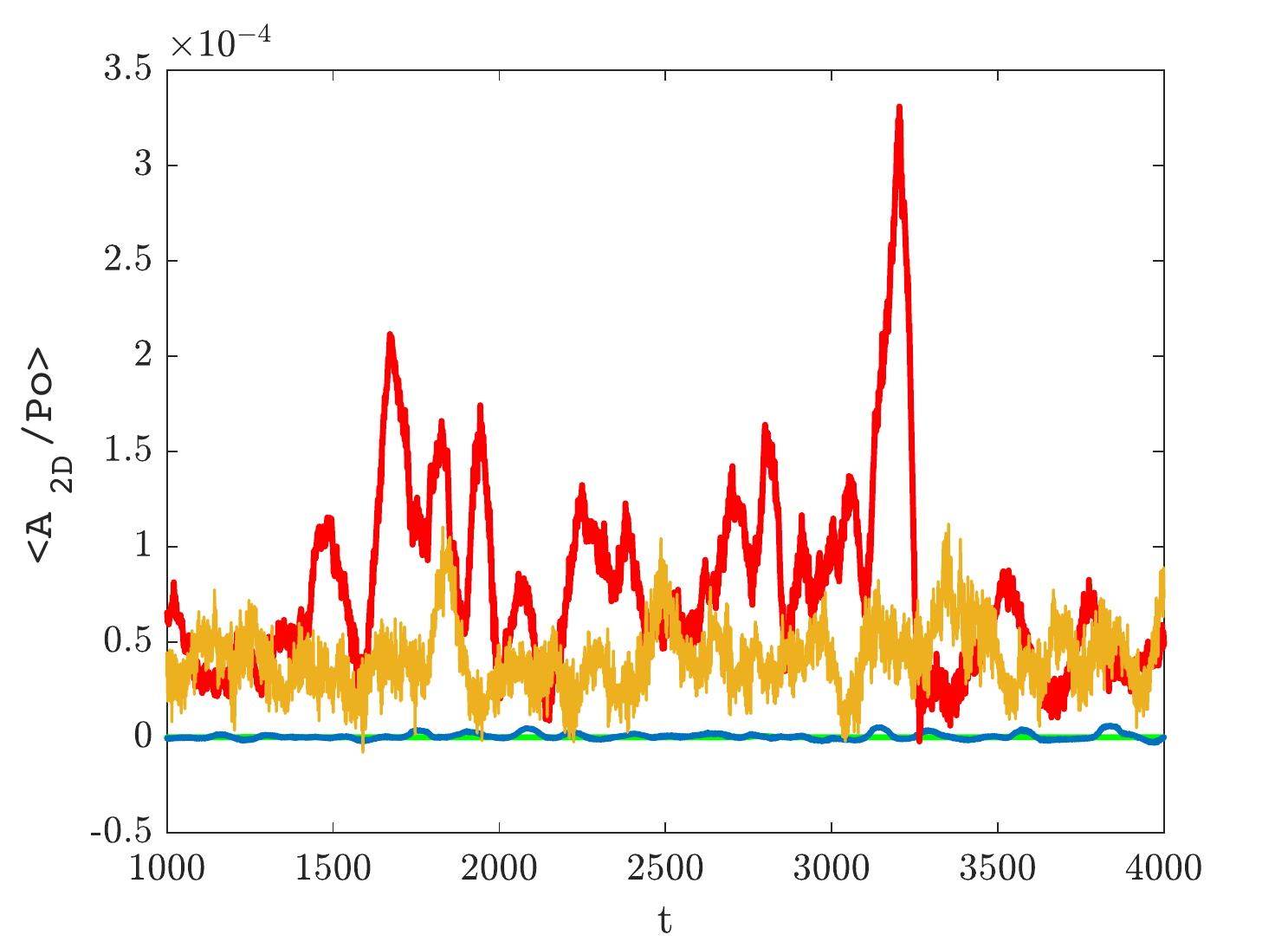}
\caption{Evolution of the volume-averaged dynamical terms -- energy injection $\langle A\rangle$ for 3D waves (top) and 2D vortices (bottom) together with nonlinear transfer $\langle NL_{2D3D}\rangle$ between these two modes (middle) in corresponding Eqs. (\ref{eq:e2d}) and (\ref{eq:e3d}) for different $Po$ and  $Re=10^{4.5}$.}\label{fig:a2d_nl23_a3d}
\end{figure}

\begin{figure*}
\includegraphics[scale=0.43]{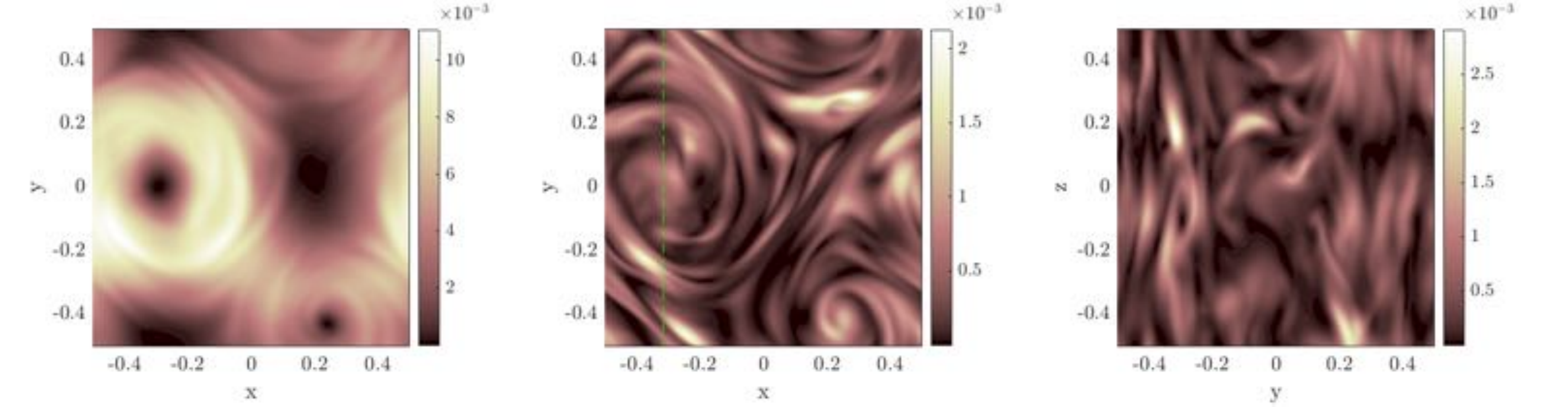}
\includegraphics[scale=0.43]{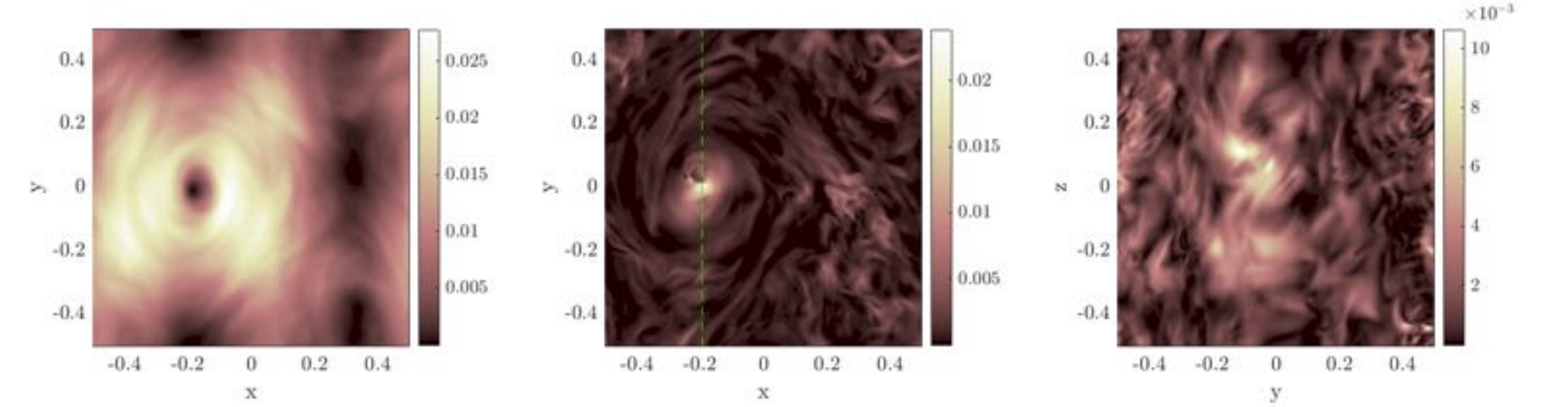}
\caption{Kinetic energy density of 2D vortices (left panels) and 3D wave modes (taken at $z=0$) in the $(x,y)-$plane (middle panels) and in $(y,z)-$plane (right panels) in the saturated state at $Po=0.2$ (top row), $Po=0.3$ (bottom row) both with $Re=10^{4.5}$. The green dotted line, which is at $x=-0.33$ for $Po=0.2$ and at $x=0.2$ for $Po=0.3$, marks that $(y,z)-$section where 3D energy is plotted.}\label{fig:contour_2d_3d}
\end{figure*}

A first comparison between the 2D and 3D mode dynamics is shown in Fig. \ref{fig:1a} where the evolution of the total kinetic energies for 2D modes, $\langle E_{2D}\rangle$, and 3D modes, $\langle E_{3D}\rangle$, are plotted for three precession parameters. For larger $Po\gtrsim 0.1$ the energy of 2D modes is more than one order of magnitude larger than that of 3D modes. However, the saturated value of $\langle E_{3D}\rangle$ tends to increase more than that of $\langle E_{2D}\rangle$ with increasing $Po$, implying that the waves, as they should be, are more affected and intensified by precession rather than the vortices. The 2D vortices are linearly stable against precession instability and hence cannot grow due to the latter \cite{Kerswell1993}. They are driven and energetically supplied by waves via nonlinear transfers \cite{Barker2016}, which will be examined in detail below using the spectral analysis. On the other hand, for the lowest precession parameter $Po=0.075$, corresponding to the bursty regime, the energy of 3D waves periodically dominates over the 2D vortical mode energy during the growth (burst) phase (referred to as State 1). In this burst phase, waves excited by the precessional instability, lose their energy to 2D vortices due to nonlinearity. As a consequence, the energy of the waves drops by about an order of magnitude (affected additionally by viscous dissipation) relative to the 2D mode energy (State 2). After that it starts to increase again due to precessional instability, closing the cycle. Although 2D mode energy also decreases at this stage, it does so much slower, on a viscous time \cite{Barker2016}. This cyclic behavior of both components is remarkable, indicating the quasi-periodic nature of evolution due to weak precessional forcing ($Po \lesssim 0.1$), which is relevant to astrophysical and geophysical regimes \cite{Barker2016, Lebars2015, Cebron2019}. We explore this behavior in more detail in the spectral analysis section below.

Figure \ref{fig:snapshot_wz} shows the structure of the vertical component of vorticity, $\omega_{z}=\left( \boldsymbol{\nabla} \times \boldsymbol{u} \right)_z$, in physical space well after the saturation for the above regimes of weak, moderate and strong precessions. The top row shows the case $Po=0.075$ characterized by bursts. In State 1, which corresponds to the burst of 3D wave energy dominating over that of 2D modes in Fig. \ref{fig:1a}, we therefore observe pronounced 3D wave structures varying along the $z-$axis. By contrast, in the State 2, where the wave energy quickly decays afterwards and 2D modes dominate, only vertically uniform columnar vortical structures aligned with the rotation $z$-axis are present. At larger $Po=0.2$ and $Po=0.3$ shown, respectively, in bottom left and right panels of Fig. \ref{fig:snapshot_wz}, the nonlinear states consist of vortices embedded in 3D waves, coexisting at all times. At medium $Po=0.2$, two columnar vortices aligned with the rotation axis with opposite vorticity (cyclonic/anticyclonic) are clearly seen in the small scale waves, whose strength with respect to vortices has increased compared to that in the above bursty regime. At even higher $Po=0.3$, the contribution of 3D wave energy is somewhat larger (Fig. \ref{fig:1a}) and therefore small-scale turbulent wave structures are more pronounced with respect to a single 2D vortex.

The regime diagram in Fig. \ref{fig:0a} summarizes the properties of all the runs for different pairs $(Po,Re)$. The colored dots represent the ratio of the time- and volume-averaged energy of 2D vortices, $\langle E_{2D}\rangle$, to the total energy of all the modes, $\langle E\rangle$, in the statistically steady turbulent state, as shown in Figs. \ref{fig:0b} and \ref{fig:1a}. The empty points represent the cases where the energy drops to negligible value meaning that the local flow in the box is stable against precessional instability. The colors show that at given $Re$, the fraction of 2D mode energy vs. total energy initially increases with $Po$ when $Po\lesssim 0.1$, then reaches a maximum at medium precessions $Po\sim 0.1$ and decreases at larger $Po\gtrsim 0.1$. The maximum shifts towards smaller $Po$ with increasing $Re$. We will carry out the analysis distinguishing the latter three groups in the following sections.

Having analyzed the time-development of the mode energies, next in Fig. \ref{fig:a2d_nl23_a3d} we plot the evolution of the volume-averaged (or equivalently integrated in Fourier space) dynamical terms in Eqs. (\ref{eq:e2d}) and (\ref{eq:e3d}), i.e., the energy injection, $\langle A\rangle=\sum_{\boldsymbol{k}}A$, for 3D waves (top) and 2D vortical (bottom) modes together with nonlinear transfer terms between them, $\langle NL\rangle=\sum_{\boldsymbol{k}}NL$ (middle). To obtain a better visualization and a clear trend of temporal evolution, we have smoothed these volume-averaged terms additionally over short time intervals around each time moment, thereby removing fast temporal oscillations and getting meaningful time-averages. For 3D waves, energy injection occurs due to the precession instability and hence increases with increasing precession strength $Po$, while the 2D modes are excited by the nonlinear interaction among these waves via $NL_{3D2D}$ term, which also increases with $Po$. This term is overall positive in time, implying transfer of energy from 3D waves to 2D vortices (see also Ref.\cite{Barker2013}). The excited 2D modes in turn also extract energy from the basic flow via positive $\langle A_{2D}\rangle>0$ term. Note that $\langle A_{2D} \rangle$ is not necessarily zero, unlike for the 2D linear modes in the background precessional flow. As a result, the evolution of $\langle A_{2D}\rangle$ is determined by the nonlinear transfer term $\langle NL_{2D3D}\rangle$ and hence follows the latter, as it is seen in Fig. \ref{fig:a2d_nl23_a3d} where the peaks of both these functions nearly coincide. For $Po=0.2$, $\langle A_{2D}\rangle \approx \langle A_{3D} \rangle$ (red), while for $Po=0.3$, $A_{3D} > A_{2D}$ (orange curves), indicating that the precession instability feeds the waves, while the waves in turn feed vortices via nonlinear cascade. Below we will see how this process occurs scale by scale in Fourier space. 

\begin{figure*}
\centering
\includegraphics[scale=0.55]{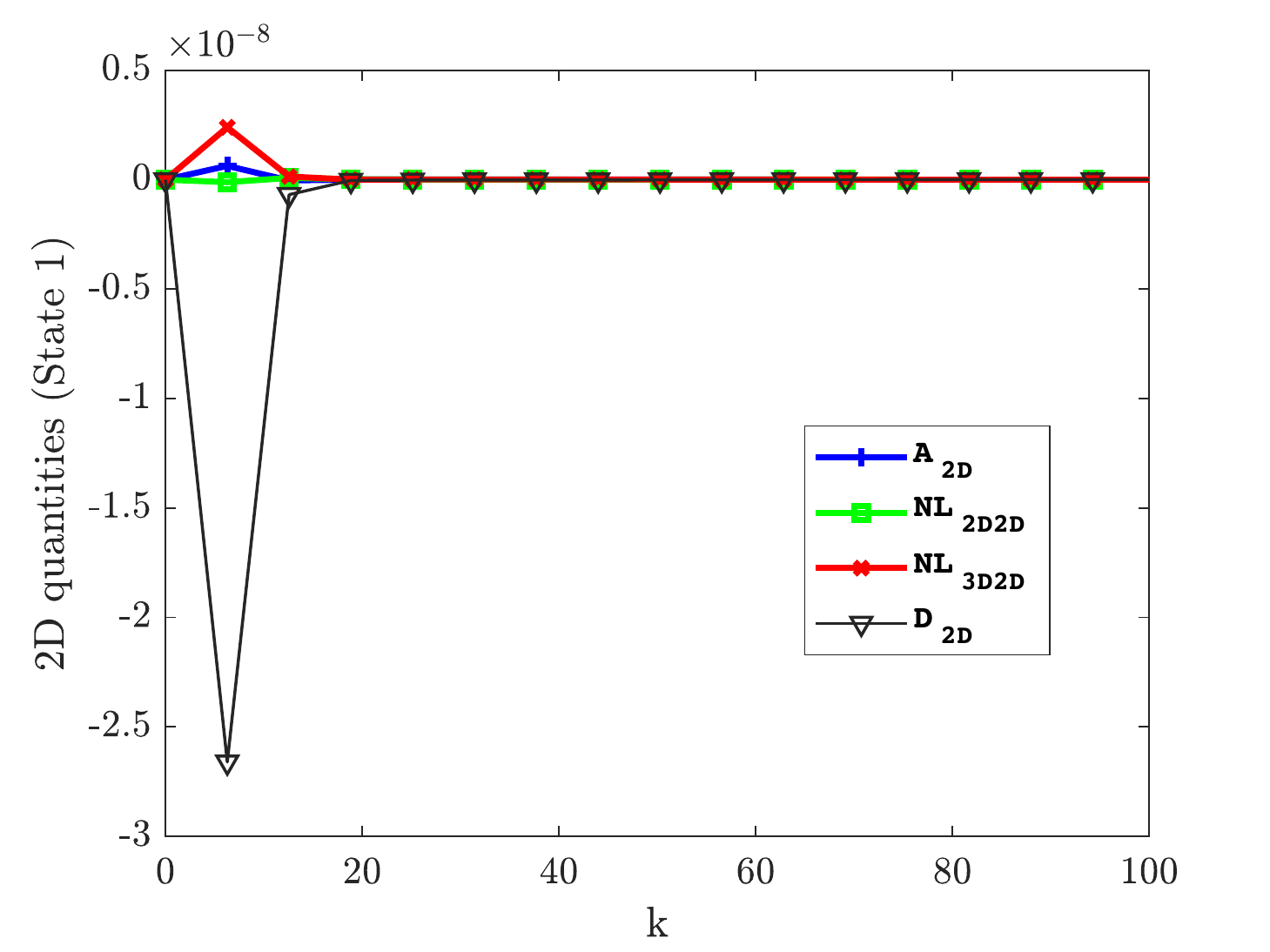}
\includegraphics[scale=0.55]{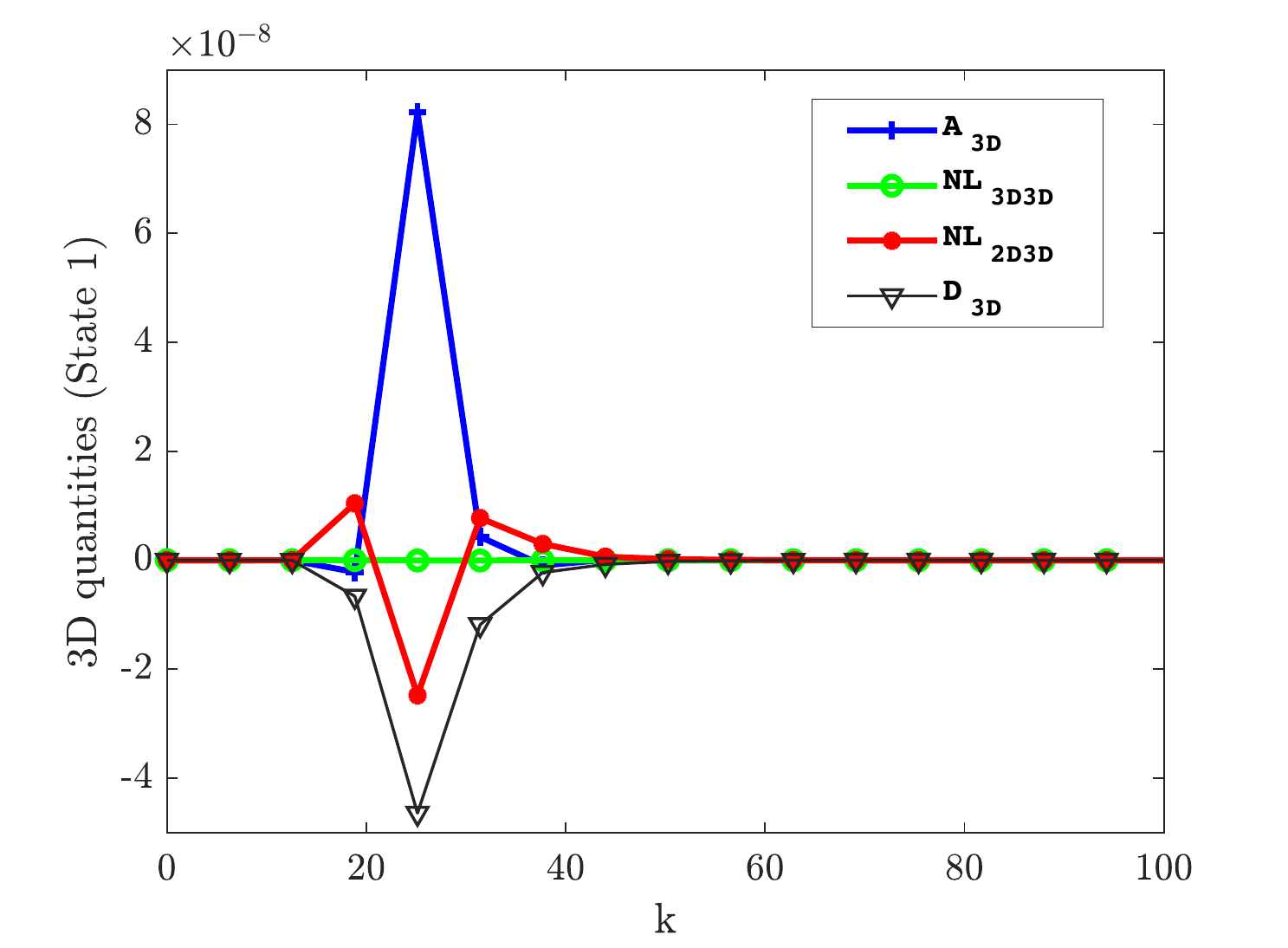}
\includegraphics[scale=0.55]{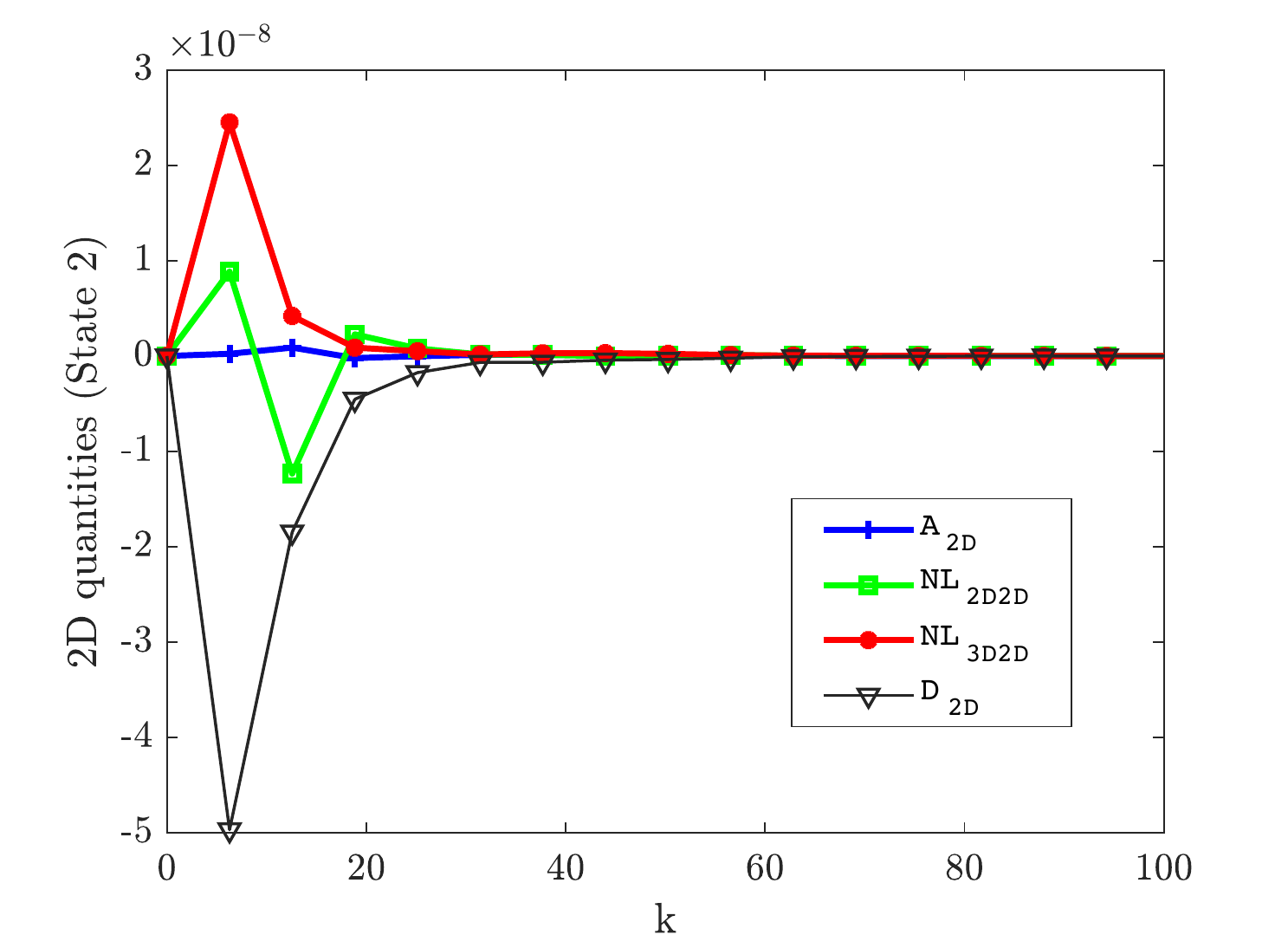}
\includegraphics[scale=0.55]{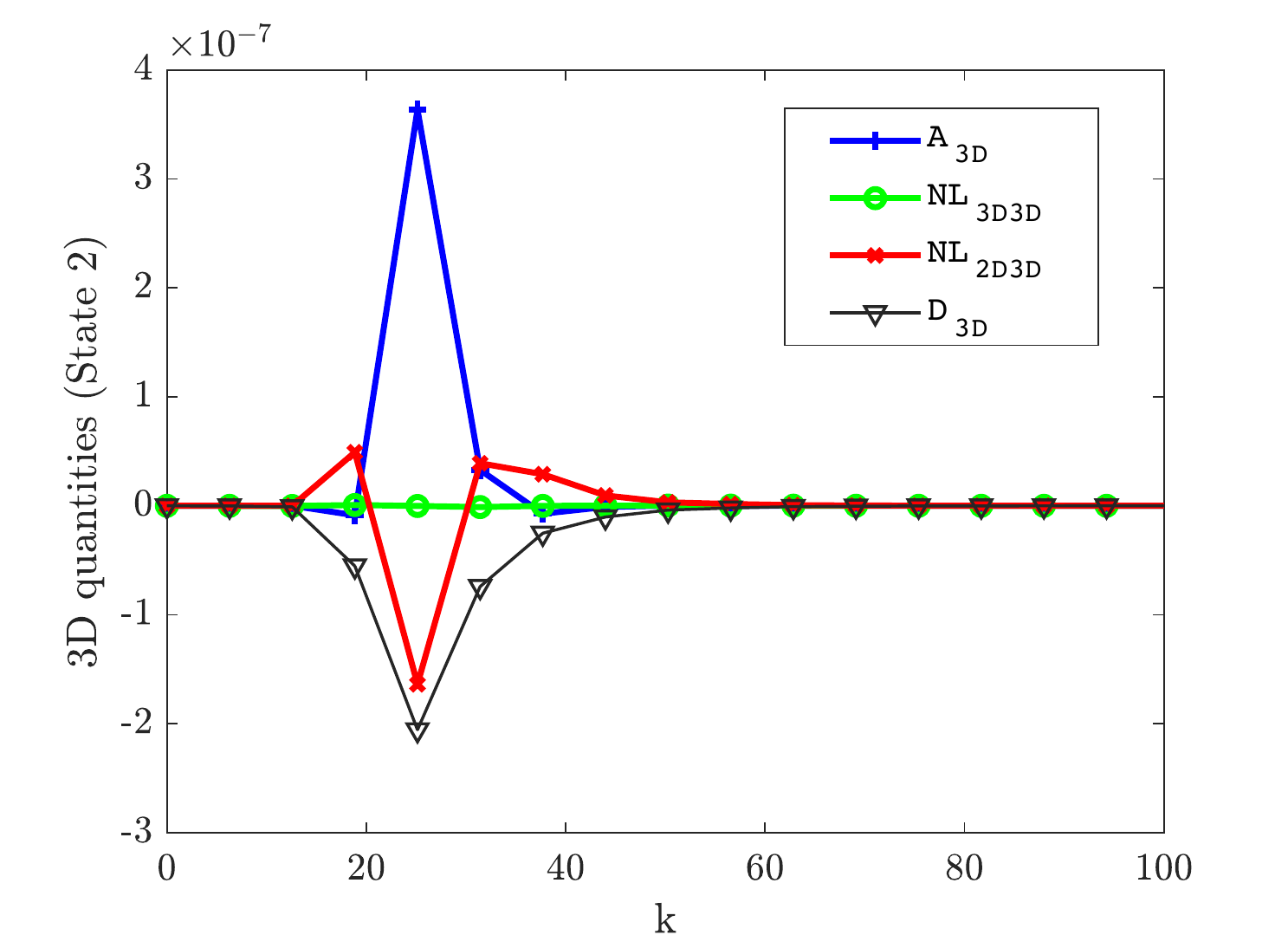}
\caption{Shell-averaged spectra for the injection $A$ (blue), viscous dissipation $D$ (black), and nonlinear transfers among modes inside the 2D manifold, $NL_{2D2D}$ (green, left panels), inside 3D manifold $NL_{3D3D}$ (green, right panels) and cross transfers $NL_{3D2D}$ (red, left panels) and $NL_{2D3D}$ (red, right panels) between the modes in these two manifolds. For 2D vortical modes (left panels) and 3D wave modes (right panels) in the State 1 (upper row) and State 2 (bottom row) at $Po=0.075$ and $Re=10^{4.5}$.} \label{fig:state_1_2_po0075}
\end{figure*}

In Fig. \ref{fig:snapshot_wz} we have shown the total vorticity field including both 2D vortices and 3D waves. To better visualize these fields, we computed the inverse Fourier transforms from $\bar{\boldsymbol{u}}_{2D}$ and $\bar{\boldsymbol{u}}_{3D}$ and show respective energy densities in physical space in the saturated regime in Fig. \ref{fig:contour_2d_3d}. The left panels of this figure show energy of 2D modes, where now we can clearly distinguish two vortices for $Po=0.2$ (top row) and a single vortex for $Po=0.3$ (bottom row). The middle panels show the small-scale 3D mode energies in the $(x,y)-$plane at the central height ($z=0$) of the box. There is a noticeable difference between the $Po=0.2$ and $Po=0.3$ cases: for $Po=0.2$ we observe larger-scale wave structures, whereas for $Po=0.3$ the wave field is more fluctuating and rich in smaller scales, implying that increasing precession parameter intensifies first of all 3D waves and indirectly vortices due to their nonlinear coupling with the former. Note also that the 3D wave structures are concentrated around the vortices -- a feature observed experimentally in a precessing spherical containers \cite{Horimoto2017}. The right panels show the vertical structure of 3D mode energy in the $(y,z)-$plane at the center of vortices (located at $x=-0.33$ and $x=-0.2$, respectively, for $Po=0.2$ and 0.3, which are marked with a green dotted line in the middle row). Again, the $Po=0.3$ case shows more fluctuating behavior with fine scales surrounding the column. So, the main dynamical picture consists of the coexisting columnar (geostrophic) vortices and waves whose magnitudes and length-scales depend on the precession strength.

\section{Spectral Dynamics of Precession-driven turbulence: vortices, waves and their interplay}\label{subsec:general}

So far the study has been mainly conducted in the physical (coordinate) space. However, a deeper insight into the precession-driven turbulence dynamics can be gained by investigating the dynamical processes -- energy injection, nonlinear transfers and viscous dissipation -- in Fourier space, where a much richer dynamical picture unfolds and becomes more accessible to analysis. Following the approach of Refs. \cite{George2014, George2016, Buzzicotti2018}, we compute and visualize the individual injection $A$, viscous dissipation $D$ and various nonlinear transfer $NL$ terms entering spectral energy Eqs. (\ref{eq:e2d}) and (\ref{eq:e3d}) in Fourier space using the simulation data,  and analyze their interplay in different regimes with respect to the precession parameter identified above. 

\begin{figure*}[t]
\centering
\includegraphics[scale=0.58]{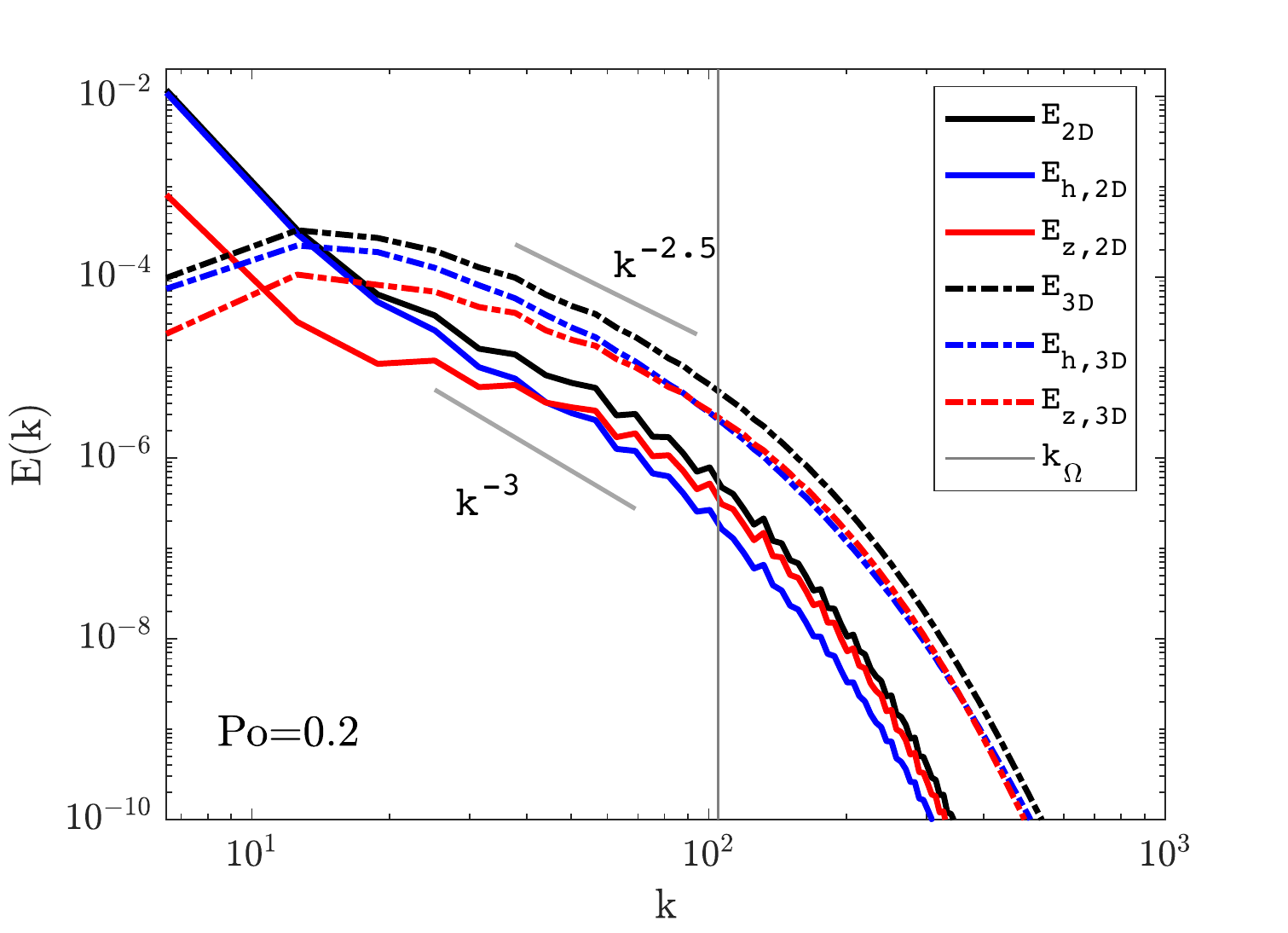}
\includegraphics[scale=0.58]{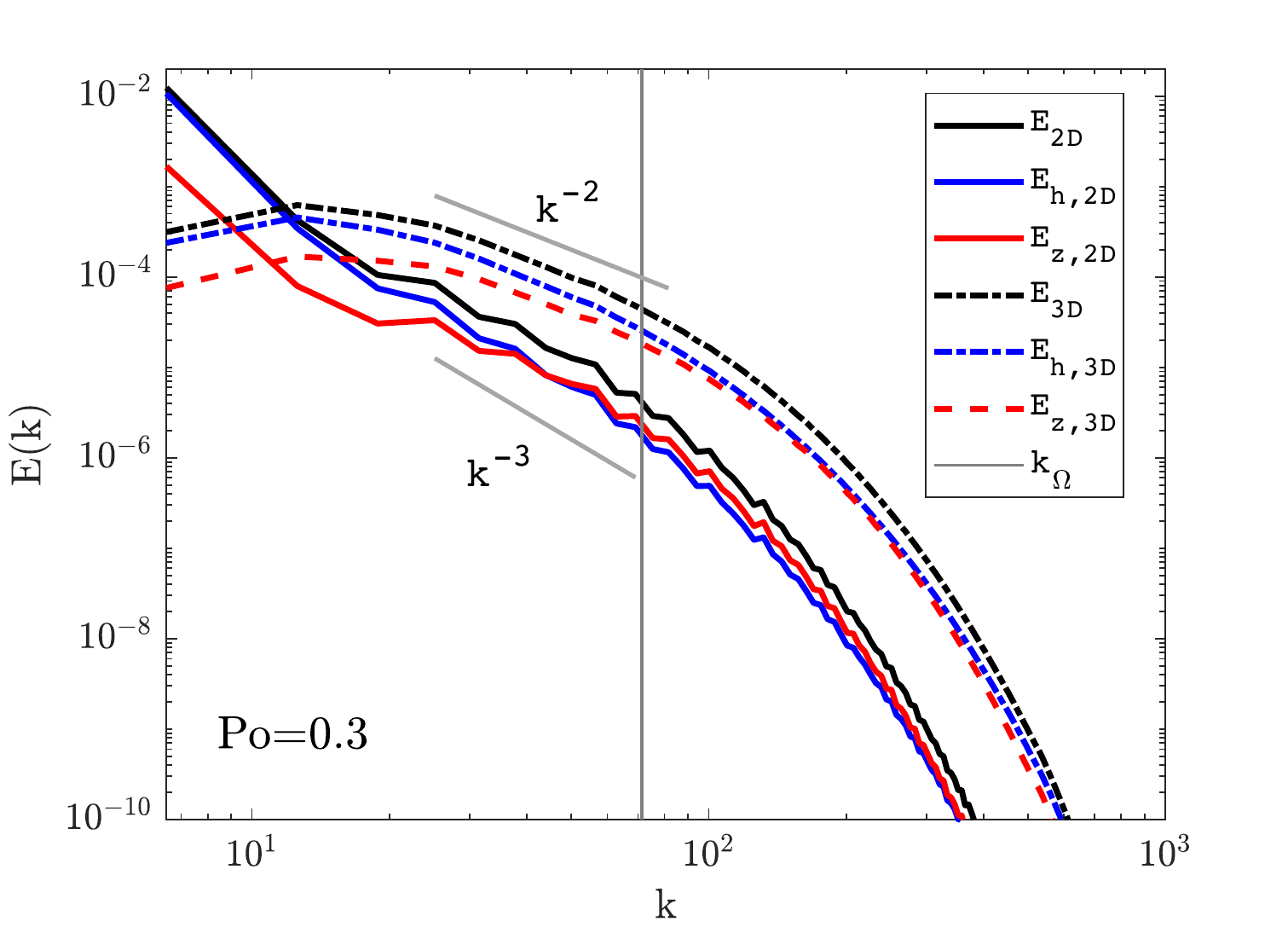}
\caption{Shell-averaged energy spectra in a statistically steady turbulent state at $Po=0.2$ (left), 0.3 (right) and $Re=10^{4.5}$. In both cases, we distinguish between 2D mode energies (solid lines) and 3D mode energies (dashed lines), while the colors represent the total $E$ (black), horizontal, $E_{h}=(|\bar{u}_x|^2+|\bar{u}_y|^2)/2$ (blue) and vertical $E_z=|\bar{u}_z|^2/2$ (red) components. Grey vertical line shows the location of Zeman scale $k_{\Omega}$.} \label{fig:spectra_re1e4d5}
\end{figure*}

\subsection{Quasi-periodic bursts: $Po=0.075$}\label{sec:bursts}

As we have seen above, the case with weak precession forcing is characterized by cyclic bursts, where the system alternates between State 1 and State 2 and the purpose of this section is to understand the underlying mechanisms of this behavior. With this goal, we analyze and compare the dynamics in two different intervals shown in Fig. \ref{fig:1a}: $\Delta t_{1}$ corresponding to State 1, when the energy of 3D wave modes initially increases due to precessional instability, while the energy of 2D modes is still at its minimum, and $\Delta t_{2}$ corresponding to State 2, when both 3D and 2D mode energies drop.

The shell-averaged spectra of the linear injection $A$ and dissipation $D$ terms for 2D and 3D modes as well as nonlinear transfer terms for 2D-2D, $NL_{2D2D}$, for 2D-3D, $NL_{2D3D}$ and $NL_{3D2D}$, and for 3D-3D $NL_{3D3D}$ mode interactions in these two states (also averaged over $\Delta t_1$ and $\Delta t_2$ time intervals, respectively) are shown in Fig. \ref{fig:state_1_2_po0075}. The basic dynamical picture in this regime is the following. In State 1 (top row), the most important contribution is due to $A_{3D}$ (blue), which injects energy into waves from the basic flow due to the initial development of precessional instability. This reaches a large peak at $k=8\pi$ whose value is positive and larger than the comparable effects of viscous dissipation $D_{3D}$ (black) and transfer $NL_{2D3D}$ (red), which are both negative reaching a minimum at the same wavenumber. The effect of nonlinear transfers among waves, $NL_{3D3D}$ (green) is relatively small at this time. This also implies that the viscosity is already important at the injection scale, that is, there is not a good scale separation (inertial range) between the injection and viscous scales. Nevertheless, $A_{3D}>0$ is sufficiently large to overcome both these negative (sink) terms and give rise to wave growth in State 1. Since $NL_{2D3D}<0$ at the injection wavenumbers, its counterpart for 2D modes $NL_{3D2D}>0$, indicating that the waves nonlinearly transfer their energy to and amplify 2D vortices, but at lower wavenumbers near the peak $k=2\pi$ of this term. These vortices additionally receive some energy from the basic flow due to the positive $A_{2D}$ (blue) term. However, the dissipation $D_{2D}<0$ (black curve in top left panel) for vortices is quite high, prevailing over the positive $NL_{3D2D}$ (red) and $A_{2D}$ and as a result vortices do not yet grow at these times. 

The nonlinear transfers between waves and vortices, $NL_{3D2D}$ and $NL_{2D3D}$, increase by absolute value (but retain their signs) with time and already in State 2 the mostly negative $NL_{2D3D}$, together with dissipation $D_{3D}<0$, dominates the positive injection $A_{3D}$ (bottom right plot in \ref{fig:state_1_2_po0075}). As a result, wave energy quickly drops in State 2 (see also Fig. \ref{fig:1a}). On the other hand, the 2D vortices, which now receive much more energy from waves via the term $NL_{3D2D}>0$, also develop an inverse cascade themselves described by $NL_{2D2D}$ (bottom left panel). This is manifested in the emergence of large-scale vortices in physical space in State 2 (top right plot of Fig. \ref{fig:snapshot_wz}). The injection $A_{2D}$ is relatively small/insignificant at these times. However, dissipation $D_{2D}$ is still larger than the nonlinear replenishment by $NL_{3D2D}$ and consequently the energy of vortices slowly decreases too (which is consistent with Fig.~\ref{fig:1a}). Once vortices have become weak enough, the waves can grow again due to the precessional instability and close the cycle loop. Thus, we can conclude that the bursts are caused by a quasi-periodic behavior of 3D dynamical terms which in State 1 with $\sum_{\boldsymbol{k}}(A_{3D}+NL_{3D}+D_{3D})>0$ leading to wave energy amplification, whereas in State 2 with $\sum_{\boldsymbol{k}}(A_{3D}+NL_{3D}+D_{3D})<0$ leading to energy decay.

\begin{figure*}
\includegraphics[scale=0.295]{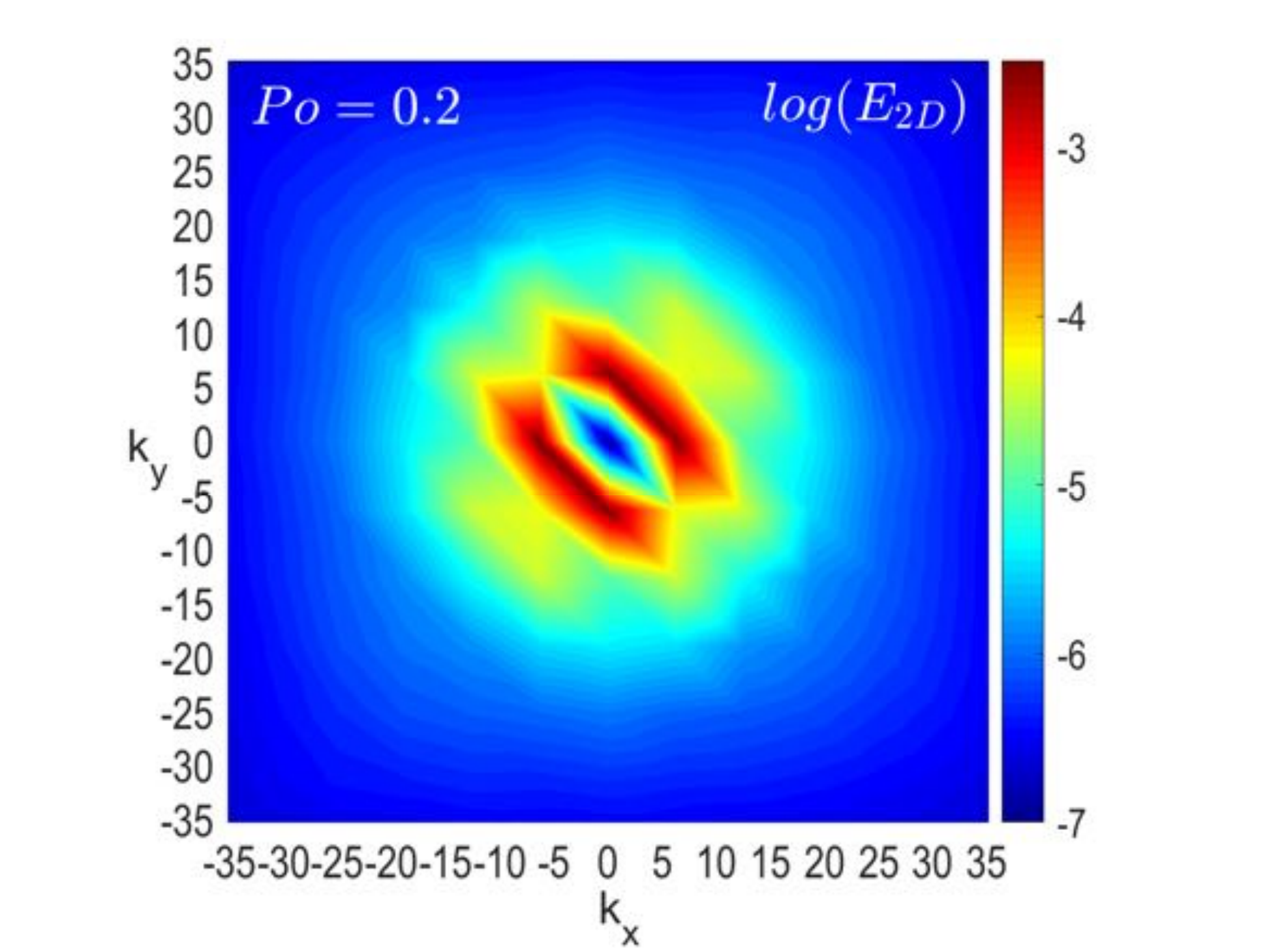}
\includegraphics[scale=0.295]{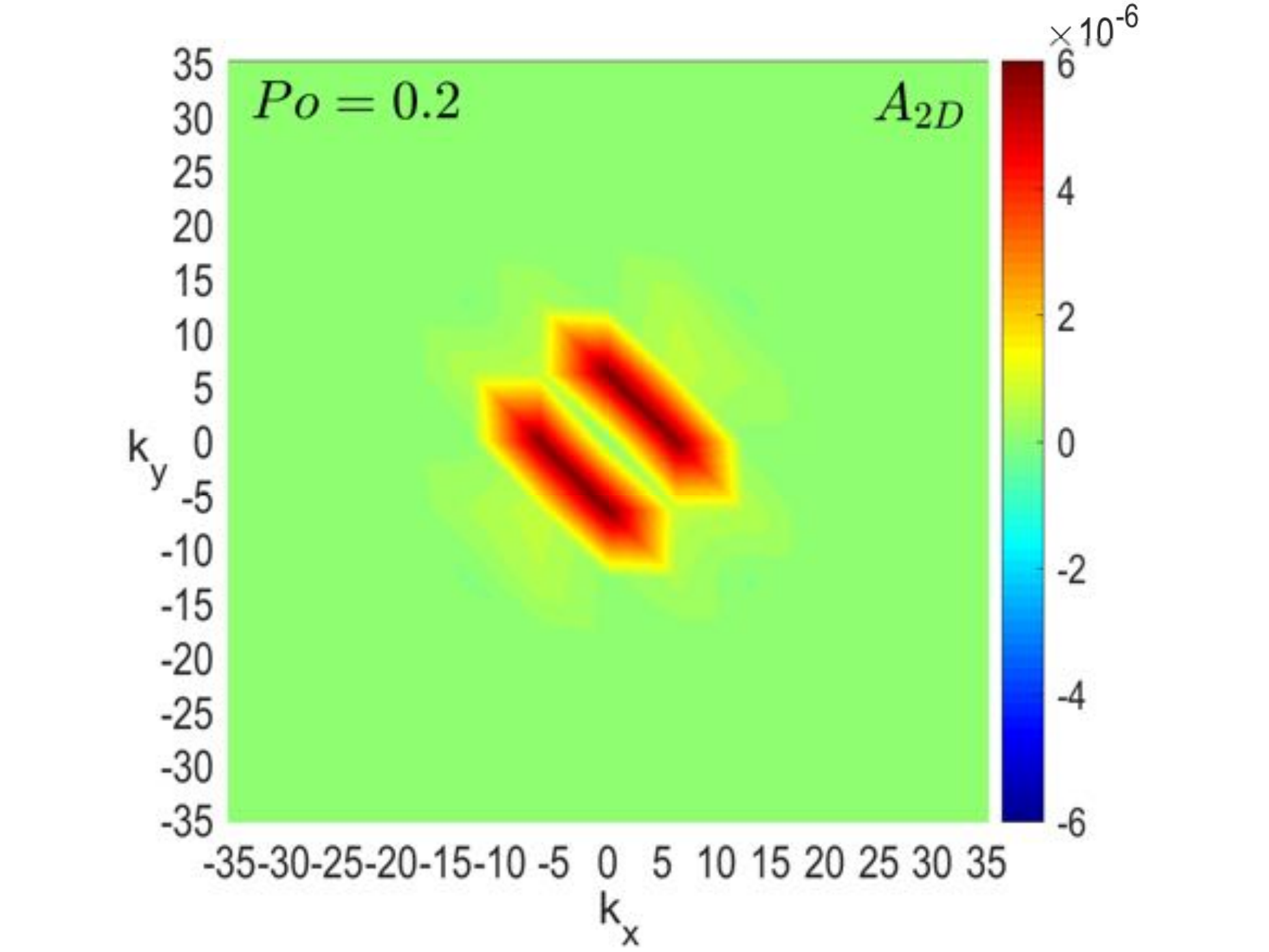}
\includegraphics[scale=0.295]{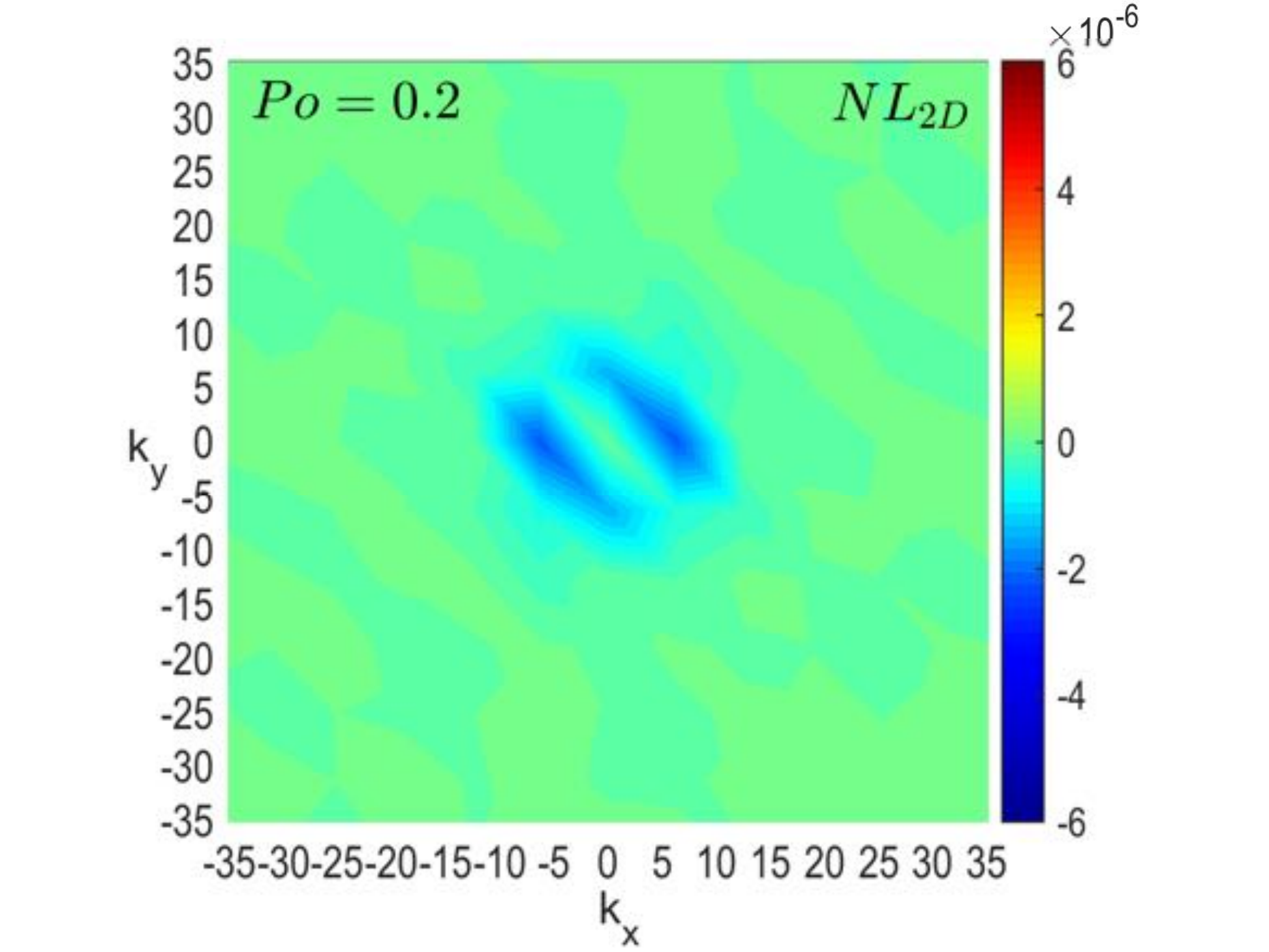}
\includegraphics[scale=0.295]{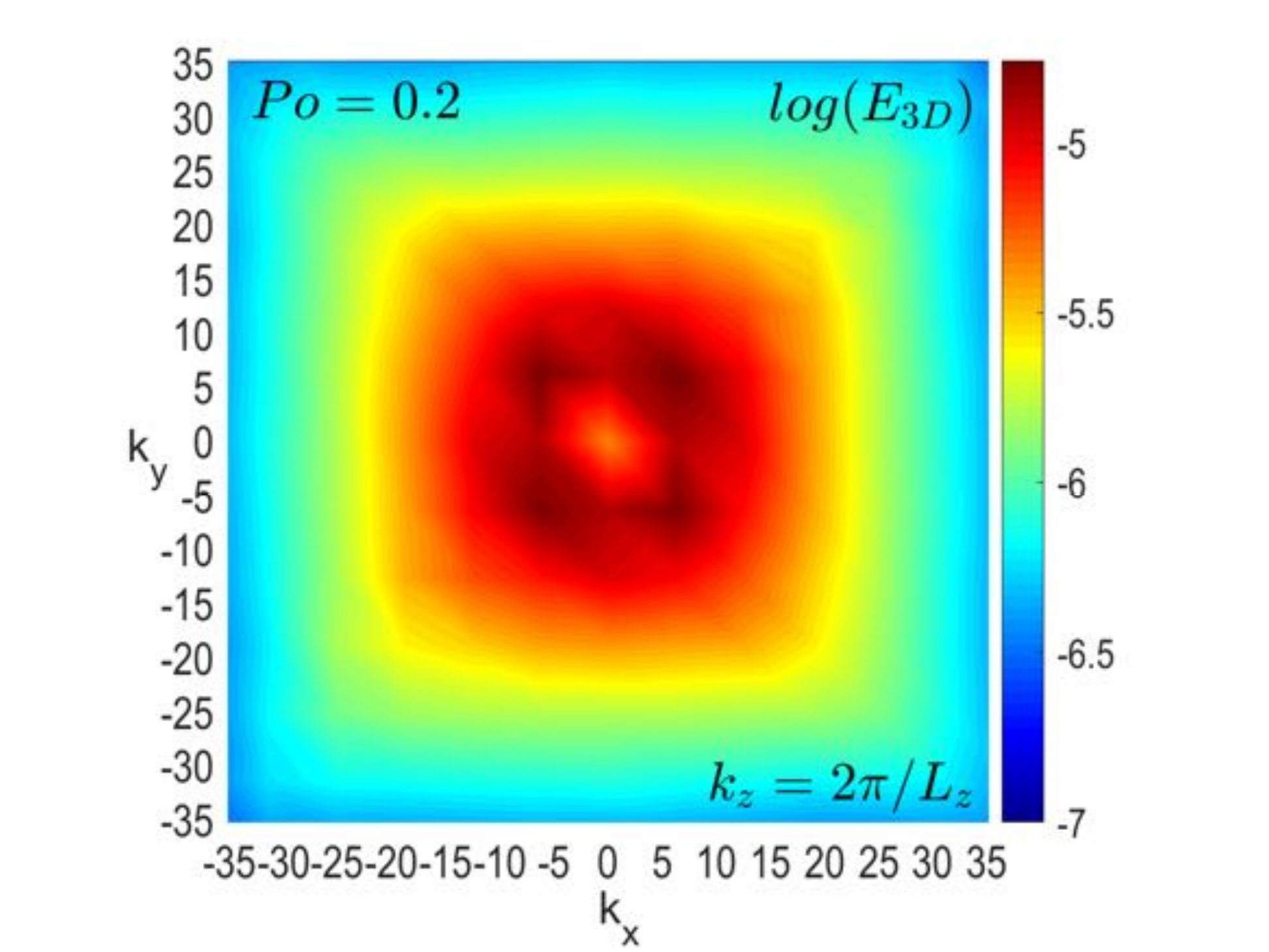}
\includegraphics[scale=0.295]{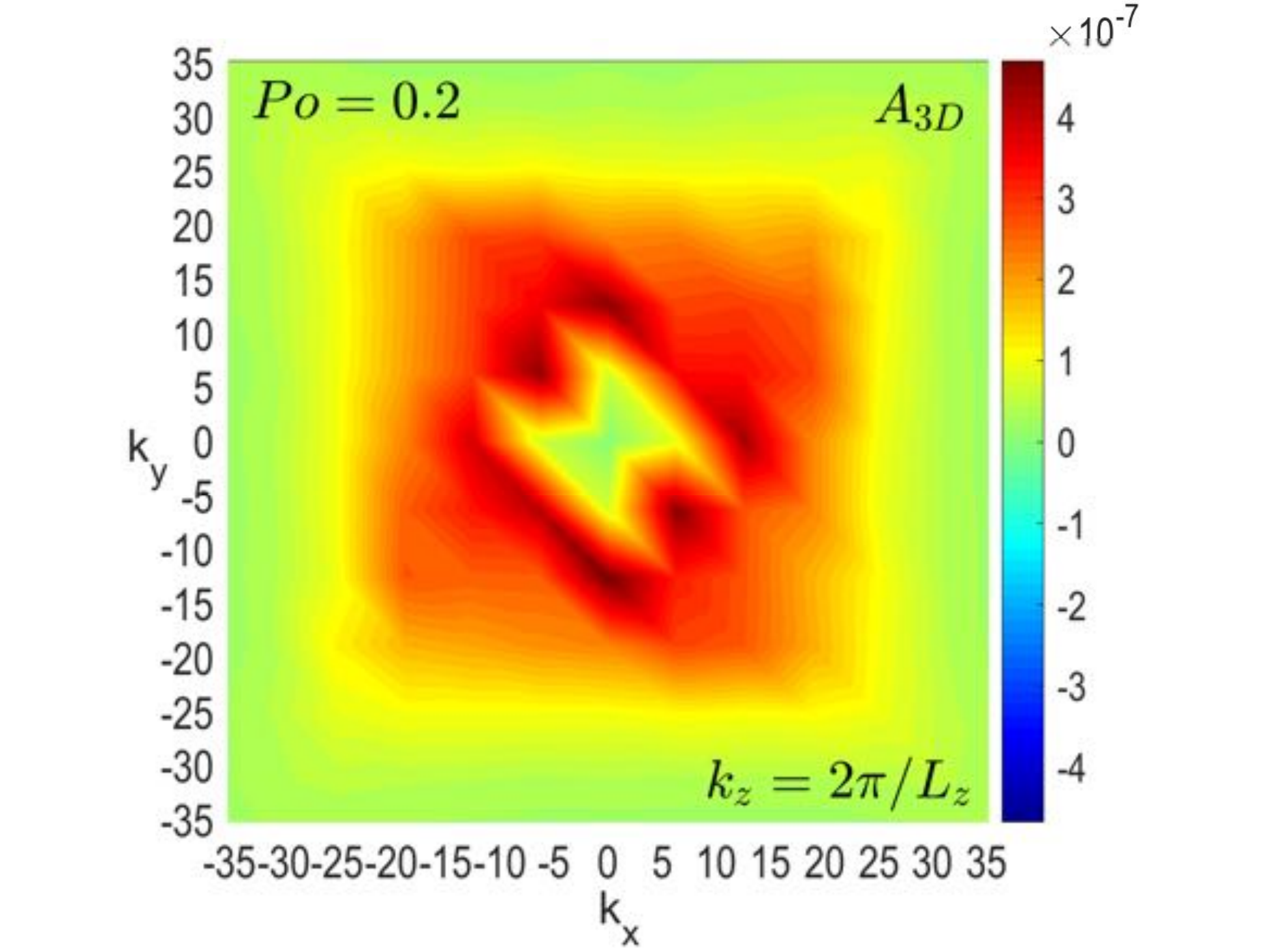}
\includegraphics[scale=0.295]{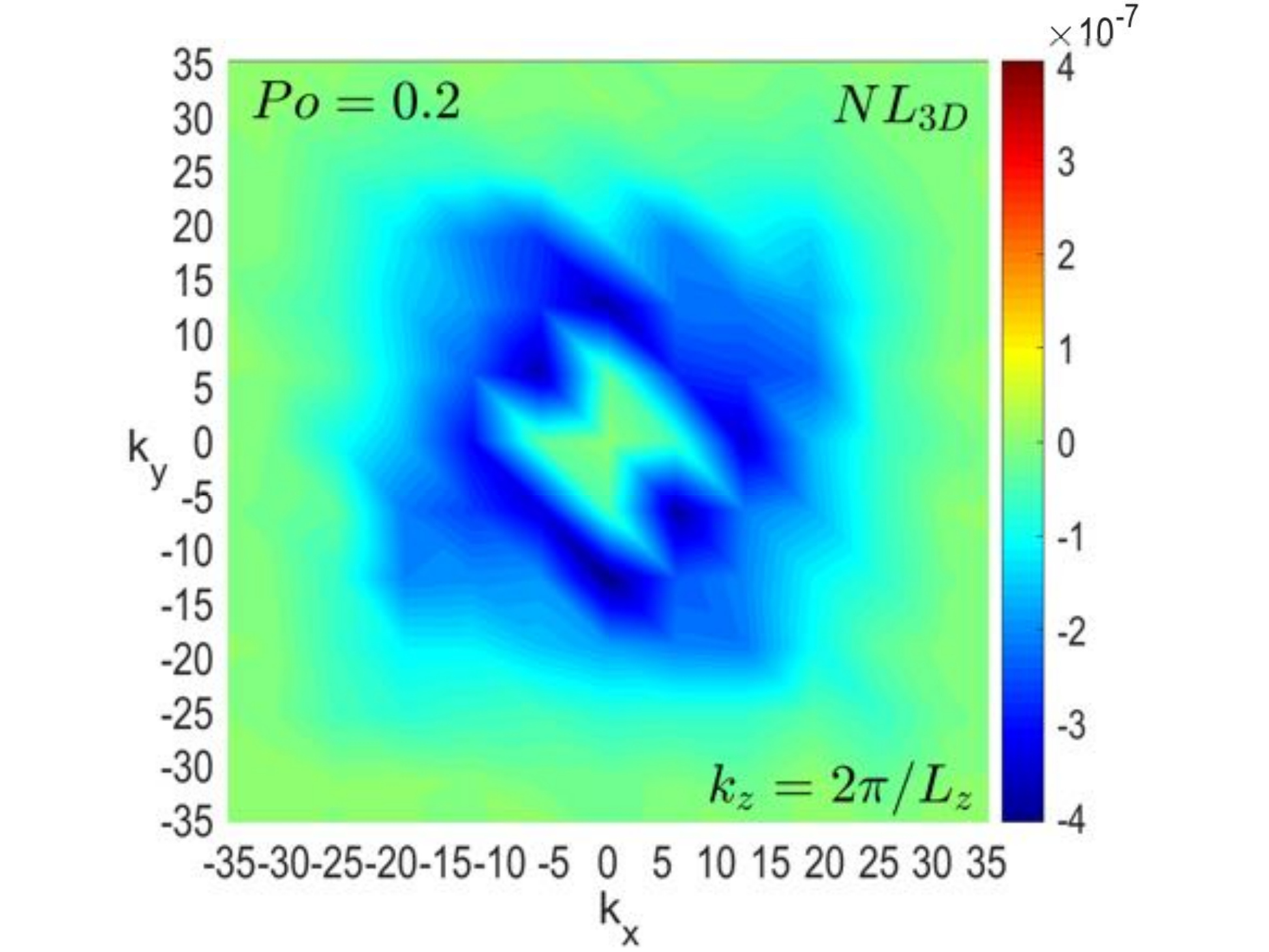}
\caption{Maps of the time-averaged spectral energy $E$ (left panels), injection $A$ (middle panels) and the total nonlinear transfer term $NL$ (right panels) in the $(k_x,k_y)-$plane for 2D vortical modes with $k_z=0$ (top row) and 3D wave modes at the first $k_z=2\pi/L_z$ in the box (bottom row) in the quasi-steady turbulent state with $Po=0.2$ and $Re=10^{4.5}$. Note the noticeable anisotropy of 2D manifold spectra compared with nearly isotropic spectra of 3D manifold.}\label{fig:spectra_maps_po02}
\end{figure*}
\begin{figure*}
\includegraphics[scale=0.295]{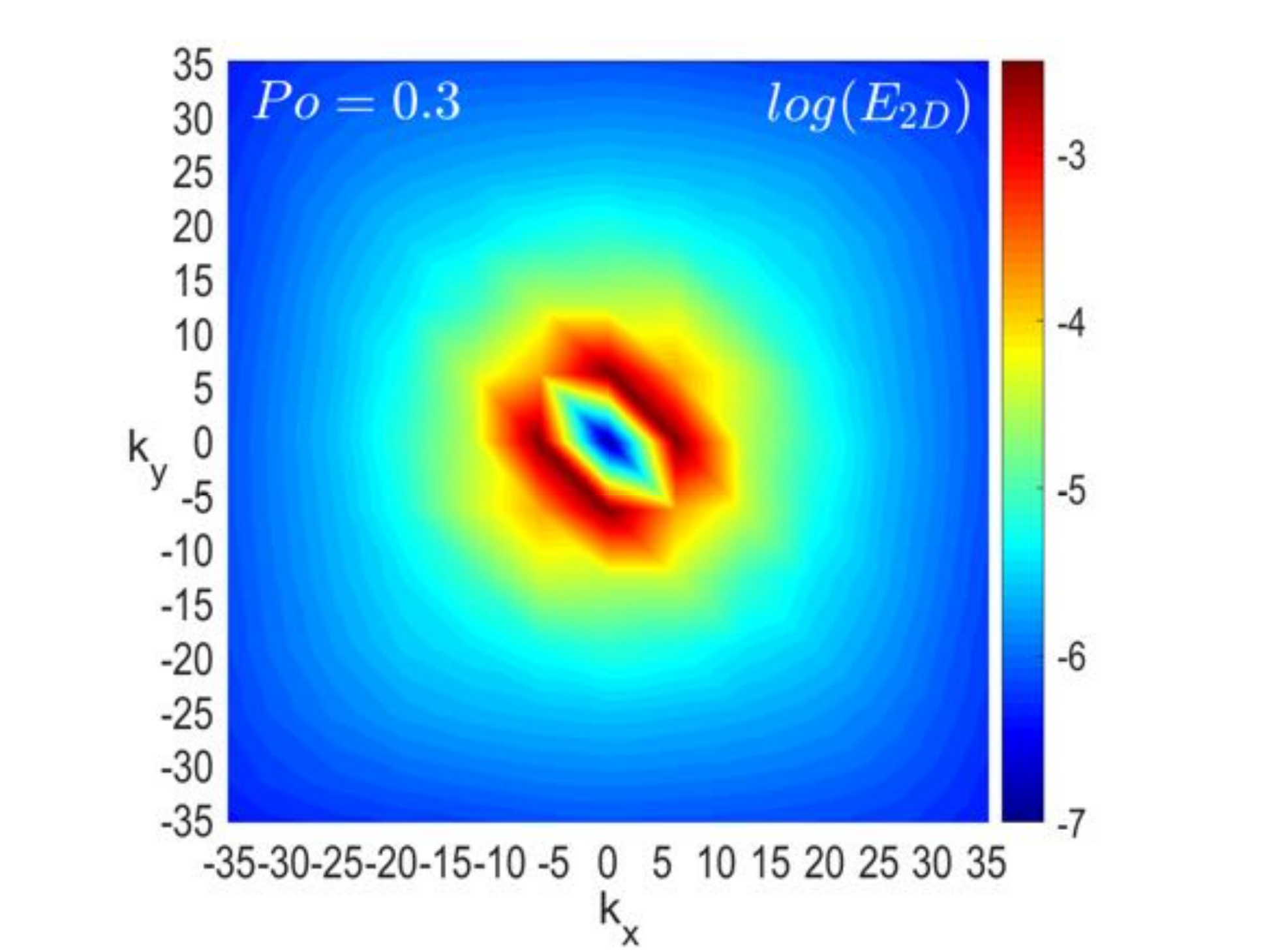}
\includegraphics[scale=0.295]{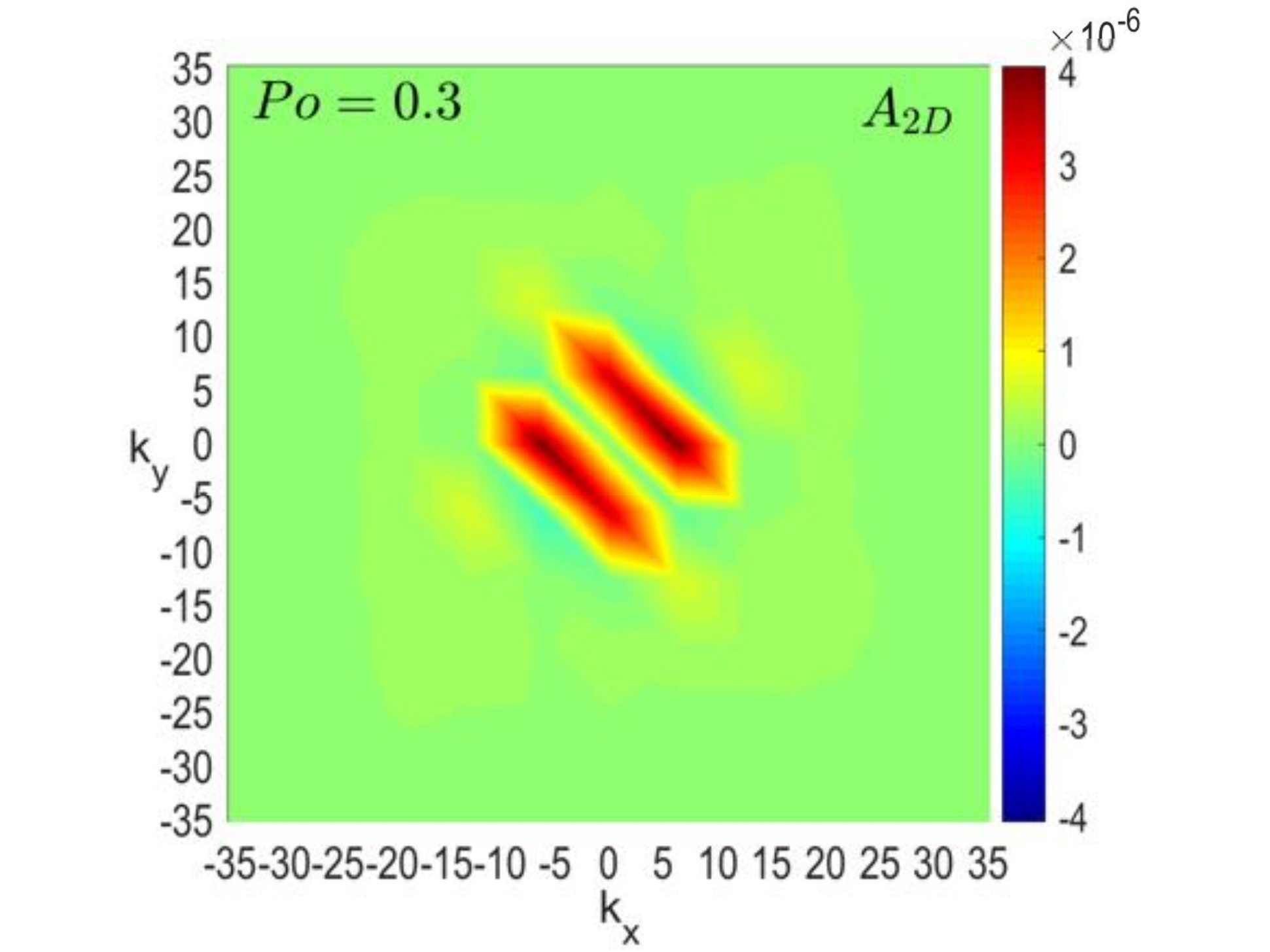}
\includegraphics[scale=0.295]{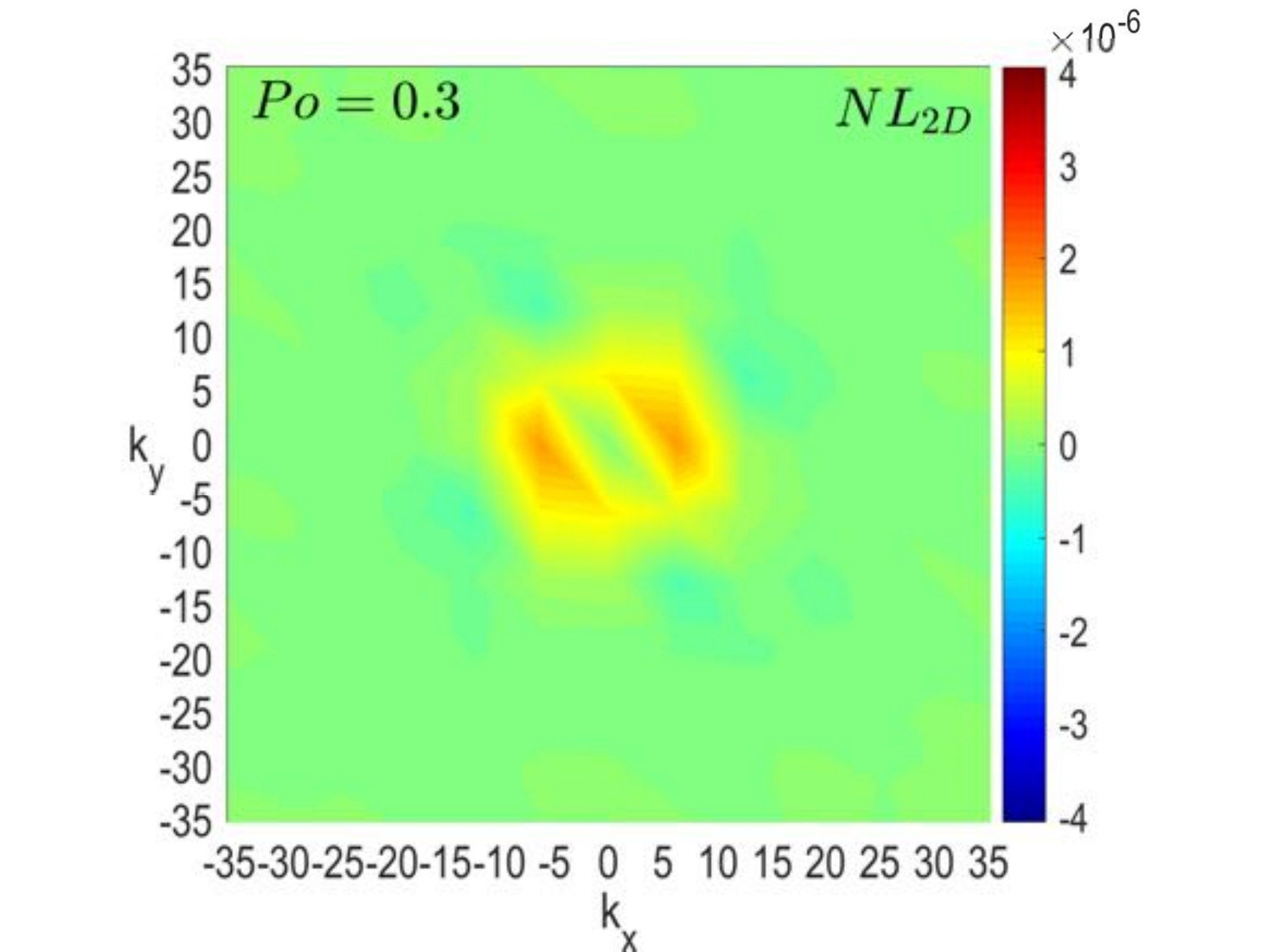}
\includegraphics[scale=0.295]{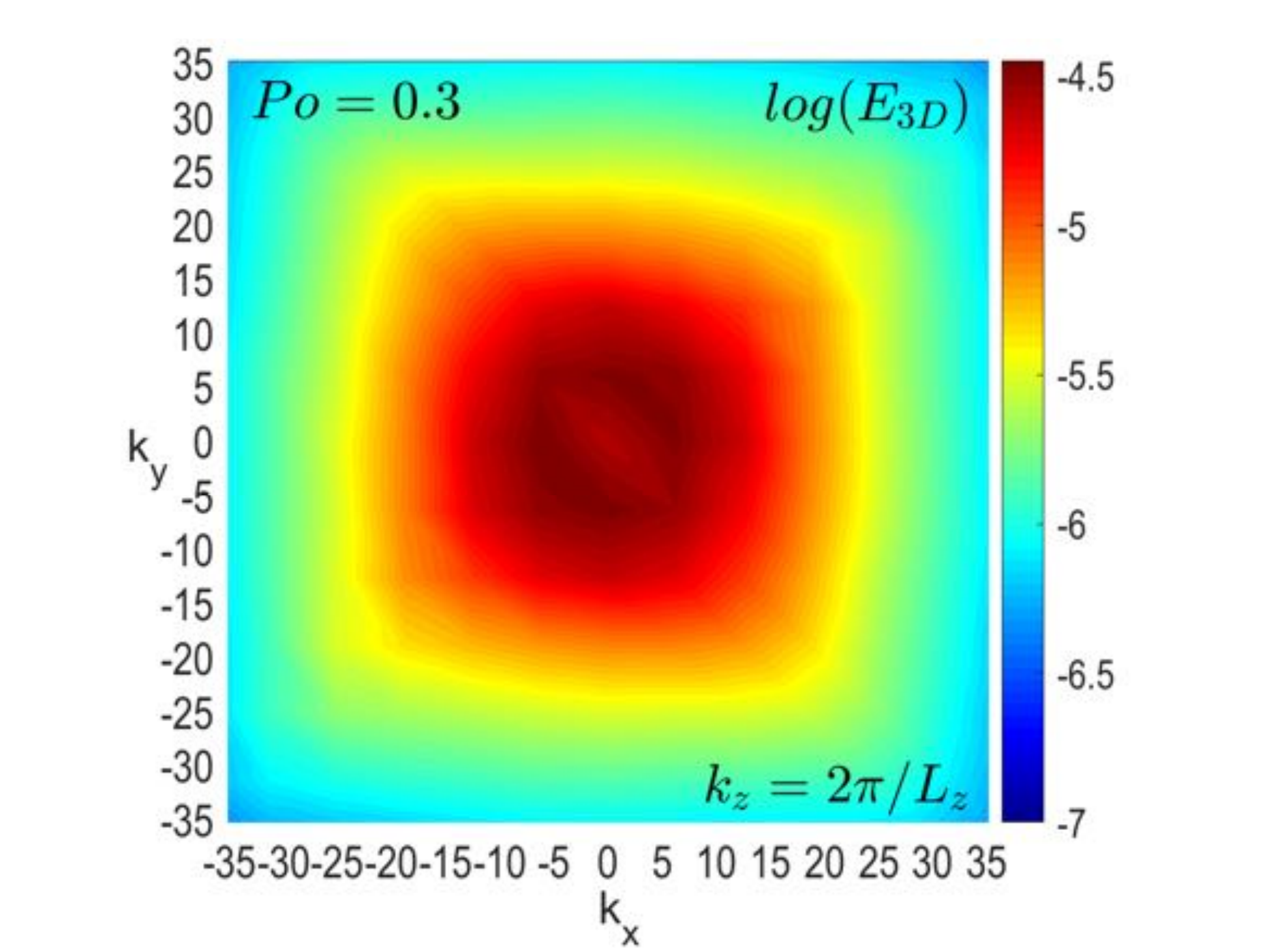}
\includegraphics[scale=0.295]{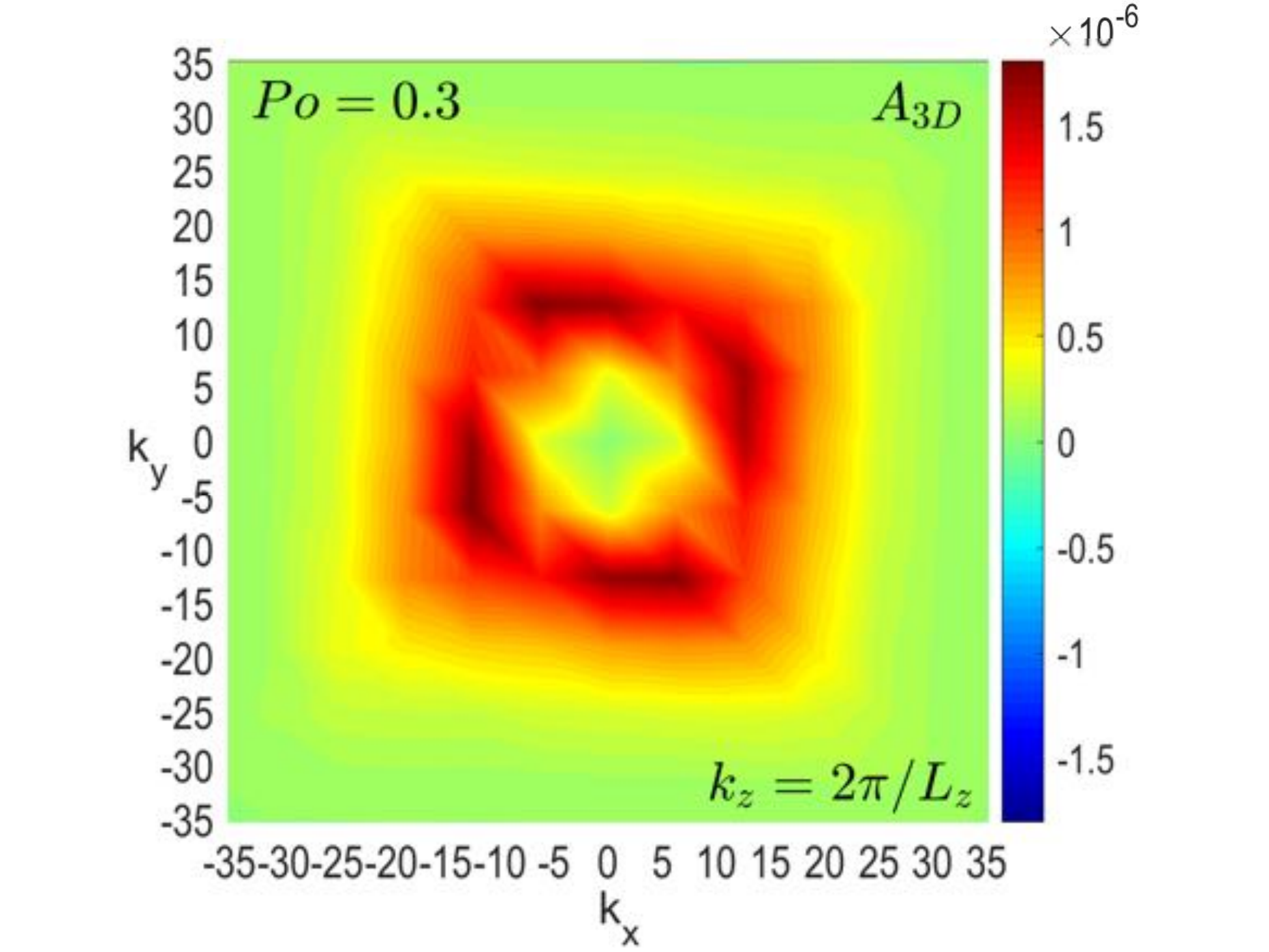}
\includegraphics[scale=0.295]{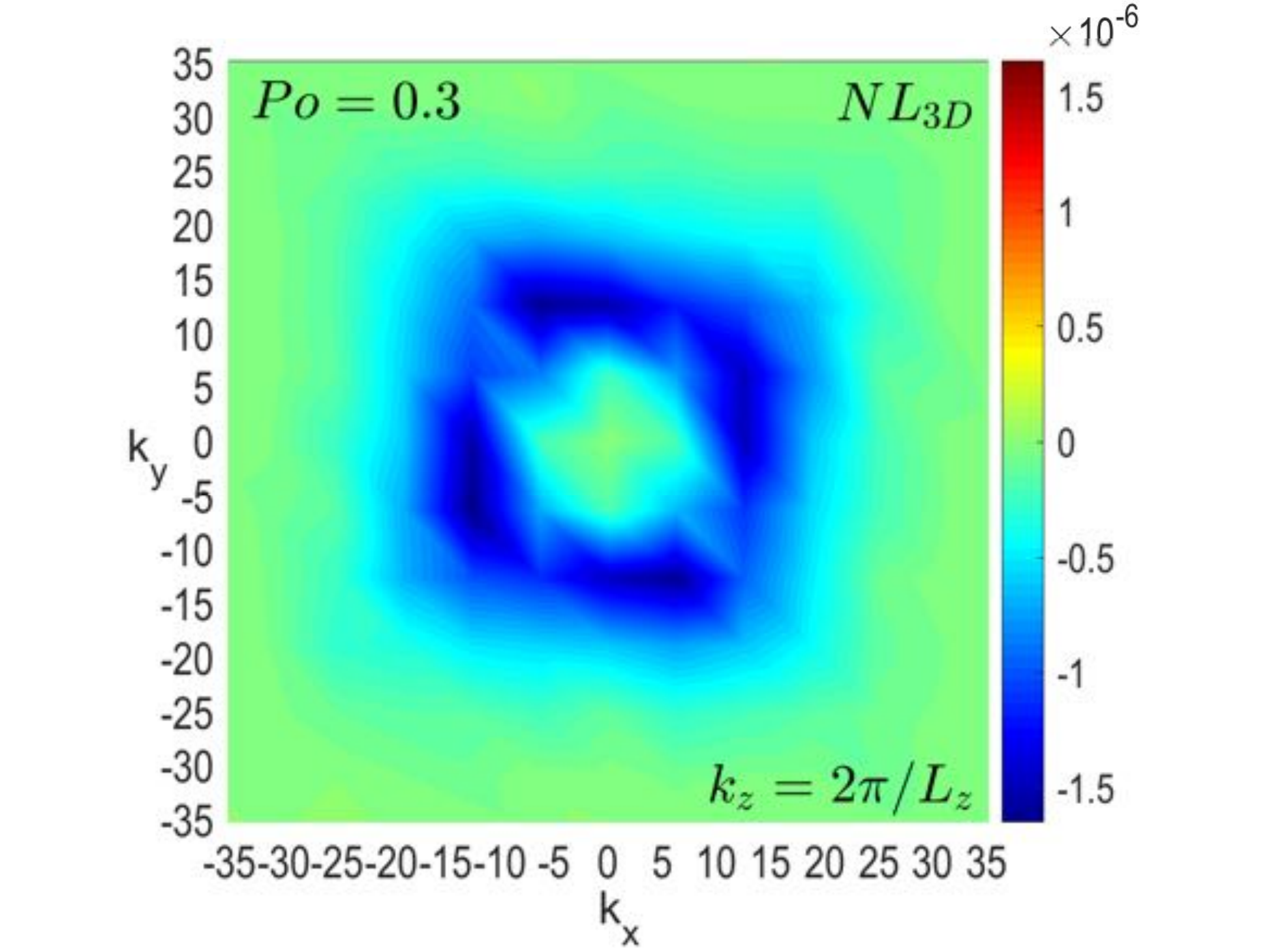}
\caption{The same as in Fig. \ref{fig:spectra_maps_po02}, but for $Po=0.3$. Note the change of sign of $NL_{2D}$ compared to the $Po=0.2$ case.}\label{fig:spectra_maps_po03}
\end{figure*}

\begin{figure*}
\centering
\includegraphics[scale=0.55]{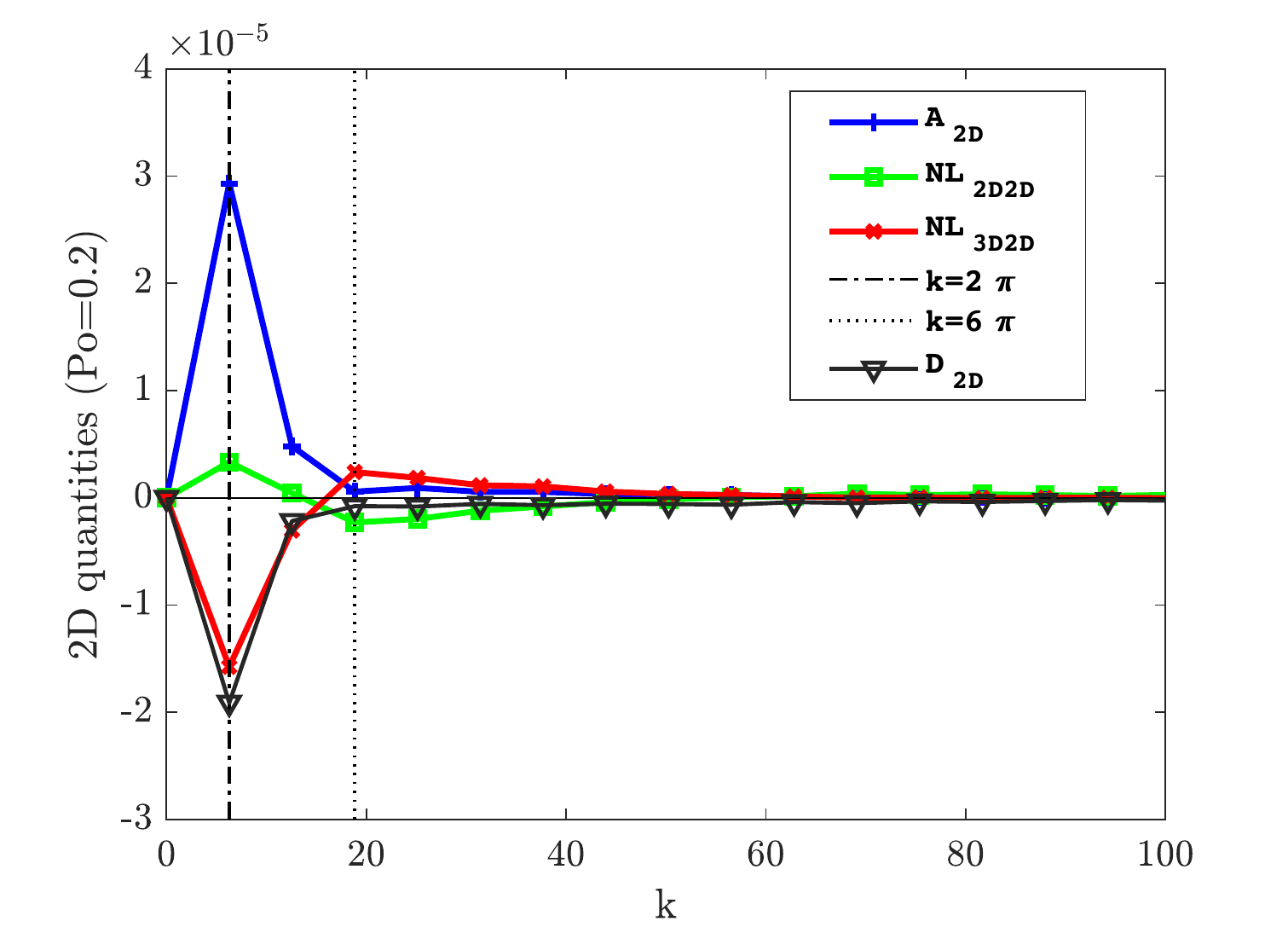}
\includegraphics[scale=0.55]{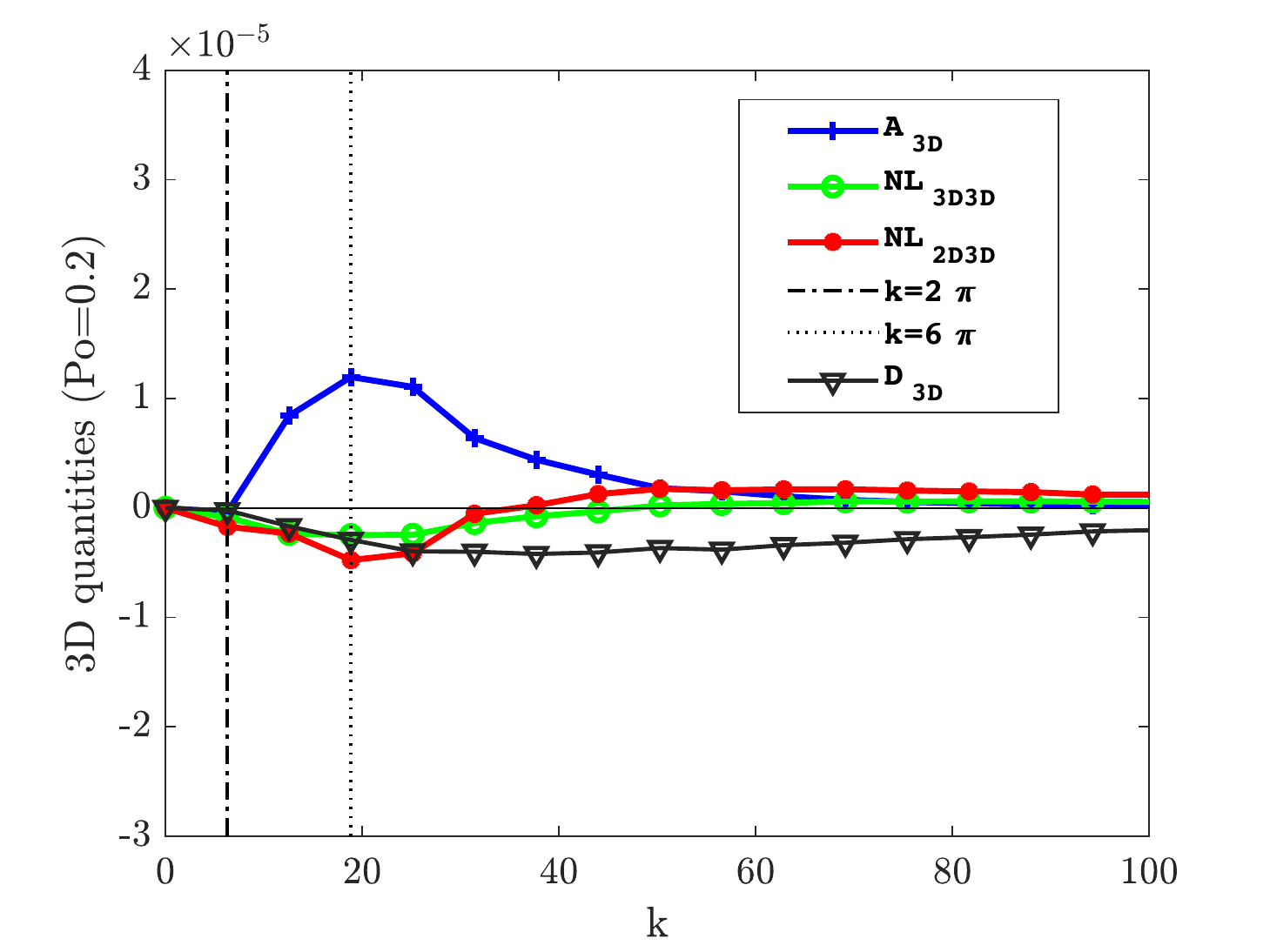}
\includegraphics[scale=0.55]{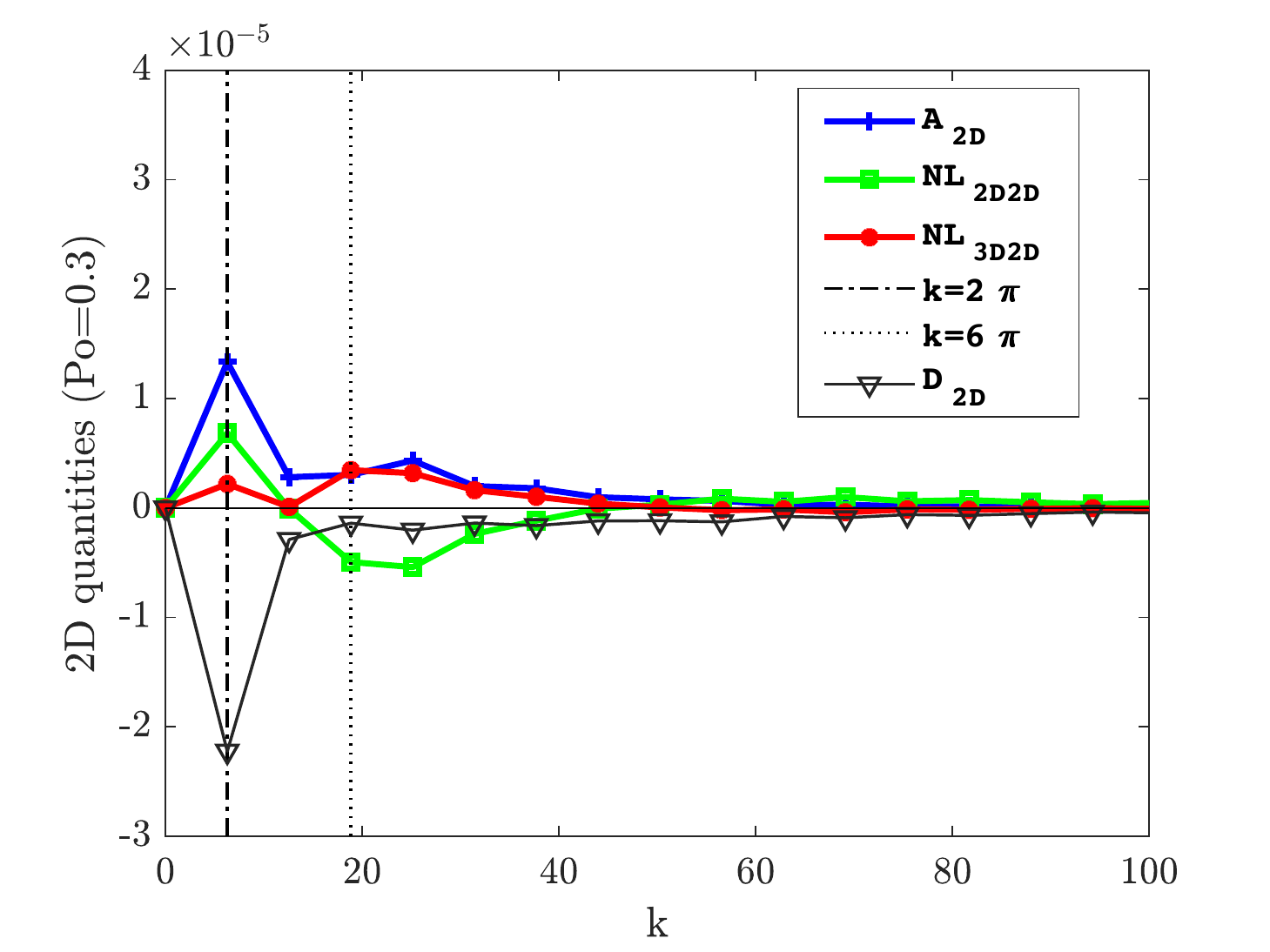}
\includegraphics[scale=0.55]{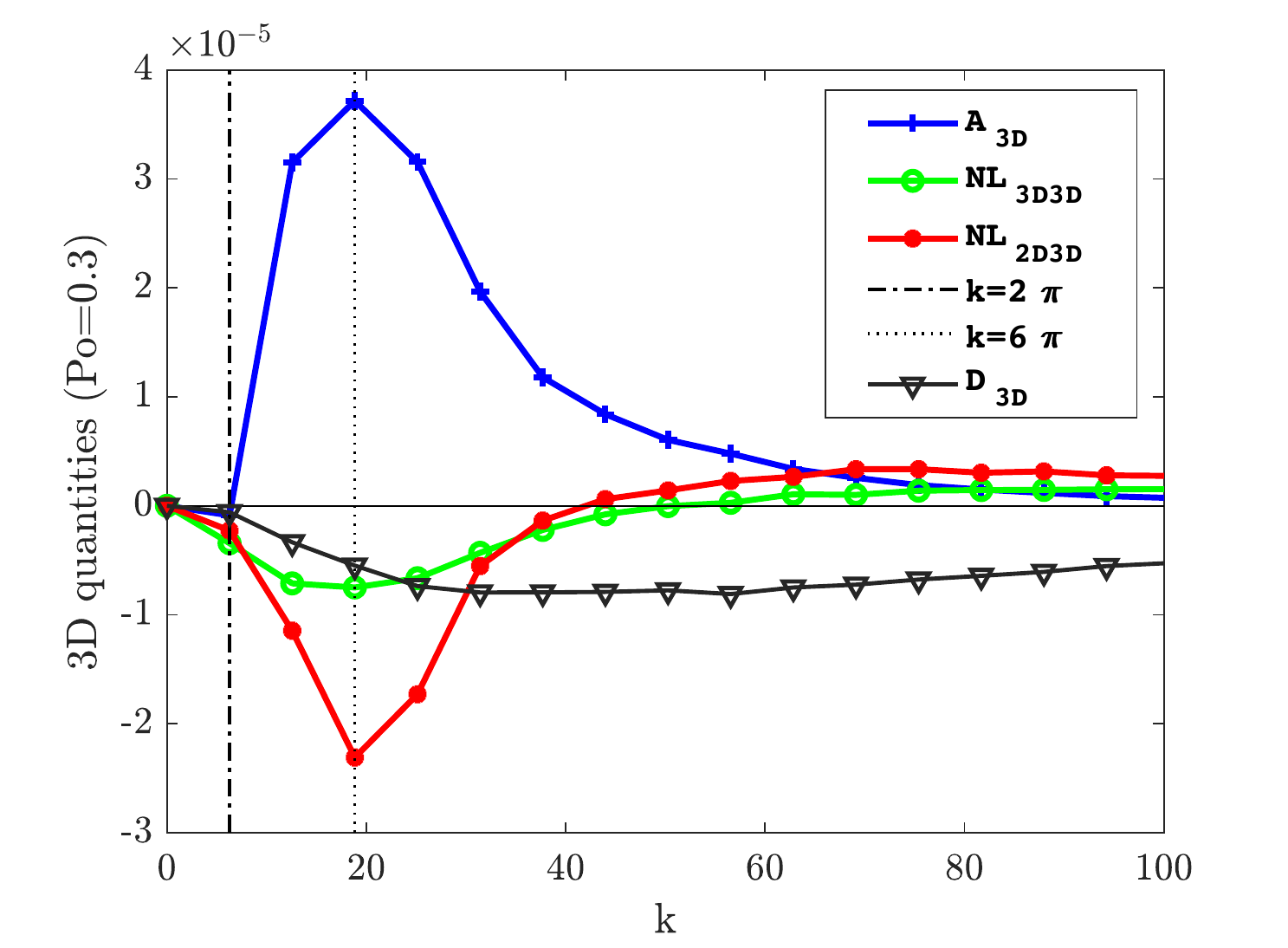}
\caption{Shell-averaged spectra for the injection $A$ (blue), viscous dissipation $D$ (black), and nonlinear transfers among modes inside 2D manifold, $NL_{2D2D}$ (green, left panels), inside 3D manifold $NL_{3D3D}$ (green, right panels) and cross transfers $NL_{3D2D}$ (red, left column) and $NL_{2D3D}$ (red, right panels) between the modes in these two manifolds. For 2D vortical modes (left panels) and 3D wave modes (right panels) in the quasi-steady turbulent state at $Po=0.2$ (top row), 0.3 (bottom row) and $Re=10^{4.5}$. Dotted black vertical line marks the peak of the injection term $A_{3D}$, while the dash-dotted line the peak of $A_{2D}$.}\label{fig:inj_nonlinear_re1e4d5}
\end{figure*}

\subsection{Quasi-steady turbulence: $Po=0.2$ and $0.3$}

In this section we present a similar analysis of the dynamical processes in Fourier space, focusing on the regime of large $Po=0.2$ and 0.3  where the saturated state is characterized by a quasi-steady turbulence, where both 3D and 2D mode energies evolve in time with only mild fluctuations in contrast to the small-$Po$ regime displaying quasi-periodic bursts (Fig. \ref{fig:1a}). In this section, we keep the Reynolds number fixed at $Re=10^{4.5}$ to focus on the impact of increasing precession on the spectral properties and dynamical balances of the turbulence.

\subsubsection{Energy spectrum}

Figure \ref{fig:spectra_re1e4d5} shows the shell-averaged kinetic energy spectra of 2D vortices and 3D waves divided further into horizontal, $E_{h}=(|\bar{u}_x|^2+|\bar{u}_y|^2)/2$, and vertical, $E_z=|\bar{u}_z|^2/2$, components at $Po=0.2$ (left panel) and 0.3 (right panel). The time-average has been done over $\Delta t \approx 1000$ in the saturated state. We have checked the robustness of the results by comparing averages over different time ranges and found a good agreement. In this figure, the grey vertical line marks the location of Zeman wavenumber $k_{\Omega}$ defined in the presence of the energy injection due to the precessional instability as $k_{\Omega}=(\Omega^3/\langle A_{2D}+A_{3D}\rangle)^{1/2}$ (equal to $1/\langle A_{2D}+A_{3D}\rangle^{1/2}$ in non-dimensional units), where $\langle A_{2D}+A_{3D}\rangle$ is the volume-averaged total injection term introduced. This definition of $k_{\Omega}$ differs from the usual one used in the rotating turbulence theory \cite{Zeman1994, Muller2007, Alexakis2018} in that the energy injection rate, $\varepsilon$, due to an external forcing is replaced here by the injection due to the instability but these definitions are consistent in steady state.\\
The most remarkable aspect is the different shape and scaling of energy spectra for the 2D and 3D modes. The 2D mode energy dominates over the 3D one at small wavenumbers $k\lesssim 10$ where it increases with decreasing $k$, reaching a maximum at the largest box scale, with its horizontal component being about an order of magnitude larger than the vertical one. This corresponds to large-scale horizontal vortical motions in physical space, as is seen in Figs. \ref{fig:snapshot_wz} and \ref{fig:contour_2d_3d}. At higher $10\lesssim k \lesssim k_{\Omega}$, the horizontal and vertical components are comparable in the $E_{2D}$ spectrum and its slope is close to $k^{-3}$, exhibiting the same power-law dependence of rotating geostrophic 2D turbulence \cite{Smith1996, Smith1999, Sen2012, Pouquet2013, Buzzicotti2018, Khlifi2018}, which does not appear to change with precession parameter $Po$.  

The energy spectrum of 3D waves, $E_{3D}$, has a peak at larger $k\approx 15$ than that of $E_{2D}$ (which approximately coincides with the peak of injection $A_{3D}$ in Fig. \ref{fig:inj_nonlinear_re1e4d5}). $E_{3D}$ decreases then at lower wavenumbers, while at higher wavenumbers $15 \lesssim k \lesssim k_{\Omega}$ follows a scaling $\sim k^{-2 \pm 0.5}$ which has been typically observed in forced rotating turbulence of inertial waves \cite{Galtier2003, Muller2007, Biferale2016, Khlifi2018, Alexakis2018}. However, in contrast to these papers using a forcing in a very narrow wavenumber band, we do not prescribe the forced wavenumber a priori, instead the injection wavenumbers are determined by the background flow itself through the precessional instability and extend over a broad range (see below). Precession influences the scaling exponent of the $E_{3D}$ spectrum: its slope seems to become shallower with increasing $Po$, as is seen in Figs. \ref{fig:spectra_re1e4d5} (compare left and right panels) and \ref{fig:3d_kolmogorov} below. Like for 2D mode energy, also for 3D mode energy, horizontal and vertical components are comparable at higher wavenumbers, but the horizontal one dominates at lower wavenumbers. 

It is seen in Fig. \ref{fig:spectra_re1e4d5} that the observed power-law scalings of both 2D and 3D mode energy spectra occur at $k<k_{\Omega}$ and therefore are strongly influenced by rotation and precession, deviating from the classical Kolmogorov $k^{-5/3}$ spectrum. However, as it is seen in this figure, with increasing precession strength $Po$, the Zeman wavenumber $k_{\Omega}$ (grey vertical lines) decreases, that is, the effect of rotation becomes increasingly weaker for smaller and smaller $k$. We will see below that in this case the energy spectrum at $k>k_{\Omega}$ indeed approaches the Kolmogorov spectrum.  

\subsubsection{Dynamical balances in Fourier space}

To see the structure of spectra of energy and dynamical terms in the quasi-steady precessional turbulence, in Figs.~\ref{fig:spectra_maps_po02} and \ref{fig:spectra_maps_po03} we show the time-averaged spectra of the kinetic energy $E$, energy injection $A$ and the total nonlinear transfer $NL$ in two different horizontal $(k_x, k_y)-$planes: at $k_z=0$ for the 2D modes and at $k_z=2\pi/L_z$ for the 3D wave modes. We have chosen $k_z=2 \pi/L_z$ because it corresponds to the maximum injection along $k_z$-axis, for which therefore the precession instability reaches the largest growth rate in the box \cite{Salhi2009}.
The most striking observation is the anisotropic nature of the 2D manifold in Fourier space (top row), for $A_{2D}$ and $NL_{2D}$ and hence for the kinetic energy spectrum $E_{2D}$ determined by the joint action of these terms, which are all localized at smaller wavenumbers, with a clear inclination towards the $k_x$ axes. The injection term $A_{2D}$ is always positive, implying some energy injection from the basic flow into vortices, however, $NL_{2D}$ is negative at the same wavenumbers for $Po=0.2$, but changes sign at $Po=0.3$. As a result, the dynamical balances for 2D modes are different for these two values of $Po$, which will be discussed below. In contrast, the 3D manifold exhibits a quasi-isotropic distribution (bottom row, similarly at larger $k_z>2\pi/L_z$ not shown here) for both $Po=0.2$ and 0.3, whose range extends over larger wavenumbers than that of 2D quantities. Notice the impact of $Po$ on the 3D manifold: for $Po=0.2$ a weak preferential direction (anisotropy) is present in the center of the wave-plane which tends to isotropize increasing $Po$. Comparing the $A_{3D}$ and $NL_{3D}$, the first injection term, which is due to the precessional instability, is always positive and appreciable at $5<|k_x|, |k_y|<25$ (yellow/red area), while the second nonlinear term is negative (blue) and also appreciable at these wavenumbers. The similar shape of these two functions in Fourier space and comparable absolute values, imply that these two processes are in balance: 3D modes receive energy from the precessional background flow predominantly in the range $5<|k_x|, |k_y|<25$, while nonlinearity, counteracting injection at these wavenumbers, transfer this energy to other 3D and 2D modes with different wavenumbers. 

From Figs.~\ref{fig:spectra_maps_po02} and \ref{fig:spectra_maps_po03} showing the distribution of total nonlinear terms $NL_{2D}$ and $NL_{3D}$ in Fourier space, one cannot establish specifically what kind of transfer mechanisms operate, that is, whether the cascades inside a given manifold are direct or inverse or if there are transfers of energy between these two manifolds, since these terms encapsulate nonlinear interactions among all kinds of modes. To get insight into the details of linear (energy injection) and nonlinear cascade processes in the precessional turbulence, in Fig. \ref{fig:inj_nonlinear_re1e4d5} we show the shell- and time-averaged spectra of all the dynamical terms -- injection $A$, nonlinear $NL$ and dissipation $D$ terms -- entering Eqs. (\ref{eq:e2d}) and (\ref{eq:e3d}) again for $Po=0.2$ and $Po=0.3$, as we did for the bursty case $Po=0.075$ in the above section.
\begin{figure}
\centering
\includegraphics[scale=0.4]{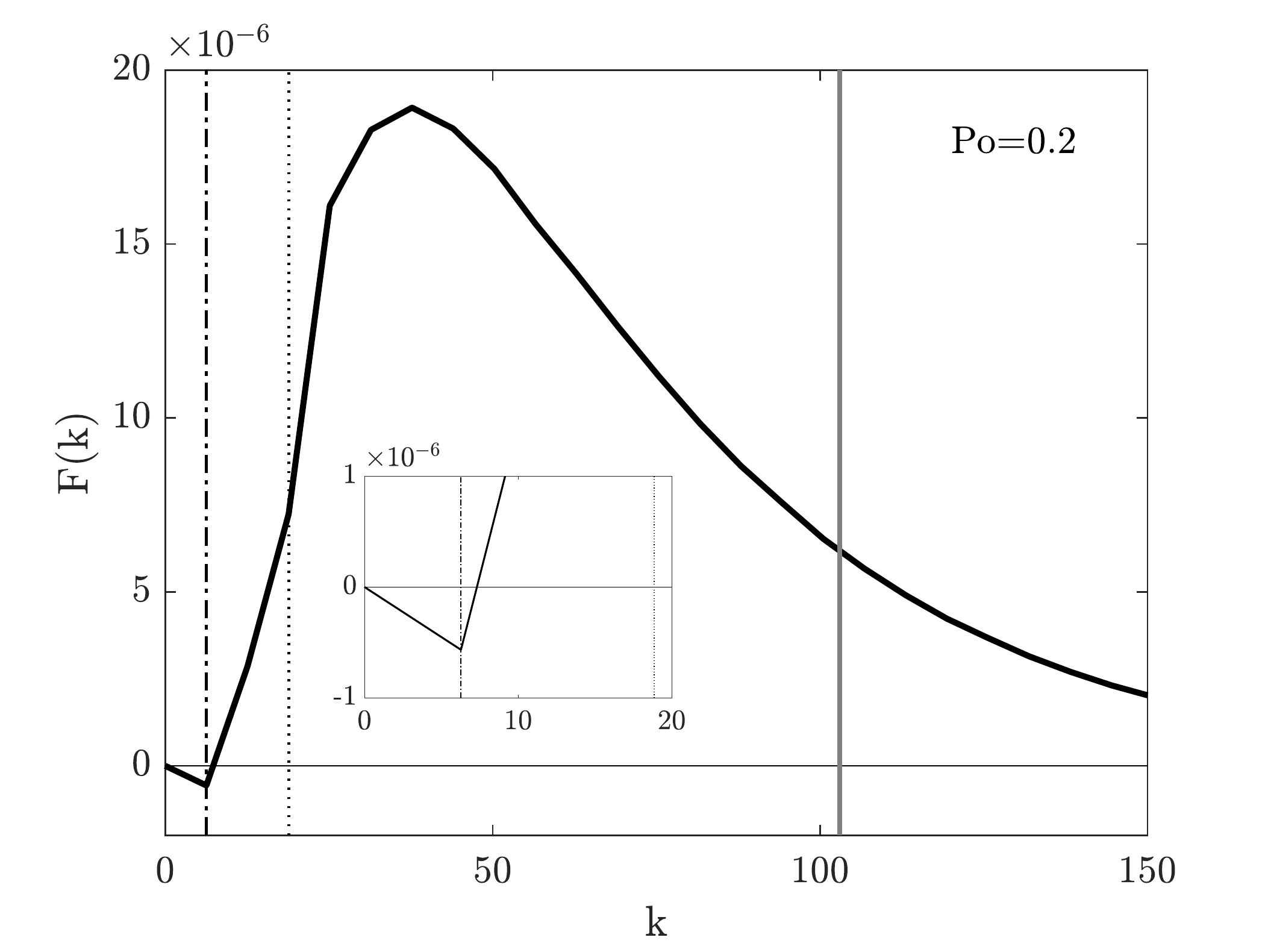}
\includegraphics[scale=0.4]{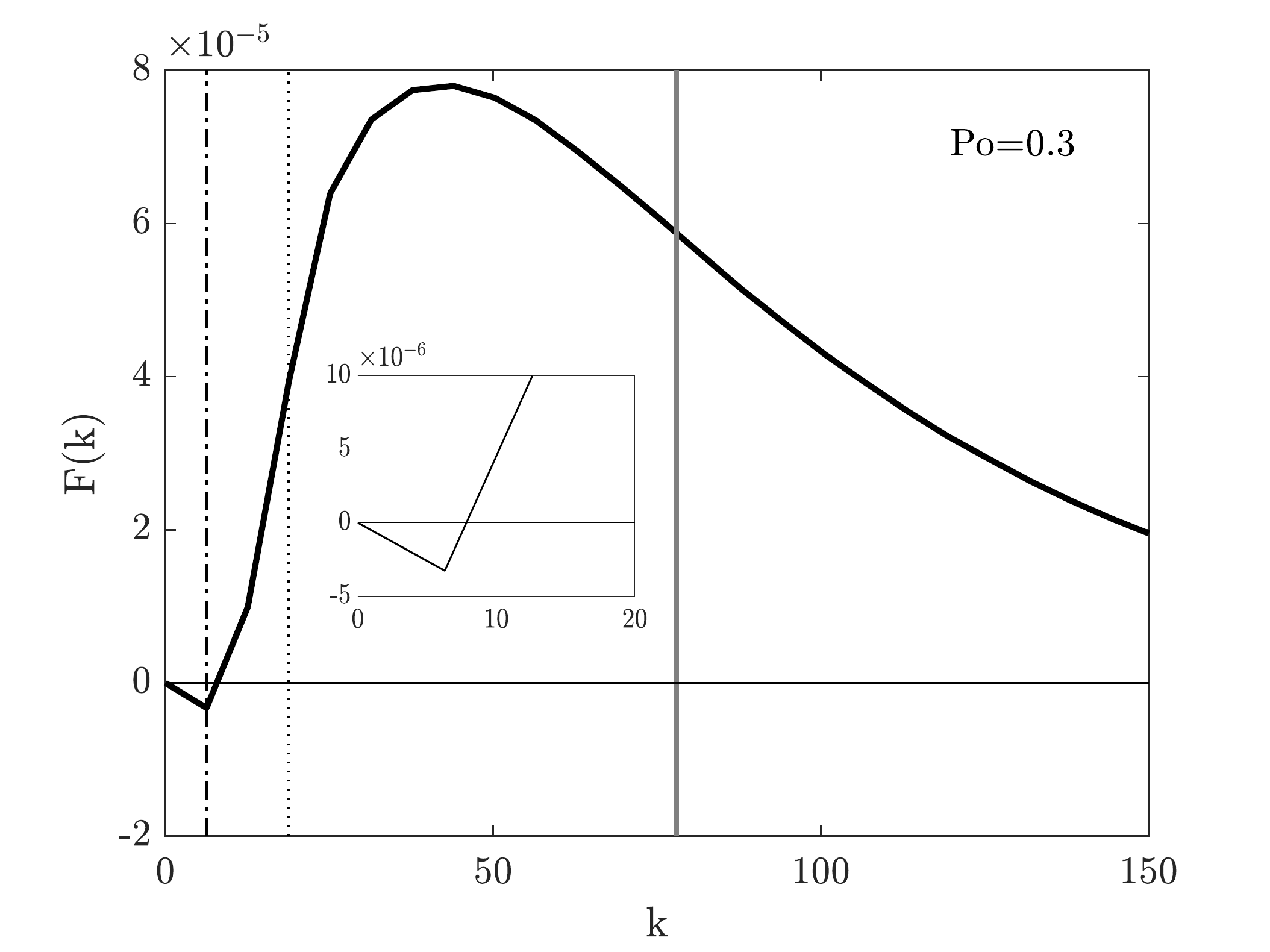}
\caption{Total energy flux $F(k)$ with the vertical lines representing, for reference, the wave numbers $k_{in,2D}=2\pi$ (dashed-dotted) and  $k_{in,3D}=6\pi$ (dotted) at the peak of the injection, respectively, for 2D and 3D modes as well as the Zeman wavenumber $k_{\Omega}$ (solid grey). Top panel is for $Po=0.2$ and bottom for $Po=0.3$. The flux is predominantly positive, $F>0$, for larger wavenumbers $k>k_{in,2D}$ corresponding to forward cascade. Inset zooms into the inverse cascade range at small $k$, where $F<0$.}\label{fig:flux_po02_03}
\end{figure}
\begin{figure}[t]
\centering
\includegraphics[scale=0.62]{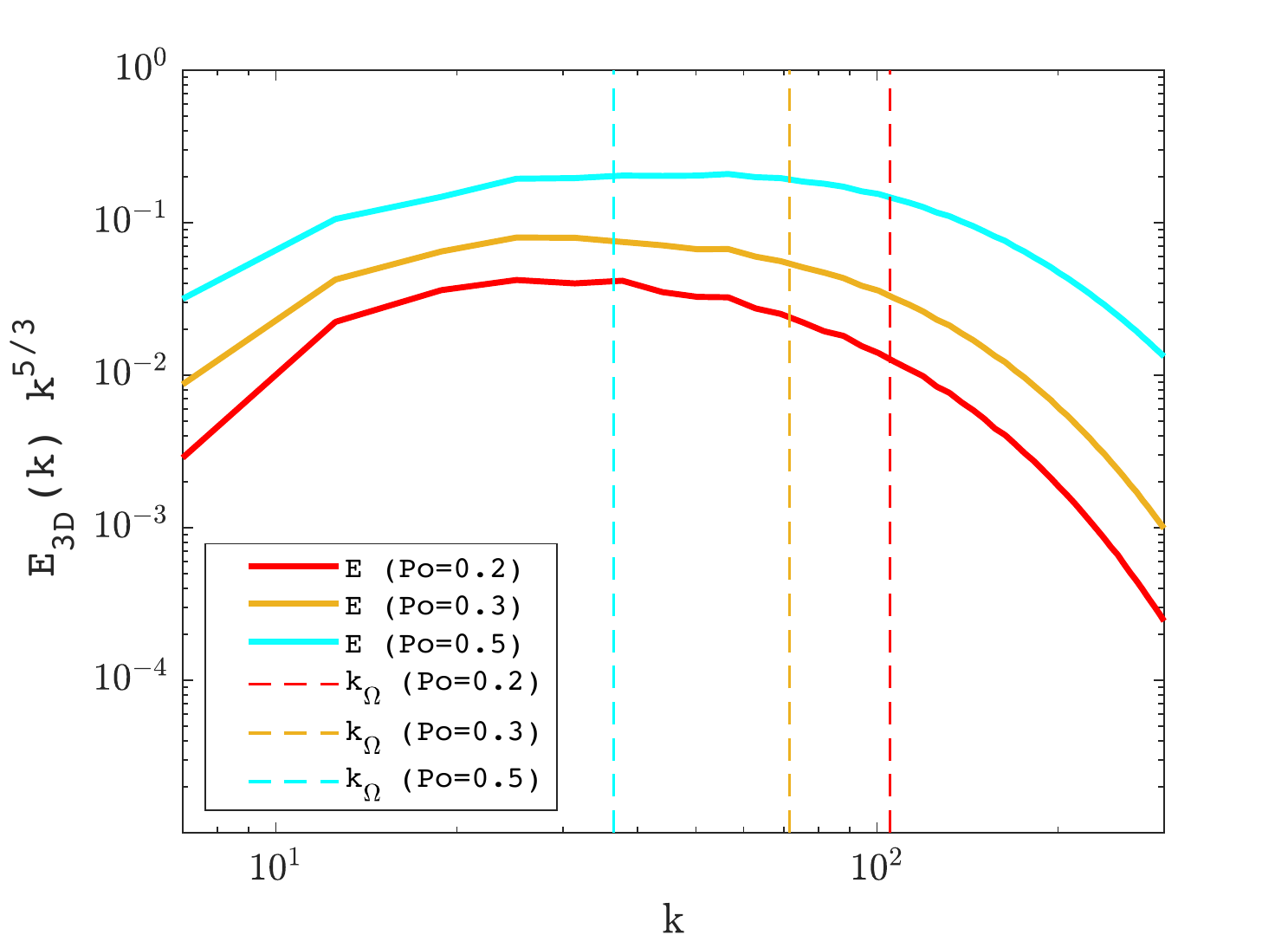}
\caption{Kinetic energy spectra for 3D wave modes compensated by the Kolmogorov law $k^{-5/3}$. Solid lines represent $E_{3D}$ for three different precession parameters $Po=0.2$, 0.3 and 0.5 at $Re=10^{4.5}$, while the vertical dashed lines represent the corresponding (in terms of colors) Zeman wavenumbers $k_{\Omega}$. As $Po$ is increased, $k_{\Omega}$ decreases and the spectrum approaches Kolmogorov scaling at $k>k_{\Omega}$.}\label{fig:3d_kolmogorov}
\end{figure}

The precession instability injects energy into 3D waves, which is described by positive $A_{3D}$ (blue, right panels). Unlike the case of a forcing localized about certain wavenumber \cite{Buzzicotti2018}, the injection due to the instability extends over a range of wavenumbers, reaching a peak at $k=6\pi$, and increases with increasing $Po$. Since it is a quasi-steady state, the energy injection is balanced by nonlinear transfers, $A_{3D}+NL_{3D3D}+NL_{2D3D}\approx 0$, at those dynamically active wavenumbers where $A_{3D}>0$ is appreciable (the role of viscous dissipation $D_{3D}$ (black, right panels) is not important at these wavenumbers). Therefore, being negative at those injection wavenumbers, $NL_{3D3D}<0$ (green, left panels) and $NL_{2D3D}<0$ (red, left panels) drain energy from the active 3D modes there and transfer it, respectively, to smaller-scale 3D waves due to positive $NL_{3D3D}>0$ at $k > 40$ (forward/direct cascade) and to 2D vortical modes. The latter process is mediated by positive $NL_{3D2D}>0$ (red, left panels) at $k > 18$, peaking at the same $k=6\pi$. At these wavenumbers, 2D-2D transfer term is negative $NL_{2D2D}<0$ (green, left panels), with a minimum also at $k=6\pi$, and causes inverse cascade of 2D mode energy to even smaller wavenumbers, where it is positive $NL_{2D2D}>0$ and reaches a maximum at $k=2\pi$ that corresponds to the the largest box scale. These vortices draw some energy from the basic flow as well due to $A_{2D}$ term, which has a peak at the same $k=2\pi$ as $NL_{2D2D}>0$. It is seen that $A_{2D}$ decreases, whereas $NL_{2D2D}$ increases with $Po$. Notice that in the linear regime, idealized steady 2D vortices would be stable against precessional instability (i.e., $A_{2D}=0$) in the basic flow $\boldsymbol{U}_b$ \cite{Kerswell1993}, so the positive $A_{2D}>0$ in this case can be attributed to the fact that these vortices are nonlinearly generated by the waves. The 2D-3D nonlinear interaction term $NL_{2D3D}$ is positive at large wavenumbers $k>40$, redistributing part of energy of 2D large-scale vortices back to smaller-scale 3D waves (forward cascade). Thus, as it is shown from Fig.\ref{fig:inj_nonlinear_re1e4d5} (left panels), similar to that for 3D modes, also for 2D modes, there is a balance among production of these modes by 2D-3D transfers, energy extraction, 2D-2D transfers, and viscous dissipation, $A_{2D}+NL_{2D2D}+NL_{3D2D}+D_{2D}\approx 0$. Note that viscous dissipation for 2D and 3D modes have completely different behavior (compare black curves in left and right panels). It is stronger and more significant for the 2D vortices: $D_{2D}$ has a clear a minimum at low wavenumber $k=2\pi$, which coincides with the peak of $A_{2D}$ and $NL_{3D2D}$, and counteracts these terms, indicating a dissipative nature of the vortices. On the other hand, for 3D modes viscosity is important only at higher $k>40$, i.e., small scales are dissipative, in contrast to that in the bursty regime, where viscous and injection scales coincide (right panels in Fig. \ref{fig:state_1_2_po0075}).

Note that the strength of all the dynamical processes depicted in Fig. \ref{fig:inj_nonlinear_re1e4d5} increases with increasing $Po$.
In all cases, the peaks of energy injections into 3D wave modes due to the precessional instability are concentrated at wavenumbers smaller than the corresponding Zeman wavenumber, $k_{in,3D}=6\pi < k_{\Omega}$ (grey vertical lines in Fig.~\ref{fig:spectra_re1e4d5}). As we have seen in the left panels of Fig. \ref{fig:inj_nonlinear_re1e4d5}, this injection affects the 2D-3D nonlinear transfer $NL_{3D2D}$, which describes driving of 2D vortices by 3D wave modes, and also has a maximum at the same $k_{in}$. This is in agreement with the general condition for the upscale/inverse energy cascade of 2D vortices towards wavenumbers smaller than the injection one, i.e. $k < k_{in}$, in rotating turbulence \cite{Buzzicotti2018}. However, in the present case of precessional driving, in the inverse cascade regime, the energy spectrum of 2D modes is slightly steeper than a $k^{-3}$ slope (Fig. \ref{fig:spectra_re1e4d5}), which is usually observed in the same regime in a purely rotating case. 

\begin{figure}[t]
\centering
\includegraphics[scale=0.45]{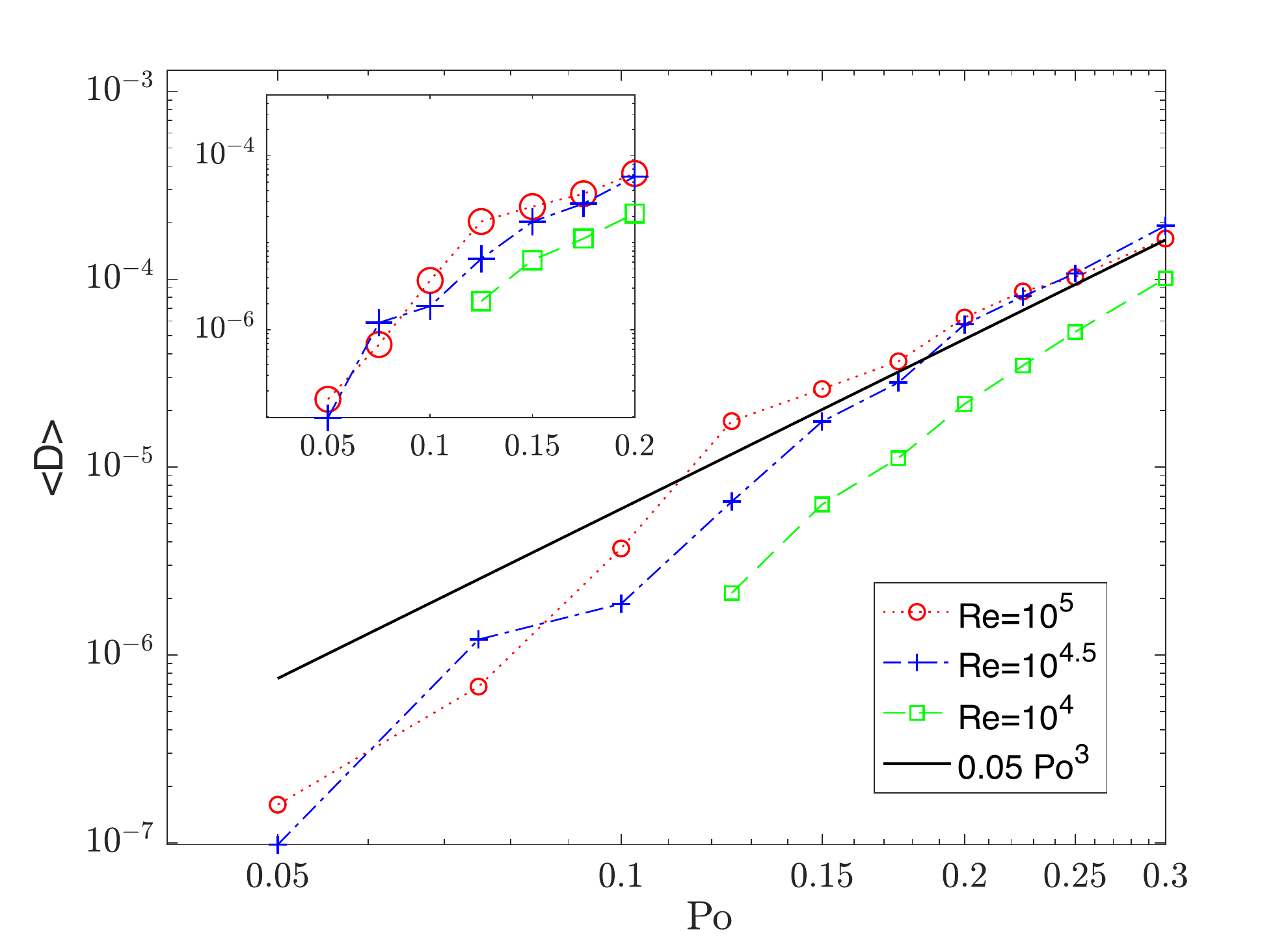}
\caption{Time and volume averaged dissipation $\langle D\rangle$ as a function of the precession parameter. The various curves represent three different Reynolds numbers and the black solid line is the scaling $\sim Po^{3}$ given for reference. Inset plot zooms into the jump around $Po \approx 0.1$ in the linear scale associated with the transition from bursty to statistically steady turbulent regimes.} \label{fig:dissipation}
\end{figure}

Overall the above-described processes of nonlinear transfers inside the 3D manifold, inside the 2D manifold, and the coupling between these two manifolds are consistent with previous spectral analysis of turbulence dynamics under rotation and an imposed external forcing \cite{Biferale2016, Buzzicotti2018, Alexakis2018}. In particular, in Fig. \ref{fig:inj_nonlinear_re1e4d5} we observe the \textit{split} (simultaneous inverse and forward) cascade of energy typical of rotating turbulence as demonstrated in those papers, that is, forward cascade of 3D wave mode energy to high wavenumbers (small-scales) due to $NL_{3D3D}$ and $NL_{2D3D}$, and inverse cascade of 2D modes to small wavenumbers (large-scales) due to $NL_{2D2D}$.

To confirm the overall type (direction) of the nonlinear cascades inferred above on the basis of the nonlinear transfers $NL$ as a function of $k$, we also analyze shell-to-shell flux of the total energy defined as \cite{Alexakis2018}
\begin{eqnarray}
F(k)= \sum_{k'\geq k} NL \left(k'\right).
\end{eqnarray}
Figure \ref{fig:flux_po02_03} shows the resulting flux function $F(k)$ and, for reference, the wavenumbers $k_{in,2D}=2\pi$ and $k_{in,3D}=6\pi$ at which the injection terms $A_{2D}$ and $A_{3D}$, respectively, reach their maximum (Fig. \ref{fig:inj_nonlinear_re1e4d5}). The grey line in this figure represents the Zeman wavenumber $k_{\Omega}$. The shape of the total fluxes are qualitatively similar for $Po=0.2$ and 0.3 and indeed display split/dual cascade: they are positive, $F>0$, at $k>k_{in,2D}$ with a maximum value around $k \approx 50$, indicating a forward cascade of energy, and negative, $F<0$, at small wavenumbers $k<k_{in,2D}$, indicating inverse cascade. These forward/inverse cascade regimes deduced from the behavior of the energy flux function $F(k)$ in fact confirm those found above based on the behavior of the transfer functions in Fig. \ref{fig:inj_nonlinear_re1e4d5}. Specifically, the forward cascade at $k>k_{in,2D}$, is related to the transfer of 3D wave mode energy to higher-$k$, while the inverse cascade at $k<k_{in,2D}$ is related to the transfer of 2D vortical mode energy to smaller-$k$.

\subsubsection{Precession forcing: a way to isotropic Kolmogorov turbulence}

In this section, we draw conclusions on the properties of 3D wave modes which are the ones directly influenced and driven by the precession instability. We have already seen clear indications that these modes exhibit characteristics of isotropicity, direct cascade and decreasing the wavenumber range where the rotation is substantially dominant, that is, decreasing the Zeman wavenumber $k_{\Omega}$, with increasing precession intensity. In order to confirm and generalize these concepts, we run another simulation for quite high precession parameter $Po=0.5$ and with the same $Re=10^{4.5}$ to check this trend.

Figure \ref{fig:3d_kolmogorov} shows the spectra of 3D mode energy for the three precession parameters. This time we compensated $E_{3D}$ spectra with the Kolmogorov spectrum $k^{-5/3}$ to better see if the energy spectrum approaches the Kolmogorov one. Indeed, it is seen in this figure that increasing $Po$, the Zeman wavenumber decreases and the compensated spectrum at $k>k_{\Omega}$ becomes gradually flatter, indicating approach to the Kolmogorov one $k^{-5/3}$ already at $Po=0.5$, that is, the regime of isotropic homogeneous turbulence. Thus, we showed that the rotating-dominated range of wavenumbers $k<k_{\Omega}$ is narrowed as $Po$ increases because of dramatic decrease in Zeman wavenumbers (e.g., $k_{\Omega} = 103$ for $Po=0.2$ reducing to $k_{\Omega}=38$ for $Po=0.5$). This is also reflected in the increase of Rossby number at the injection scales $Ro=\left(A_{2D}k_{in,2D}^2+A_{3D}k_{in,3D}^2\right)^{1/3}/{\Omega}$. Specifically, we have $Ro \approx 0.176$ for $Po=0.2$, $Ro \approx 0.24$ for $Po=0.3$ and $Ro=0.32$ for $Po=0.5$.

\subsubsection{Turbulent dissipation}

We examine the dissipative nature of the precession-driven turbulent flow. Dissipation rate is an important quantity used in both experiments and numerical works to check global changes in the flow behavior such as hysteresis cycles or transition to turbulence, resulting in noticeable increase of this quantity. Figure \ref{fig:dissipation} plots time- and volume-averaged dissipation term $\langle D\rangle$ as a function of $Po$ at different $Re$. It is seen in this figure that the turbulent dissipation more depends on $Po$ and changes only weakly with $Re$. This result is in agreement with the observations by Goto et al. \cite{Goto2014} according to which turbulence properties are mainly governed by $Po$ rather than $Re$. At larger $Po\gtrsim 0.1$, the turbulent dissipation scales with  $Po^3$ in accordance with Ref.\cite{Barker2016}. Moreover, around $Po \approx 0.1$ we observe a jump which is consistent with the global simulation results in cylindrical geometry \cite{Kong2014a, Pizzi2021a, Pizzi2021b} as well as with the local analysis \cite{Barker2016}. This jump is associated with the transition of the bursty regime at $Po\lesssim 0.1$, dominated by large-scale columnar vortices where waves and hence turbulent dissipation are relatively weak, to the quasi-steady turbulence regime at $Po\gtrsim 0.1$, where the contribution of small-scale waves is larger leading to efficient dissipation. Therefore, the well-known transition observed in precessing fluid filled cylinders (connected with the hysteresis regime \cite{Herault2015}) can be interpreted in light of the results of this work.

\begin{figure}
\includegraphics[scale=0.45]{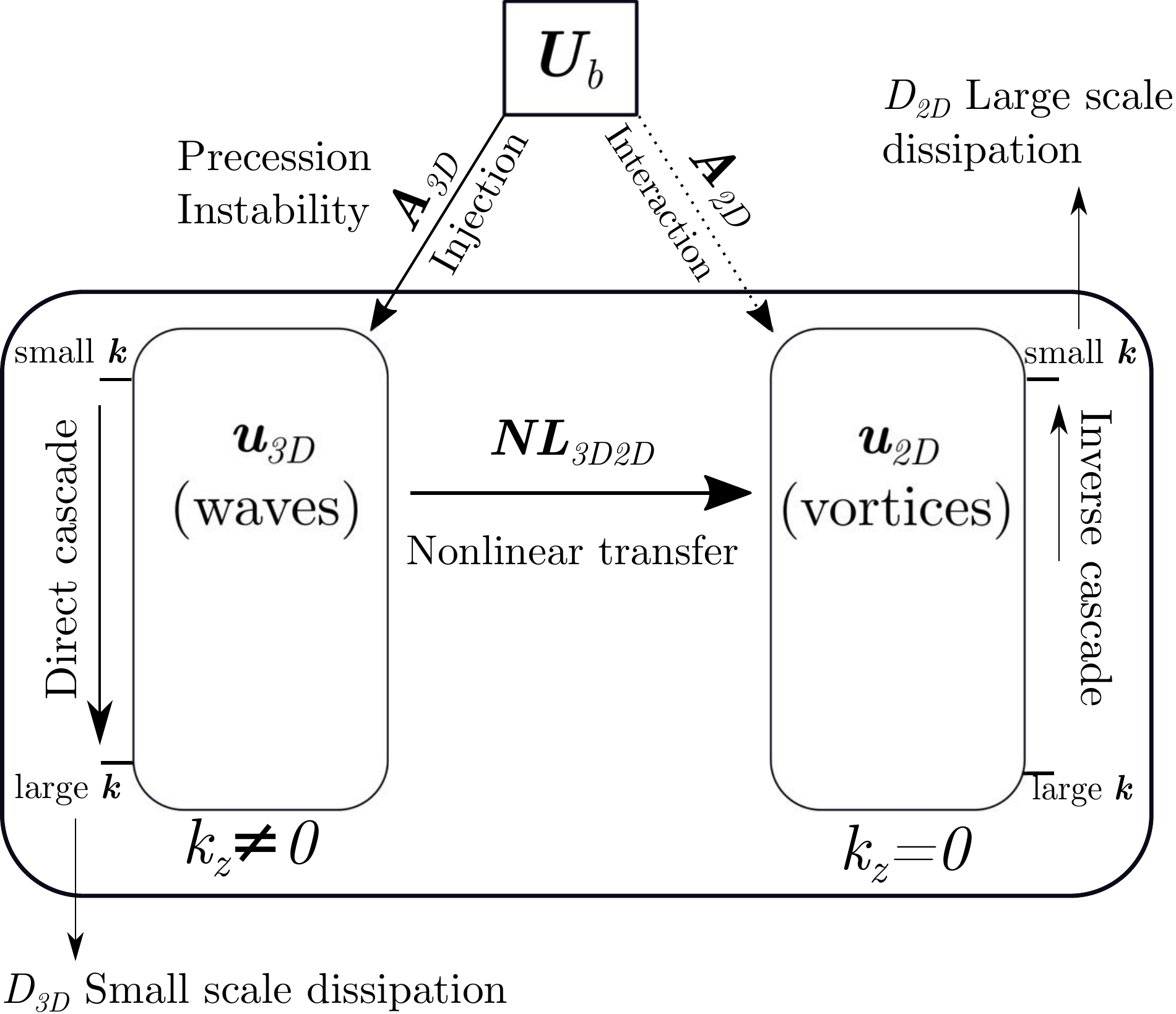}
\caption{Sketch of the main mechanism underlying precession-driven turbulence.} \label{fig:skectch_final}
\end{figure}

\section{Summary and discussions}\label{sec:conclusions}

In the present work, we have studied the properties of turbulence developed in precession-driven flows by using numerical simulations in the local model with a periodic box. Through an extended data-set of simulations, new results have been obtained concerning the role of precessional forcing modeled as a background flow which injects energy into our local patch. In this case, the precession ratio, or Poincar\'{e} number, is a crucial parameter to trigger and sustain a turbulent state in the flow, as observed in global simulations \cite{Pizzi2021b, Pizzi2021a, Kong2014a, Cebron2019} as well as in experiments \cite{Goto2007, Herault2015, Horimoto2017}. Our detailed analysis was motivated mainly by the works of Barker \cite{Barker2013, Barker2016} and it was developed both in physical and mainly in Fourier (wavenumber $\boldsymbol{k}$-) space. Precessional turbulence is a type of rotating turbulence, where energy injection comes from the precessional instability instead of an externally imposed forcing, and shares several common features with those in the presence of other forcing mechanisms such as the concurrence of waves and geostrophic structures. We have used the 2D-3D manifold decomposition method (where 2D modes have $k_{z}=0$ while 3D ones $k_{z} \neq 0$) to distinguish and quantify the vortices and the waves as used by several authors for other external forcings localized in a narrow band of wavenumbers \cite{Smith1999, Khlifi2018, Salhi2020, Buzzicotti2018, Biferale2016}. By contrast, precessional instability injects energy into turbulence over a broad range of wavenumbers which also modifies the character of nonlinear transfers compared with that in the case of external forcing. We quantified the nonlinear interactions between 3D waves and 2D geostrophic vortices, obtain the spectral scalings and determine the types of cascades in each manifold. We extended the study not just limiting to the shell-averaged approach, generalizing spectral analysis in Fourier space. In this way, we identified the anisotropic structure of these modes.
Each phenomena deserves a more extended discussion:
\begin{enumerate}
\item \textit{Different states observed}: precession forcing is responsible for the appearance of a turbulent state and the magnitude (i.e Poincar\'{e} number) determines the flow response. We have identified three different states: quasi-periodic states, characterized  by the competition between geostrophic vortex columns and 3D inertial waves at small $Po\lesssim 0.1$; intermediate states at $Po\sim 0.1$ with coexisting 2D vortices and 3D waves with comparable amplitudes, and the state dominated by smaller scale waves with some mixture of vortices at high-$Po$. At very small $Po < 0.05$, our models (with numerically accessible $Re$) are linearly stable against precessional instability and hence turbulence has not been observed. 

\item \textit{Bursting evolution}: for small precession parameters we observed a cyclic trend of the flow where the vortices appear and disappear periodically. In this regime, the precession instability injects energy in the 3D waves which they transfer directly to vortices due to nonlinearity. However, they decay due to their large dissipative character which is not counteracted by the energy taken from the 3D waves. This explanation is consistent with the analogous behavior observed for tidal elliptical instability. Indeed the bursty nature of vortices due to viscosity disappears when a hyperviscosity model is adopted, i.e., when dissipation is concentrated only at large wavenumbers \cite{Barker2013, Barker2016}.  

\item \textit{Quasi-steady turbulent states}: at moderate and large precession parameters the essential dynamical picture and balances in the precessional-driven turbulence is described in Fig. \ref{fig:skectch_final} and can be summarized as follows. The precession background flow is unstable to precession instability, whose nonlinear development causes transition to sustained turbulence. In this state, the instability injects energy in the 3D waves, which in turn, is transferred partly to 2D vortices and partly dissipated at small scales through a forward cascade. The 2D vortices receive energy from 3D waves and at the same time they interact with the background flow. These vortices are  subjected to inverse cascade which is balanced by dissipation at large scales. Their energy spectra scales as $E(k) \sim k^{-3}$ reminiscent of the typical geostrophic turbulence while 3D waves have $E(k) \sim k^{-2 \pm 0.5}$ as found in several works on the forced turbulence \cite{Muller2007, Salhi2020}. The small differences in this scalings can be attributed to the influence of precession. Overall, we observe a so-called split, or dual cascade: inverse cascade for 2D vortices and direct cascade for 3D waves. The borderline between these two types of cascade occurs near the peak of energy injection for 2D vortices (see e.g., Fig. \ref{fig:flux_po02_03}). \\
In any case, the 2D vortices represent \textit{condensates} that gain energy from smaller-scale waves without dissipating it at the same rate \cite{Boffetta2011}. Consistent with what was observed by Smith et al. \cite{Smith1999} the vortices are produced mainly by the energy transfer from 3D waves and grow in size by the 2D inverse cascade; this is a clear indication of strongly nonlinear phenomena at moderate Rossby numbers, $Ro\sim 0.1$. By contrast, the weakly nonlinear wave theory at small $Ro\ll 1$ prohibits the interaction of geostrophic vortical mode and waves \cite{Galtier2003} allowing only resonant triads between fast 3D wave modes. This scenario, sometimes called \textit{Greenspan's theorem} \cite{Greenspan1969} has led to the idea that the geostrophic flows in precessing cylinders can arise only by the nonlinear interaction in the Ekman layers at the endcaps, that is a purely boundary effect \cite{Kong2104b, zhang_liao_2017, Meunier2008} Our local model, which by definition has no boundary layers, proves that this condition in fact is not necessary, since vortices can arise also in unbounded precessional flows for moderate $Ro$. In this regard, our results are also important in relation to the recent work by LeReun \cite{Lereun_2019} who showed that the inertial waves can excite the geostrophic mode through an instability driven by near-resonant triadic nonlinear interactions.\\
The anisotropic nature of 2D vortices is demonstrated by two aspects: they have a preferential direction with the substantial part of energy being horizontal; from a spectral point of view the kinetic energy, injection and nonlinear transfer have a preferential direction in $(k_x,k_y)-$plane.

\item \textit{The role of precession parameter $Po$}: the precession, as other forcing mechanisms, counteracts the effects of rotation. This fact has been shown through several phenomena: the larger the precession ratio the stronger the 3D waves, thereby the flow is more isotropic and the vortices are weaker. The Zeman scale decreases with the precession ratio and this means that the range of rotationally-dominated wavenumbers is reduced, extending the inertial range (characterized by the direct cascade), while the range of wavenumbers where inverse cascade occurs shrinks. Finally, the increase of precession parameter brings the spectral law for 3D energy from $k^{-2}$ to the classical Kolmogorov $k^{-5/3}$. This kind of shift was proposed initially by Zhou \cite{Zhou1994} but for the transition from strong rotation to non-rotation.      

\end{enumerate}

\subsection{Importance of this study in the context of global simulations, laboratory experiments and geophysical applications}
Even though our simple model is based on a local version of the Poincar\'{e} flow (which is typical of spheroidal containers), some similarities with the global cases has been found. For instance, the dissipation shows a quite steep jump around an intermediate precession parameter $Po\sim 0.1$ consistent with the transition to turbulence observed in global simulations \cite{Goto2014, Kong2014a, Cebron2019, Pizzi2021a, Pizzi2021b} and experiments \cite{Malkus1968, Herault2015}. Moreover, the presence of a geostrophic flow which dominates the bulk region is a hallmark of precessing cylinders at rather large $Po$ \cite{Kong2014a, Kobine1996, Jiang2015} and could correspond to 2D vortices in our local model. At large precession parameters $Po\gtrsim 0.1$, the turbulent dissipation scales as $Po^3$ as found already in Ref. \cite{Barker2016} and this fact somehow challenges the analytical results by Kerswell \cite{Kerswell1996} where an upper bound on dissipation for fluid filled precessing container has been claimed to be independent of $Po$. The knowledge of the dissipation behavior at strong precessions is crucial to predict the power required to drive the experimental facilities.
The coexistence of vortices and small-scale 3D waves has been observed in a precessing sphere by Horimoto et al. \cite{Horimoto2017}, however, they argue that large vortices sustain small-scale eddies through a forward cascade. This conflicts in part with our scenario since the precession injects energy directly into small-scale 3D waves, which in turn nonlinearly transfer energy to 2D vortices. Therefore, in our model, precession itself sustains small-scale waves.\\
One of the main goals of this work was to put a theoretical basis for the analysis of turbulence properties in precession-driven flows in the context of the upcoming DRESDYN (DREsden Sodium facility for DYNamo and thermohydraulic studies) precession experiment \cite{Stefani2012, stefani2012dresdyn, Stefani2019}. This motivates the interest in the  moderate to large $Po$, which are different from the ones of geophysical and astrophysical objects. However, some speculations can be made since the different regimes observed here at $Re=10^{4.5}$ may carry over to large-$Re$ regime too. Because of normally weak precession of geo- and astrophysical objects, we can speculate that they would be also characterized by the bursty behavior as described in Section \ref{sec:bursts}. If this is the case, it would influence the planetary evolution, producing a series of formation and destruction events (bursts) of vortices due to the nonlinear transfer between 2D and 3D flows and oscillating dissipation.

Let us finish with some discussions related to the magnetohydrodynamic (MHD) dynamo effect. It has been demonstrated numerically that precession can in general drive dynamo \cite{Tilgner2005, Wu2009, Nore2011, Cappanera2016, Giesecke2018, Goepfert2016,Lin2016}. Within our local model, we can further investigate the properties of MHD turbulence and related dynamo action and how the magnetic field, when sufficiently strong, influences the studied here 2D and 3D flows. The work by Barker \cite{Barker2016} indicates that the precession instability is able in principle to drive dynamo action locally and the turbulent flow dynamics changes completely due to the back-reaction of the magnetic field.

\begin{acknowledgments} This project has received funding from the European Research Council (ERC) under the European Union's Horizon 2020 research and innovation program (grant agreement No 787544). AJB was funded by STFC grants ST/S000275/1 and ST/W000873/1.
\end{acknowledgments}

\bibliography{aipsamp}

\end{document}